\documentclass[12pt]{article}
\pdfoutput=1

\usepackage{putex}
\usepackage{graphicx}
\usepackage{caption}
\usepackage{subcaption}
\usepackage{epstopdf}
\usepackage{enumerate}
\usepackage{cite}
\usepackage{tensor}
\usepackage{slashed}
\usepackage[aligntableaux=center]{ytableau}
\usepackage{bbm} 
\usepackage{amsthm}
\usepackage{array}
\usepackage{mathtools}
\usepackage{dsfont}
\usepackage{multirow}

\numberwithin{equation}{section}

\newcommand{\abs}[1]{\left\lvert #1 \right\rvert}

\newcommand {\be} {\begin {equation}}
\newcommand {\ee} {\end {equation}}

\newcommand {\bes} {\begin {equation*}}
\newcommand {\ees} {\end {equation*}}

\newcommand{\es}[2] {\begin{equation} \label{#1} \begin{split} #2 \end{split} \end{equation}}

\newcommand{\beq}{\begin{equation}}
\newcommand{\eeq}{\end{equation}}

\def\<{\langle}
\def\>{\rangle}

\newcommand{\veps}{\varepsilon}

\newcommand{\ed}{\,.}
\newcommand{\ec}{\,,}
\newcommand{\ecq}{\ec\quad}

\newcommand{\red}[1]{\textcolor{red}{#1}}

\newcommand{\bR}{\ensuremath{\mathbb{R}}}

\newcommand{\cA}{\ensuremath{\mathcal{A}}}
\newcommand{\cB}{\ensuremath{\mathcal{B}}}
\newcommand{\cC}{\ensuremath{\mathcal{C}}}
\newcommand{\cD}{\ensuremath{\mathcal{D}}}

\newcommand{\cF}{\ensuremath{\mathcal{F}}}
\newcommand{\cG}{\ensuremath{\mathcal{G}}}
\newcommand{\cH}{\ensuremath{\mathcal{H}}}

\newcommand{\cN}{\ensuremath{\mathcal{N}}}
\newcommand{\cO}{\ensuremath{\mathcal{O}}}

\newcommand{\cQ}{\ensuremath{\mathcal{Q}}}
\newcommand{\cR}{\ensuremath{\mathcal{R}}}
\newcommand{\cS}{\ensuremath{\mathcal{S}}}

\begin{document}

\preprint{PUPT-2465\\ MIT-CTP-4559}

\institution{Princeton}{Joseph Henry Laboratories, Princeton University, Princeton, NJ 08544, USA}
\institution{MITCTP}{Center for Theoretical Physics, Massachusetts Institute of Technology, Cambridge, MA 02139, USA}

\title{The ${\cal N} = 8$ Superconformal Bootstrap in Three Dimensions}

\authors{Shai M.~Chester,\worksat{\Princeton}\footnote{e-mail: {\tt schester@Princeton.EDU}} Jaehoon Lee,\worksat{\MITCTP}\footnote{e-mail: {\tt jaehlee@MIT.EDU}} Silviu S.~Pufu,\worksat{\Princeton}\footnote{e-mail: {\tt spufu@Princeton.EDU}} and Ran Yacoby\worksat{\Princeton}\footnote{e-mail: {\tt ryacoby@Princeton.EDU}}}

\abstract{We analyze the constraints imposed by unitarity and crossing symmetry on the four-point function of the stress-tensor multiplet of $\cN=8$ superconformal field theories in three dimensions.  We first derive the superconformal blocks by analyzing the superconformal Ward identity. Our results imply that the OPE of the primary operator of the stress-tensor multiplet with itself must have parity symmetry.  We then analyze the relations between the crossing equations, and we find that these equations are mostly redundant. We implement the independent crossing constraints numerically and find bounds on OPE coefficients and operator dimensions as a function of the stress-tensor central charge. To make contact with known $\cN=8$ superconformal field theories, we compute this central charge in a few particular cases using supersymmetric localization. For limiting values of the central charge, our numerical bounds are nearly saturated by the large $N$ limit of ABJM theory and also by the free $U(1)\times U(1)$ ABJM theory.  }
\date{December, 2014}

\maketitle

\tableofcontents

\setlength{\unitlength}{1mm}

\newpage
\section{Introduction}
\label{intro}

The conformal bootstrap \cite{Polyakov:1974gs, Ferrara:1973yt, Mack:1975jr} is an old idea that uses the associativity of the operator algebra to provide an infinite set of constraints on the operator dimensions and the operator product expansion (OPE) coefficients of abstract conformal field theories (CFTs).  For two-dimensional CFTs, this idea was used to compute the correlation functions of the minimal models \cite{Belavin:1984vu} and of Liouville  theory \cite{Zamolodchikov:1995xb}. In more than two dimensions, conformal symmetry is much less restrictive, and as a consequence it is difficult to extract such detailed information from the bootstrap.

Recently, it has been shown by the authors of \cite{Rattazzi:2008pe} that the constraints arising from the conformal bootstrap can be reformulated as a numerical problem.\footnote{See also \cite{Gliozzi:2014jsa,Gliozzi:2013ysa} for a different recent method.} This provides a new method to exclude CFTs with a large enough gap in the operator spectrum and to obtain non-perturbative bounds on certain OPE coefficients \cite{Rychkov:2011et, ElShowk:2012ht, Nakayama:2014lva, Rattazzi:2008pe, Rychkov:2009ij, Rattazzi:2010gj, Poland:2010wg, Rattazzi:2010yc, Vichi:2011ux, Poland:2011ey, Nakayama:2014yia, Kos:2013tga, El-Showk:2013nia, El-Showk:2014dwa, Berkooz:2014yda, Beem:2013qxa, Alday:2013opa, Alday:2014qfa, Alday:2013bha, Beem:2013hha, Bashkirov:2013vya}. In addition, the operator spectrum and OPE coefficients of CFTs that saturate these bounds can be determined numerically \cite{ElShowk:2012hu}. The CFTs analyzed through this method so far consist of non-supersymmetric CFTs with various global symmetries in dimensions three \cite{Rychkov:2011et, ElShowk:2012ht, Nakayama:2014lva, El-Showk:2014dwa}, four \cite{Rattazzi:2008pe, Rychkov:2009ij, Rattazzi:2010gj, Poland:2010wg, Rattazzi:2010yc, Vichi:2011ux, Poland:2011ey}, five \cite{Nakayama:2014yia}, more general fractional dimensions \cite{Liendo:2012hy,Kos:2013tga, El-Showk:2013nia}, and also of boundary CFTs \cite{Liendo:2012hy,Gaiotto:2013nva}.  The numerical bootstrap was also applied to 4-d superconformal field theories (SCFTs) with (minimal) ${\cal N} =1$ supersymmetry \cite{Poland:2010wg, Vichi:2011ux, Poland:2011ey, Berkooz:2014yda} and with (maximal) ${\cal N} = 4$ supersymmetry \cite{Beem:2013qxa, Alday:2013opa, Alday:2014qfa, Alday:2013bha, Beem:2013hha}, and to 3-d SCFTs with ${\cal N} = 1$ supersymmetry \cite{Bashkirov:2013vya}.

The goal of this paper is to set up and develop the conformal bootstrap program in three-dimensional SCFTs with ${\cal N} = 8$ supersymmetry, which is the largest amount of supersymmetry in three dimensions.  There are only a few infinite families of such theories that have been constructed explicitly, and they can all be realized as Chern-Simons (CS) theories with a product gauge group $G_1 \times G_2$, coupled to two matter hypermultiplets transforming in a bifundamental representation.  These families are:\footnote{It is believed that non-trivial $\cN=8$ SCFTs such as some of the ones listed here can be obtained by taking the infrared (IR) limit of $\cN=8$ supersymmetric Yang-Mills (SYM) theories.  In particular, the $U(N)$ and $O(2N)$ SYM theories are believed to flow in the IR to the $U(N)\times U(N)$ ABJM theories with levels $k=1$ and $k=2$, respectively. In addition, $SO(2N+1)$ SYM is believed to flow to the $U(N+1)_2\times U(N)_{-2}$ ABJ theory.  See, for instance, \cite{Kim:2009wb,Kapustin:2010xq,Bashkirov:2010kz,Bashkirov:2011pt,Gang:2011xp,Bianchi:2012ez} for evidence of the above relations between ABJ(M) theories and $\cN=8$ SYM\@.  We thank O.~Aharony for emphasizing this point to us.}  the $SU(2)_k \times SU(2)_{-k}$ reformulation \cite{VanRaamsdonk:2008ft, Bandres:2008vf} of the theories of Bagger-Lambert-Gustavsson (BLG) \cite{Bagger:2007vi,Bagger:2007jr,Bagger:2006sk, Gustavsson:2007vu}, which are indexed by an arbitrary integer Chern-Simons level $k$;  the $U(N)_k \times U(N)_{-k}$ theories of Aharony-Bergman-Jafferis-Maldacena (ABJM) \cite{Aharony:2008ug}, which are labeled by the integer $N$ and $k = 1, 2$; and the $U(N+1)_2 \times U(N)_{-2}$ theories \cite{Bashkirov:2011pt} of Aharony-Bergman-Jafferis (ABJ) \cite{Aharony:2008gk}, which are labeled by the integer $N$.\footnote{The invariance of the ABJM and ABJ theories under the ${\cal N} = 8$ superconformal algebra, $\mathfrak{osp}(8|4)$, is not visible at the classical level, but an enhancement to ${\cal N} = 8$ is expected at the quantum level. The arguments for this symmetry enhancement are based partly on M-theory \cite{Aharony:2008ug} and partly on field theory \cite{Benna:2009xd, Gustavsson:2009pm,Kwon:2009ar,Bashkirov:2010kz}.} 

That there should exist SCFTs with $\mathfrak{osp}(8|4)$ global symmetry had been anticipated from the $AdS_4$/CFT$_3$ correspondence.  Indeed, the $AdS_4 \times S^7$ background of eleven-dimensional supergravity was conjectured to be dual to an ${\cal N} = 8$ SCFT in three dimensions, describing the infrared limit of the effective theory on $N$ coincident M2-branes in flat space, in the limit of large $N$.  The ABJM theory with CS level $k=1$ is an explicit realization of this effective theory\footnote{The ABJM and ABJ theories have M-theory interpretation for any $N$ and $k$. For some special values of $k$ the BLG theories were argued to be isomorphic to ABJM and ABJ theories with $N=2$ \cite{Lambert:2010ji,Bashkirov:2011pt}. However, for general $k$ the BLG theories have no known M-theory interpretation.} that is believed to be correct for any $N$.  When $N=1$ the theory becomes free, as the interaction potential between the matter fields vanishes and the gauge interactions are trivial \cite{Klebanov:2009sg}.  At large $N$, ABJM theory is strongly coupled, but it can be studied through its supergravity dual, which is weakly coupled in this limit. The duality with the supergravity description has passed impressive tests, such as a match in the $N^{3/2}$ behavior of the number of degrees of freedom \cite{Drukker:2010nc, Herzog:2010hf}.  At finite $N > 1$, both ABJM theory and its supergravity dual are strongly interacting and not much detailed information is available. In this paper we aim to uncover such information by using the conformal bootstrap. Indeed, our bootstrap study provides us, indirectly, with non-perturbative information about M-theory.

At some level, our work parallels that of \cite{Beem:2013qxa}, who developed the numerical bootstrap program in four-dimensional theories with ${\cal N} = 4$ superconformal symmetry.   The authors of \cite{Beem:2013qxa} studied the implications of unitarity and crossing symmetry on the four-point function of the superconformal primary operator ${\cal O}_{\bf 20'}$ of the ${\cal N} = 4$ stress-tensor multiplet.\footnote{The OPE of the stress-tensor multiplet in $\cN=4$ SYM was first analyzed in \cite{Arutyunov:2000ku,Arutyunov:2000im,Arutyunov:2001mh,Arutyunov:2001qw}.}  This superconformal primary is a Lorentz scalar that transforms in the ${\bf 20'}$ irrep under the $\mathfrak{so}(6)$ R-symmetry.  In the present work, we study the analogous question in three-dimensional ${\cal N} = 8$ SCFTs. In particular, we analyze the four-point function of the superconformal primary ${\cal O}_{{\bf 35}_c}$ of the ${\cal N} = 8$ stress-tensor multiplet.  This superconformal primary is a Lorentz scalar transforming in the ${{\bf 35}_c}$ irrep of the $\mathfrak{so}(8)$ R-symmetry.\footnote{That this operator transforms in the ${\bf 35}_c$ as opposed to ${\bf 35}_v$ or ${\bf 35}_s$ is a choice that we make.   See the beginning of Section~\ref{conventions}.}

Upon using the OPE, the four-point function of ${\cal O}_{{\bf 35}_c}$ can be written as a sum of contributions, called superconformal blocks, coming from all superconformal multiplets that appear in the OPE of ${\cal O}_{{\bf 35}_c}$ with itself.  In addition, this four-point function can be decomposed into the six R-symmetry channels corresponding to the $\mathfrak{so}(8)$ irreps that appear in the product ${\bf 35}_c \otimes {\bf 35}_c$. Generically, each superconformal multiplet contributes to all six R-symmetry channels.  These superconformal blocks can be determined by analyzing the superconformal Ward identity written down in \cite{Dolan:2004mu}.   Crossing symmetry then implies six possibly independent equations that mix the R-symmetry channels amongst themselves.  The situation described here is analogous to the case of 4-d ${\cal N} = 4$ theories where one also has six R-symmetry channels and, consequently, six possibly independent crossing equations.

There are a few significant differences between our work and that of  \cite{Beem:2013qxa} that are worth emphasizing:
 \begin{itemize}
   \item In the case of 4-d ${\cal N} = 4$ theories, the crossing equations contain a closed subset that yields a ``mini-bootstrap'' program, which allows one to solve for the BPS sector of the theory. In the 3-d ${\cal N} = 8$ case, we do not know of any such closed subset of the crossing equations that might allow one to solve for the BPS sector. 
   
   \item  At a more technical level, in the case of 4-d ${\cal N} = 4$ theories, the solution to the superconformal Ward identity involves algebraic relations between the six R-symmetry channels. As a consequence, after solving for the BPS sector, it turns out that the six crossing equations reduce algebraically to a single  equation. In 3-d, the solution to the superconformal Ward identity can be written formally in terms of non-local operators acting on a single function \cite{Dolan:2004mu}.   As we will show, despite the appearance of these non-local operators, the various R-symmetry channels can be related to one another with the help of local second order differential operators.  These relations show that the six crossing equations are mostly redundant, but still no single equation implies the others, as was the case in 4-d.
   
   \item As in the 4-d case, we will parameterize our abstract 3-d theories by the central charge $c_T$, which is defined as the coefficient of the stress-tensor two-point function (in some normalization). In 4-d, $c_T$ is the Weyl anomaly coefficient, which allows one to determine it rather easily in particular realizations of the 4-d $\cN=4$ theory.  In 3-d, there is no Weyl anomaly, so in order to connect our bootstrap study to more conventional descriptions in terms of BLG and/or  ABJ(M) theory, we calculate $c_T$ for several of these theories using the supersymmetric localization results of \cite{Jafferis:2010un, Closset:2012vg}.  

 \end{itemize}

The remainder of this paper is organized as follows.  In Section~\ref{conventions} we set up our conventions and review the constraints on the four-point function of ${\cal O}_{{\bf 35}_c}$.  In Section~\ref{CROSSING} we write the crossing equations and describe the differential relations that they satisfy.  Section~\ref{scblocks} is devoted to the derivation of the superconformal blocks building on the results of \cite{Dolan:2004mu}.    In preparation for our numerical results, in Section~\ref{centralcharge} we calculate the coefficient $c_T$ for several explicit ${\cal N} = 8$ SCFTs using supersymmetric localization.  In Section~\ref{numerics} we study the crossing equations using the semi-definite programing method introduced in \cite{Poland:2011ey} and present our findings.  We obtain quite stringent non-perturbative bounds on scaling dimensions of operators belonging to long multiplets and on OPE coefficients.  We provide several checks of our results in the free theory (namely the $U(1)_k \times U(1)_{-k}$ ABJM theory) and in the limit of large $c_T$.  We end with a discussion of our results in Section~\ref{CONCLUSIONS}.  Several technical details are delegated to the Appendices.

\section{Constraints from Global Symmetry}
\label{conventions}

Let us start with a short review of some general properties of the four-point function of the stress-tensor multiplet in an ${\cal N} = 8$ SCFT, and of the constraints imposed on it by the $\mathfrak{osp}(8|4)$ superconformal algebra.

In any ${\cal N} = 8$ SCFT, the stress tensor sits in a half-BPS multiplet, whose members are listed in Table~\ref{Stress}.  These include the spin-$3/2$ super-current, which in our convention\footnote{We use the notation $\mathbf{8}_v = [1000]$, $\mathbf{8}_c = [0010]$, and $\mathbf{8}_s = [0001]$.  In our convention, the supercharges transform in the $\mathbf{8}_v$ of $\mathfrak{so}(8)$.} transforms (like the supercharges) in the ${\bf 8}_v$ of the $\mathfrak{so}(8)$ R-symmetry; and the spin-$1$ R-symmetry current, which transforms in the adjoint (i.e.~the ${\bf 28}$) of $\mathfrak{so}(8)_R$.   In addition, the multiplet contains a spin-$1/2$ operator transforming in the ${\bf 56}_v$, and two spin-$0$ operators with scaling dimension $1$ and $2$, which transform (in our conventions) in the ${\bf 35}_c$ and ${\bf 35}_s$, respectively.  
\begin{table}[htdp]
\begin{center}
 \begin{tabular}{c|c|r}
  dimension & spin & $\mathfrak{so}(8)$ irrep \\
  \hline
  $1$ & $0$ & ${\bf 35}_c = [0020]$ \\
  $3/2$ & $1/2$ & ${\bf 56}_v = [0011]$ \\
  $2$ & $0$ & ${\bf 35}_s = [0002]$ \\
  $2$ & $1$ & ${\bf 28} = [0100]$ \\
  $5/2$ & $3/2$ & ${\bf 8}_v = [1000]$ \\
  $3$ & $2$ & ${\bf 1} = [0000]$
 \end{tabular}
\end{center}
\caption{The operators comprising the ${\cal N} = 8$ stress-tensor multiplet along with their scaling dimension, spin, and R-symmetry representation.}
\label{Stress}
\end{table}%

The dimension-one operator in the ${\bf 35}_c$ is the superconformal primary, which we will denote by $\cO_{\mathbf{35}_c}$.  In the rest of this paper we will only consider the four-point function of this superconformal primary.  The four-point functions of other members of the stress-tensor multiplet can be obtained from the one of $\cO_{{\bf 35}_c}$ with the help of the $\mathfrak{osp}(8|4)$ algebra. The constraints of superconformal invariance on the four-point functions of $1/2$-BPS operators in various dimensions were analyzed in detail in~\cite{Dolan:2004mu}.  The rest of this section reviews some details of~\cite{Dolan:2004mu} that are relevant for our case of interest.

\subsection{Constraints from Conformal Symmetry and R-symmetry}

Let us start by reviewing the constraints arising from the maximal $\mathfrak{so}(3,2)\oplus \mathfrak{so}(8)_R$ bosonic subalgebra of $\mathfrak{osp}(8|4)$.

The ${\bf 35}_c$ of $\mathfrak{so}(8)_R$ can be identified as the rank-two symmetric traceless product of the ${\bf 8}_c$.  It is convenient to analyze any such symmetric traceless products by introducing polarization vectors $Y^i$, $i = 1, \ldots, 8$, whose indices we can contract with the ${\bf 8}_c$ indices to form $\mathfrak{so}(8)_R$ invariants.  For instance, for an ${\bf 8}_c$ vector $\psi_i$ we should define $\psi = \psi_i Y^i$;  for a rank-two tensor ${\cal O}_{ij}$, as is the case for our operator, we should consider
 \es{cO35}{
  {\cal O} = {\cal O}_{ij} Y^i Y^j \,;
 }
and so on.  

The $Y^i$ should be thought of as a set of auxiliary commuting variables, and they are required to satisfy the null condition $Y \cdot Y \equiv \sum_{i=1}^8 Y^i Y^i = 0$.  Commutativity is related to the fact that the tensor ${\cal O}_{ij}$ is symmetric, while the null condition is connected to its tracelessness.  The advantage of introducing the polarization vectors is that instead of keeping track of the various $\mathfrak{so}(8)_R$ tensor structures that can appear in correlation functions, one can just construct all possible $\mathfrak{so}(8)_R$ invariants out of the polarizations. 

Invariance of our SCFT under $\mathfrak{so}(3,2)$ implies that the four-point function of ${\cal O}_{{\bf 35}_c}$ evaluated at space-time points $x_m$, with $m = 1, \ldots, 4$, should take the form
\es{so32}{
  \langle {\cal O}_{i_1 j_1 }(x_1) {\cal O}_{i_2 j_2}(x_2)  {\cal O}_{i_3 j_3}(x_3)  {\cal O}_{i_4 j_4}(x_4)\rangle = \frac{1}{x_{12}^2 x_{34}^2} G_{i_1\cdots i_4 j_1\cdots j_4} (u, v) \,,
 }
where $x_{mn}^2 = (x_m - x_n)^2$, and $u$ and $v$ are the conformally-invariant cross-ratios
 \es{uvDef}{
  u\equiv \frac{x_{12}^2 x_{34}^2}{x_{13}^2 x_{24}^2}\,, \qquad v\equiv \frac{x_{14}^2 x_{23}^2}{x_{13}^2 x_{24}^2} \,.
 } 
The four-point function of $\cO_{{\bf 35}_c}(x,Y) ={\cal O}_{ij}(x) Y^i Y^j$ can be written as a quadratic polynomial in each $Y$ coordinate. Furthermore, invariance under $\mathfrak{so}(8)_R$ implies that these polynomials should only depend on the $\mathfrak{so}(8)_R$ invariant combinations $Y_n \cdot Y_m$, which are non-zero when $n \neq m$.  In other words,
 \es{4pnt}{
    \langle \cO_{{\bf 35}_c}(x_1,Y_1) \cO_{{\bf 35}_c}(x_2,Y_2) 
      \cO_{{\bf 35}_c}(x_3,Y_3) \cO_{{\bf 35}_c}(x_4,Y_4)\rangle 
      = \frac{(Y_1\cdot Y_2)^2 (Y_3\cdot Y_4)^2}{x_{12}^2 x_{34}^2} \cG(u,v;U,V) \,,
 }
where ${\cal G}$ is a quadratic polynomial in $1/U$ and $V/U$, $U$ and $V$ being the cross-ratios
 \es{UV}{
   U\equiv \frac{Y_1\cdot Y_2 \, Y_3\cdot Y_4}{Y_1\cdot Y_3 \, Y_2\cdot Y_4}\,, \qquad   V\equiv \frac{Y_1\cdot Y_4 \, Y_2\cdot Y_3}{Y_1\cdot Y_3 \, Y_2\cdot Y_4} \,.
 }

Being a quadratic polynomial in $1/U$ and $V/U$, ${\cal G}(u, v; U, V)$ contains six distinct functions of $u$ and $v$. It is helpful to exhibit explicitly these six functions by writing 
 \es{GExpansion}{
  {\cal G}(u, v; U, V) = \sum_{a=0}^2 \sum_{b = 0}^a A_{ab}(u, v) Y_{ab}(1/U, V/U) \,,
 }
where the quadratic polynomials $Y_{ab}(\sigma, \tau)$ are defined as
 \es{polyns}{
   Y_{00}(\sigma, \tau) &= 1 \,, \\
   Y_{10}(\sigma, \tau) &= \sigma - \tau \,, \\
   Y_{11}(\sigma, \tau) &= \sigma + \tau -\frac{1}{4} \,, \\
   Y_{20}(\sigma, \tau) &= \sigma^2 + \tau^2 - 2\sigma\tau - \frac{1}{3}(\sigma + \tau) + \frac{1}{21} \,,\\
   Y_{21}(\sigma, \tau) &= \sigma^2 - \tau^2 - \frac{2}{5}(\sigma - \tau) \,,\\
   Y_{22}(\sigma, \tau) &= \sigma^2 + \tau^2 + 4\sigma\tau - \frac{2}{3}(\sigma+\tau) + \frac{1}{15} \,.
 }
The definition \eqref{polyns} could be regarded simply as a convention.  It has, however, a more profound meaning in terms of the $\mathfrak{so}(8)_R$ irreps that appear in the $s$-channel of the four-point function \eqref{4pnt}.  We have 
 \es{35times35}{
  {\bf 35}_c \otimes {\bf 35}_c = {\bf 1} \oplus {\bf 28} \oplus {\bf 35}_c \oplus {\bf 300} \oplus {\bf 567}_c \oplus {\bf 294}_c \,.
 }
The six polynomials\footnote{The polynomials in \eqref{polyns} are harmonic polynomials, which are eigenfunctions of the $\mathfrak{so}(8)_R$ Casimir. More details on these polynomials can be found in \cite{Dolan:2003hv,Nirschl:2004pa}.} in \eqref{polyns} correspond, in order, to the six terms on the right-hand side of \eqref{35times35}.  In terms of Dynkin labels, the indices $(a,b)$ correspond to the irrep $[0\, (a-b)\, (2b)\, 0]$.  

The irreps ${\bf 28} = [0100] = (1, 0)$ and ${\bf 567}_c = [0120] = (2, 1)$ are in the anti-symmetric product of the two copies of ${\bf 35}_c$, while the other irreps are in the symmetric product.  Therefore only operators belonging to the $\cO_{{\bf 35}_c}(x_1, Y_1) \times \cO_{{\bf 35}_c}(x_2, Y_2)$ OPE with odd integer spin can contribute to the $[0100]$ and $[0120]$ channels.  The other R-symmetry channels receive contributions only from operators with even integer spin.

\subsection{Constraints from Supersymmetry}

The full $\mathfrak{osp}(8|4)$ superconformal algebra imposes additional constraints on \eqref{GExpansion}.  For the purpose of writing down these constraints, it is convenient to introduce the following parameterization of the cross-ratios in terms of the variables $x,\bar{x}$ and $\alpha, \bar{\alpha}$:
\begin{alignat}{3}
u &= x\bar{x} \ecq &\qquad  v &= (x-1)(\bar{x}-1) \ec \label{xvars}\\
U &= \frac{1}{\alpha\bar{\alpha}} \ecq &\qquad V &= \frac{(\alpha-1)(\bar{\alpha}-1)}{\alpha\bar{\alpha}} \label{alphavars}\ed
\end{alignat}
In this parameterization, the function $\cG(x,\bar{x}; \alpha,\bar{\alpha})$ appearing in \eqref{4pnt} as well as the function $A_{ab}(x, \bar x)$ appearing in \eqref{GExpansion} should be taken to be symmetric under the interchanges $x\leftrightarrow\bar{x}$ and $\alpha\leftrightarrow\bar{\alpha}$.  As shown in \cite{Dolan:2004mu}, in terms of the variables \eqref{xvars} and \eqref{alphavars} the superconformal Ward identity takes a particularly neat form.

In \cite{Dolan:2004mu}, the superconformal Ward identity of the four-point function of $1/2$-BPS operators was written down for any theory with $\mathfrak{so}(n)$ R-symmetry in space-time dimension $d$ with $3 \leq d \leq 6$.  It takes the form
 \es{GenWard}{
   \left.\left( x \partial_x - \veps\, \alpha\partial_{\alpha}\right) 
      \cG(x,\bar{x};\alpha,\bar{\alpha})\right|_{\alpha = 1/x} &= 0 \,, \\
   \left.\left( \bar{x}\partial_{\bar{x}} - \veps\, \bar{\alpha}\partial_{\bar{\alpha}}\right) 
      \cG(x,\bar{x};\alpha,\bar{\alpha})\right|_{\bar{\alpha} =  1/\bar{x}} &= 0 \,,
 }
where $\veps \equiv (d-2)/2$ is the scaling dimension of a free scalar field in $d$ space-time dimensions.

The solution of the superconformal Ward identity \eqref{GenWard} depends quite significantly on the parameter $\veps$.  In the case $\veps = 1$, which would apply to four-dimensional SCFTs with ${\cal N} = 4$ supersymmetry, the solution is very simple.  Indeed, in this case, \eqref{GenWard} reduces to $\partial_x \cG(x,\bar{x};1/x,\bar{\alpha}) = \partial_{\bar x} \cG(x,\bar{x};\alpha,1/\bar{x}) = 0$, as can be seen from using the chain rule.  Therefore, $\cG(x,\bar{x};1/x,\bar{\alpha})$ is independent of $x$ and $\cG(x,\bar{x};\alpha,1/\bar{x})$ is independent of $\bar x$. 
Taking into account the fact that ${\cal G}$ is symmetric w.r.t. interchanging $x$ with $\bar x$ and $\alpha$ with $\bar \alpha$, and that ${\cal G}$ is a quartic polynomial in $\alpha$ and $\bar \alpha$, as follows from the definition \eqref{alphavars}, one can write the general solution for the four-point function as \cite{Dolan:2001tt, Nirschl:2004pa}
 \es{G4d}{
    &\cG_{d=4}(x,\bar{x};\alpha,\bar{\alpha}) 
      = (x\alpha-1)(\bar{x}\alpha-1)(x\bar{\alpha}-1)(\bar{x}\bar{\alpha}-1) A(x,\bar{x}) -C \\
     &+ \frac{ (\bar{x}\alpha-1)(x\bar{\alpha}-1) \left[F(x,\alpha)+F(\bar{x},\bar{\alpha})\right] 
        - (x\alpha-1)(\bar{x}\bar{\alpha}-1)\left[F(x,\bar{\alpha})+F(\bar{x},\alpha)\right]}
        {(x-\bar{x})(\alpha-\bar{\alpha})} \,,
 }
where $A(x,\bar{x})$, $F(x,\alpha)$, and $C$ are arbitrary.  In other words, apart from the restricted function $F(x, \alpha)$ and the constant $C$, which can be determined in terms of the anomaly coefficient $c$ \cite{Beem:2013qxa}, the whole four-point function $\cG_{d=4}(x,\bar{x};\alpha,\bar{\alpha})$ can be written in terms of a single function $A(x, \bar x)$.  The six R-symmetry channels in this case are related algebraically.

The solution of the superconformal Ward identity for arbitrary $\veps$, and in particular for $\veps = 1/2$, can be written in terms of powers of the differential operator
\begin{align}
  \cD_{\veps} &\equiv \frac{\partial^2}{\partial x\partial\bar{x}} 
   -  \frac{\veps}{x-\bar{x}}\left(\frac{\partial}{\partial x} - \frac{\partial}{\partial\bar{x}}\right) \ed \label{Deps} 
\end{align}
For non-integer $\veps$, the general solution\footnote{The solution \eqref{wardsol} corresponds to the four-point function of $1/2$-BPS operators which are rank-$2$ symmetric traceless tensors of $\mathfrak{so}(n)_R$. The solution for tensors of arbitrary rank can also be written in a similar way, but it depends on more undetermined functions. The reader is referred to \cite{Dolan:2004mu} for more details.} of \eqref{GenWard} can be written, formally, in terms of a single arbitrary function $a(x,\bar{x})$ as\footnote{The function $a(x, \bar x)$ that appears in this equation equals $(x \bar x)^{\veps -1} a(x, \bar x)$ in the notation of \cite{Dolan:2004mu}.}
 \es{wardsol}{
   \cG(x,\bar{x};\alpha,\bar{\alpha}) 
      = (x\bar{x})^{2 \veps} \left( \cD_\veps\right)^{\veps-1} 
      \left[(x\alpha-1)(\bar{x}\alpha-1)(x\bar{\alpha}-1)(\bar{x}\bar{\alpha}-1) a(x,\bar{x})\right] \,.
 }
The appearance of the operator $\left( \cD_\veps\right)^{\veps-1}$, which is non-local for non-integer $\veps$, makes using \eqref{wardsol} rather subtle.  However, we can demystify the operator $\cD_\veps$ and its non-integer powers by interpreting $\cD_\veps$ as the Laplacian in $d = 2 (\veps + 1)$ dimensions.\footnote{That ${\cal D}_\veps$ is the Laplacian in $d = 2(\veps + 1)$ dimensions was first observed by Dolan and Osborn in~\cite{Dolan:2011dv}.}

Using conformal transformations we can fix three of the coordinates of the four-point function on a line, such that:  $x_1 = 0$, $x_3 = (0,\ldots,0, 1) \equiv \hat{z}$ and $x_4=\infty$.  (We denote the unit vector $(0, \ldots, 0, 1) \in \bR^d$ by $\hat z$ because we will eventually be interested in working in three dimensions where we denote the third coordinate by $z$.)   We write the remaining unfixed point $x_2 \equiv \vec{r} \in \bR^d$ in spherical coordinates $\vec{r} = (r, \theta, \Omega_{d-2})$, where $\theta$ is the angle between $\vec{r}$ and $\hat{z}$, and $\Omega_{d-2}$ parameterizes $S^{d-2}$. The four-point function does not depend on $\Omega_{d-2}$ because of the additional rotation symmetry which fixes the line determined by $x_1$, $x_3$, and $x_4$. The cross-ratios in these coordinates are given by
\begin{alignat}{2}
u &= r^2 \ecq &\qquad v &= \abs{\hat{z} - \vec{r}}^2 = 1 + r^2 -2r\cos \theta \ec \\
x &= r e^{i\theta} \ecq &\qquad \bar{x} &= r e^{-i\theta} \label{xTor}\ed
\end{alignat}
In other words, $u$ can be interpreted as the square of the distance to the origin of $\bR^d$, while $v$ is the square of the distance to the special point $(0, \ldots, 0, 1)$.

The operator $\cD_{\veps}$ can then be written as
\begin{align}
  \cD_{\veps} = \frac{1}{4} \left[ \frac{1}{r^{2\veps+1}} 
     \partial_r\left(r^{2\veps+1}\partial_r\right) 
     + \frac{1}{r^2 \sin^{2\veps}\theta} \partial_{\theta}\left(\sin^{2\veps}\theta\partial_{\theta}\right) \right] \,.
\end{align}
Up to an overall factor of $1/4$, ${\cal D}_\veps$ is nothing but the $d$-dimensional Laplacian $\boldsymbol{\Delta}$ acting on functions that are independent of the azimuthal directions $\Omega_{d-2}\in S^{d-2}$.

In $d = 3$, the solution \eqref{wardsol} to the Ward identity can then be written formally as
 \es{wardsol3}{
   \cG(\vec{r}; \alpha, \bar \alpha)
      = r^2 \frac{2}{\sqrt{ \boldsymbol{\Delta} }} \biggl[ 
       \abs{\alpha \vec{r} - \hat z}^2 \abs{\bar \alpha \vec{r} - \hat z}^2 a(\vec{r})\biggr] \,,
 }
for some undetermined function $a(\vec{r})$.  Here, both $\cG(\vec{r}; \alpha, \bar \alpha)$ and $a(\vec{r})$ should be taken to be invariant under rotations about the $z$-axis.  This expression will become quite useful when we analyze the crossing symmetry in the next section.

\section{Constraints from Crossing Symmetry}
\label{CROSSING}

In this section we will discuss the constraints of crossing symmetry on the four-point function \eqref{4pnt}. 

In terms of $\cG(u, v; U, V)$ defined in \eqref{4pnt}, the crossing constraint corresponding to the exchange of $(x_1,Y_1)$ with $(x_3,Y_3)$ is
 \es{13crossing}{
   \cG(u, v; U, V) = \frac uv \left( \frac{V}{U} \right)^2 \cG\left(v, u; V, U \right) \,.
 }
By expanding \eqref{13crossing} in $U$ and $V$ one obtains six crossing equations, mixing the different R-symmetry channels \eqref{GExpansion}. However, these crossing equations cannot be used in the numerical bootstrap program as they stand, for the following reason. The different R-symmetry channels are related by supersymmetry, so these equations are not independent. Using these dependent equations in a semidefinite program solver like {\tt sdpa} \cite{sdpa} (as we will discuss in detail in Section~\ref{numerics}) results in a numerical instability.

To understand the dependencies between the equations \eqref{dbasis} we have to study the solution \eqref{wardsol3} of the Ward identity. In terms of \eqref{wardsol} the crossing equation \eqref{13crossing} takes the form\footnote{In deriving \eqref{wardCross} we use the fact that under crossing $\vec{r} \to \hat z - \vec{r}$ and $\boldsymbol{\Delta}$ is invariant.}
 \es{wardCross}{
    \frac{1}{\sqrt{ \boldsymbol{\Delta} }} \biggl[\abs{\alpha \vec{r} - \hat z}^2 
      \abs{\bar \alpha \vec{r} - \hat z}^2 \bigl( a(u,v) - a(v,u) \bigr)\biggr] = 0 \,.
 }
This expression seems to suggest that there is only one independent crossing equation given by $a(u,v) - a(v,u)=0$.  However, it is not easy to calculate $a(u, v) - a(v, u)$ by acting with the non-local operator $\sqrt{\boldsymbol{\Delta} }$ on \eqref{wardCross}, because currently there is too little global information available about the four point function of ${\cal O}_{{\bf 35}_c}$ and its (super)conformal block expansion.  It would be interesting to explore this avenue in future work.

Despite the appearance of a non-local operator in the solution of the superconformal Ward identity, we can in fact show that the six R-symmetry channels and, consequently the six crossing equations, satisfy certain differential equations that relate them to one another.  These relations will be crucial for the implementation of the numerical bootstrap program in Section~\ref{numerics}.

\subsection{Relations Between R-Symmetry Channels}
\label{relations}

The inverse square root of the Laplacian appearing in \eqref{wardsol3} can be defined by its Fourier transform
\begin{align}
  \frac{1}{\sqrt{\boldsymbol{\Delta}}} = \left(-p^2\right)^{-1/2} \ed
\end{align}
In expressions of the form $\boldsymbol{\Delta}^{-\frac{1}{2}}f(r,\theta) \boldsymbol{\Delta}^{\frac 12}$, we can then use the canonical commutation relation of quantum mechanics, $[x,p]=i$, to commute $ \boldsymbol{\Delta}^{\frac 12}$ through $f(r,\theta)$. For example, it is straightforward to show that
\begin{align}
\boldsymbol{\Delta}^{-\frac{1}{2}} r^2 \boldsymbol{\Delta}^{\frac{1}{2}} &= r^2 -  \boldsymbol{\Delta}^{-1} \left( 4 + 2 r \partial_r \right) \ec \label{rComm}\\
\boldsymbol{\Delta}^{-\frac{1}{2}} z \boldsymbol{\Delta}^{\frac{1}{2}} &= z - \boldsymbol{\Delta}^{-1}\partial_z \label{zComm}\ec
\end{align}
where we defined $z \equiv r\cos \theta$.

To proceed, it is convenient to decompose the solution of the Ward identity \eqref{wardsol} in the basis
\begin{alignat}{2}
  e_1 &\equiv \frac{1}{\sqrt{\boldsymbol{\Delta}}} a(u,v)  &\ecq 
  e_2 &\equiv \frac{1}{\sqrt{\boldsymbol{\Delta}}} \bigl[(u-v) \, a(u,v) \bigr] \ec \notag\\
  e_3 &\equiv \frac{1}{\sqrt{\boldsymbol{\Delta}}} \bigl[(u+v) \,  a(u,v) \bigr]   &\ecq 
  e_4 &\equiv \frac{1}{\sqrt{\boldsymbol{\Delta}}} \left[(u^2-v^2) \,  a(u,v) \right]  \ec \label{ebasis}\\
  e_5 &\equiv \frac{1}{\sqrt{\boldsymbol{\Delta}}} \left[(u-v)^2 \, a(u,v) \right]  &\ecq 
  e_6 &\equiv \frac{1}{\sqrt{\boldsymbol{\Delta}}} \left[(u+v)^2 \, a(u,v) \right] \notag\ed
\end{alignat}
These $e_i$ are simply related to the different R-symmetry channels $A_{ab}$ by
\begin{align}
\begin{pmatrix}
e_1\\e_2\\e_3\\e_4\\e_5\\e_6
\end{pmatrix} 
   = \frac{1}{u} 
     \begin{pmatrix}
1  & -1 & \frac{3}{4}  & \frac{5}{7}    & -\frac{3}{5} & \frac{2}{5}   \\
-1 & 0  & \frac{1}{4}  & \frac{20}{21}  & -1           & \frac{14}{15} \\
1  & 0  & -\frac{1}{4} & \frac{22}{21}  & -1           & \frac{16}{15} \\
-1 & -1 & -\frac{3}{4} & -\frac{5}{7}   & -\frac{3}{5} & \frac{28}{5} \\
1  & 1  & \frac{3}{4}  & -\frac{9}{7}   & -\frac{7}{5} & \frac{22}{5} \\
1  & 1  & \frac{3}{4}  & \frac{19}{7}   & \frac{13}{5} & \frac{42}{5}
\end{pmatrix} \begin{pmatrix}
A_{00} \\  A_{10} \\  A_{11} \\  A_{20} \\  A_{21} \\  A_{22} 
\end{pmatrix} \ed
\end{align}

Defining the operators
\begin{align}
 \boldsymbol{\cal D}_{\pm} \equiv \frac 14 \sqrt{\boldsymbol{\Delta}}  (u\pm v) \sqrt{\boldsymbol{\Delta}} \label{Dpm} \,,
\end{align}
it can be seen from \eqref{ebasis} that the following relations hold:
\begin{alignat}{2}
\boldsymbol{\cal D}_+ e_1 &= \boldsymbol{\Delta} e_3 \ecq &\qquad \boldsymbol{\cal D}_- e_1 &=\boldsymbol{\Delta} e_2 \ec \label{rel1}\\
\boldsymbol{\cal D}_+ e_2 &= \boldsymbol{\Delta} e_4 \ecq &\qquad \boldsymbol{\cal D}_- e_2 &= \boldsymbol{\Delta} e_5 \ec \label{rel2}\\
\boldsymbol{\cal D}_+ e_3 &= \boldsymbol{\Delta} e_6 \ecq &\qquad \boldsymbol{\cal D}_- e_3 &= \boldsymbol{\Delta} e_4 \ec \label{rel3}\\
\boldsymbol{\cal D}_+ e_4 &= \boldsymbol{\cal D}_- e_6 \ecq &\qquad \boldsymbol{\cal D}_- e_4 &= \boldsymbol{\cal D}_+ e_5 \label{rel4} \ed 
\end{alignat}

It is easy to convince oneself that these are the most general relations between the $e_i$ that can be obtained by acting with $\boldsymbol{\cal D}_{\pm}$. Moreover, instead of thinking of the solution to the Ward identity as given in terms of a single unconstrained function $a(u,v)$, we can think of it as given in terms of the six constrained functions $e_i$, with the constraints given by \eqref{rel1}--\eqref{rel4}.

The advantage of this formulation of the solution is that the constraints \eqref{rel1}--\eqref{rel4} only involve local differential operators. Indeed, using \eqref{rComm}, \eqref{zComm}, and the coordinate transformation \eqref{xTor}, we find
 \es{GotDpDm}{
   \boldsymbol{\cal D}_- &= \frac {2z-1}4 \boldsymbol{\Delta} + \frac{1}{2}\partial_z \ec\\
   \boldsymbol{\cal D}_+ &= \frac{1 + 2r^2-2z}{4} \boldsymbol{\Delta} + r\partial_r - \frac{1}{2}\partial_z+1 \ed
 }
In terms of the $x$, $\bar x$ coordinates, we have
 \es{GotDpDmAgain}{
    \boldsymbol{\cal D}_- &= (x+\bar{x}-1)\cD_{\frac{1}{2}} + \frac{1}{2}\left(\frac{\partial}{\partial x} + \frac{\partial}{\partial\bar{x}}\right) \ec\\
    \boldsymbol{\cal D}_+ &= \left(1+2x\bar{x}-x-\bar{x} \right)\cD_{\frac{1}{2}} 
      + \left(x-\frac{1}{2}\right)\frac{\partial}{\partial x} 
      + \left(\bar{x} - \frac{1}{2}\right)\frac{\partial}{\partial\bar{x}} + 1 \,,
 }
where ${\cal D}_{\frac 12}$ was defined in~\eqref{Deps}.

\subsection{Relations Between the Crossing Equations}
\label{RELCROSSING}

Define $\tilde{e}_i$ to be the same as the $e_i$ in \eqref{ebasis}, but with the factors of $a(u,v)$ replaced by ${a(u,v) -  a(v,u)}$. It is clear that the $\tilde{e}_i$ also satisfy the differential equations \eqref{rel1}--\eqref{rel4}. The crossing symmetry constraints are simply given by $\tilde{e}_i=0$. 

One can solve the differential equations \eqref{rel1}--\eqref{rel4} by using series expansions around the crossing symmetric point. In particular, define $\tilde{e}^i_{n,m}$ through the expansions
\begin{align}
\tilde{e}_i(x,\bar{x}) &= \sum_{n,m=0}^{\infty} \frac{1}{n!m!} \left(x-\frac{1}{2}\right)^n\left(\bar{x}-\frac{1}{2}\right)^m \tilde{e}^i_{n,m} \ec \label{eiexp}\\
\tilde{e}^i_{n,m} &\equiv \left.\partial^n\bar{\partial}^m e_i(x,\bar{x})\right|_{x=\bar{x}=\frac{1}{2}} \ed \label{efunc}
\end{align}
From $x\leftrightarrow\bar{x}$ symmetry and (anti-)symmetry under $u\leftrightarrow v$ we have
\begin{align}
\tilde{e}^i_{n,m} &= \tilde{e}^i_{m,n} \ec\\
\tilde{e}^i_{n,m} &= 0 \quad\mathrm{if} \,\,\begin{cases}
m+n=\mathrm{even} \ecq i=1,3,5,6 \,,\\
m+n=\mathrm{odd} \ecq  i=2,4 \,.
\end{cases} 
\end{align}

We can now plug the expansions \eqref{eiexp} into the differential equations \eqref{rel1}--\eqref{rel4} and solve for the coefficients $\tilde{e}_{n,m}^i$ order by order. The results can be stated as follows. If we assume only the crossing equation $\tilde{e}_2=0$, then equations \eqref{rel1}--\eqref{rel4} imply
\begin{align}
\tilde{e}_1 &= 0 \ec\\
\tilde{e}_3 &= 0 \ec \\
\tilde{e}^4_{n,m} &= a_{nm} \tilde{e}^4_{m+n, 0} \ec\\
\tilde{e}^5_{n,m} &= b_{nm} \tilde{e}^4_{m+n+1, 0} \ec \\
\tilde{e}^6_{n,m} &= c_{nm} \tilde{e}^4_{m+n-1, 0} \ec
\end{align}
for some constants $a_{nm}$, $b_{nm}$, and $c_{nm}$ that can be determined order by order in the expansion. We conclude that the maximal set of independent crossing equations can be taken to be $\tilde{e}^2_{n,m} = 0$ and $\tilde{e}^4_{n,0}=0$ for all integers $n,m\ge 0$.

\section{Superconformal Blocks}
\label{scblocks}

In this section we will derive the $\cN=8$ superconformal blocks of the four-point function \eqref{4pnt}.  Any given superconformal block represents the total contribution to the four-point function \eqref{4pnt} coming from all operators appearing in the ${\cal O}_{{\bf 35}_c} \times {\cal O}_{{\bf 35}_c} $ OPE that belong to a given superconformal multiplet.  Since superconformal multiplets are made of conformal multiplets, the superconformal blocks are just linear combinations of the usual conformal blocks.  Our task is to determine which conformal blocks appear in a given superconformal block and with which coefficients.

A common approach to deriving superconformal blocks involves analyzing the detailed structure of the three-point function between two $\cO_{\mathbf{35}_c}$ and a third generic superconformal multiplet.  In this approach one has to construct the most general superconformal invariants out of the superspace variables appearing in this three-point function (see e.g., \cite{Poland:2010wg}). However, it is difficult to implement this method in theories with extended supersymmetry due to complications in using superspace techniques in such theories.

In practice, we will compute the superconformal blocks in our case of interest using two different methods. One method involves expanding the solution of the Ward identity given in \eqref{wardsol3} in conformal blocks.\footnote{The superconformal blocks of $\cN=2$ and ${\cal N} = 4$ theories in $d=4$ were first derived in this way \cite{Dolan:2001tt}.} Even though this method is hard to implement due to the appearance of the non-local operator $1/\sqrt{\boldsymbol{\Delta}}$ in \eqref{wardsol}, significant progress was made in \cite{Dolan:2004mu} and we will build on it in Section~\ref{DGSblocks}.  
In the next subsection we will introduce a new strategy for computing the superconformal blocks. 
This second method relies on the fact that the superconformal Ward identity \eqref{GenWard} holds separately for each superconformal block.  As we will see momentarily in Section~\ref{OURMETHOD}, this approach is simpler and more systematic than working directly with the full solution to the Ward identity.  

Before we begin, let us quickly review the unitary irreducible representations of the $\mathfrak{osp}(8|4)$ superconformal algebra, following \cite{Dolan:2008vc}.  Unitary irreps of $\mathfrak{osp}(8|4)$ are specified by the scaling dimension $\Delta$, Lorentz spin $j$, and $\mathfrak{so}(8)$ R-symmetry irrep $[a_1\, a_2 \, a_3 \, a_4]$ of their bottom component, as well as by various shortening conditions.  There are twelve different types of multiplets that we list in Table~\ref{Multiplets}.  
\begin{table}[htdp]
\begin{center}
\begin{tabular}{|l|c|c|c|c|}
\hline
 Type     & BPS    & $\Delta$             & Spin & $\mathfrak{so}(8)_R$  \\
 \hline 
 $(A,0)$ (long)      & $0$    & $\ge \Delta_0 + j+1$ & $j$  & $[a_1 a_2 a_3 a_4]$  \\
 $(A, 1)$  & $1/16$ & $\Delta_0 + j +1$    & $j$  & $[a_1 a_2 a_3 a_4]$  \\
 $(A, 2)$  & $1/8$  & $\Delta_0 + j +1$    & $j$  & $[0 a_2 a_3 a_4]$   \\
 $(A, 3)$  & $3/16$  & $\Delta_0 + j +1$    & $j$  & $[0 0 a_3 a_4]$     \\
 $(A, +)$  & $1/4$  & $\Delta_0 + j +1$    & $j$  & $[0 0 a_3 0]$       \\
 $(A, -)$  & $1/4$  & $\Delta_0 + j +1$    & $j$  & $[0 0 0 a_4]$       \\
 $(B, 1)$  & $1/8$  & $\Delta_0$           & $0$  & $[a_1 a_2 a_3 a_4]$ \\
 $(B, 2)$  & $1/4$  & $\Delta_0$           & $0$  & $[0 a_2 a_3 a_4]$   \\
 $(B, 3)$  & $3/8$  & $\Delta_0$                  & $0$  & $[0 0 a_3 a_4]$     \\
 $(B, +)$  & $1/2$  & $\Delta_0$           & $0$  & $[0 0 a_3 0]$       \\
 $(B, -)$  & $1/2$  & $\Delta_0$           & $0$  & $[0 0 0 a_4]$       \\
 conserved & $5/16$  & $j+1$                & $j$  & $[0 0 0 0]$         \\
 \hline
\end{tabular}
\end{center}
\caption{Multiplets of $\mathfrak{osp}(8|4)$ and the quantum numbers of their corresponding superconformal primary operator. The conformal dimension $\Delta$ is written in terms of $\Delta_0 \equiv a_1 + a_2 + (a_3 + a_4)/2$.  The Lorentz spin can take the values $j=0, 1/2, 1, 3/2, \ldots$.  Representations of the $\mathfrak{so}(8)$ R-symmetry are given in terms of the four $\mathfrak{so}(8)$ Dynkin labels, which are non-negative integers.}
\label{Multiplets}
\end{table}
There are two types of shortening conditions denoted by the $A$ and $B$ families.  The multiplet denoted by $(A, 0)$ is a long multiplet and does not obey any shortening conditions.  The other multiplets of type $A$ have the property that certain $\mathfrak{so}(2, 1)$ irreps of spin $j-1/2$ are absent from the product between the supercharges and the superconformal primary.  The multiplets of type $B$ have the property that certain $\mathfrak{so}(2, 1)$ irreps of spin $j \pm 1/2$ are absent from this product, and consequently, the multiplets of type $B$ are smaller.  The stress-tensor multiplet that we encountered in Section~\ref{conventions} is of $(B, +)$ type and has $a_3 = 2$. The conserved current multiplet appears in the decomposition of the long multiplet at unitarity: $\Delta\to j+1$. This multiplet contains higher-spin conserved currents, and therefore can only appear in the free theory \cite{Maldacena:2011jn}.

We will sometimes denote the superconformal multiplets by $(\Delta,j)_{X}^{[a_1\, a_2 \, a_3 \, a_4]}$, with $(\Delta,j)$ and $[a_1\, a_2 \, a_3 \, a_4]$ representing the $\mathfrak{so}(3, 2)$ and $\mathfrak{so}(8)_R$ quantum numbers of the superconformal primary, and the subscript $X$ denoting the type of shortening condition (for instance, $X = (A, 2)$ or $X = (B, +)$).

\subsection{Superconformal Blocks from Ward Identity}
\label{OURMETHOD}

Our strategy to compute the superconformal blocks is very simple. Let $\cG_{\Delta,j}^{(a,b)}$ denote the contribution to the four-point function of a multiplet whose primary has dimension and spin $(\Delta,j)$ and transforms in the $(a,b)\equiv [0\, (a-b)\, (2b)\, 0]$ irrep of $\mathfrak{so}(8)_R$. This contribution can be written as some linear combination of a finite number of conformal blocks:
 \es{scblocks4pnt}{
    \cG_{\Delta,j}^{(a,b)}(x,\bar{x},\alpha,\bar{\alpha}) 
      = \sum_{c=0}^2\sum_{d=0}^c \left[ Y_{cd}(\alpha,\bar{\alpha}) \!\!\!\!\!\!
       \sum_{\cO\in (\Delta,j)_{a,b}} \!\!\!\!\!\! \lambda_{\cO}^2 \, 
        g_{\Delta_{\cO},j_{\cO}}(x,\bar{x})\right] \,,
 }
where $g_{\Delta, j}(x, \bar x)$ is the conformal block corresponding to the exchange of an operator with scaling dimension $\Delta$ and Lorentz spin $j$.  (We will determine precisely which conformal blocks appear in this sum shortly.)  The innermost sum runs over all conformal primaries in the superconformal multiplet $(\Delta,j)_{a,b}$ transforming in the R-symmetry channel $(c,d)$ (specified by the outer sums).

By using the OPE one can show that the superconformal Ward identity \eqref{GenWard} is satisfied on each $\cG_{\Delta,j}^{(a,b)}$ contribution independently.   We can expand \eqref{scblocks4pnt} in a Taylor series around $x = \bar x = 0$ using the known expansions of the conformal blocks  (see, for example, \cite{Hogervorst:2013sma} or Appendix~\ref{confblock}).  Plugging in this expansion in the suprconformal Ward identity \eqref{GenWard}, we can generate infinitely many equations for the undetermined coefficients $\lambda_{{\cal O}}^2$.  These equations must be consistent if in \eqref{scblocks4pnt} we sum over all the operators ${\cal O}$ belonging to a given superconformal multiplet.

Before we can apply this strategy outlined above concretely, we need to determine which superconformal multiplets can appear in the OPE\@.   In addition, we should also determine the spectrum of conformal primaries in each of those superconformal multiplets. The first task was preformed in \cite{Ferrara:2001uj}, and we list their results  in Table~\ref{opemult}. 

\begin{table}
\centering
\begin{tabular}{|c|c|r|}
\hline
Type    & $(\Delta,j)$     & $\mathfrak{so}(8)_R$ irrep   \\
\hline
$(B,+)$ &  $(2,0)$         & ${\bf 294}_c = [0040]$ \\ 
\red{$(B,2)$} &  \red{$(2,0)$}         & \red{${\bf 567}_c = [0120]$} \\
$(B,2)$ &  $(2,0)$         & ${\bf 300} = [0200]$ \\
$(B,+)$ &  $(1,0)$         & ${\bf 35}_c = [0020]$ \\
$(A,+)$ &  $(j+2,j)$       & ${\bf 35}_c = [0020]$ \\
\red{$(B,2)$} &  \red{$(1,0)$}         & \red{${\bf 28} = [0100]$} \\
$(A,2)$ &  $(j+2,j)$       & ${\bf 28} = [0100]$ \\
$(A,0)$ &  $\Delta\ge j+1$ & ${\bf 1} = [0000]$ \\
\hline
\end{tabular}
\caption{The possible superconformal multiplets in the $\cO_{\mathbf{35}_c}\times\cO_{\mathbf{35}_c}$ OPE\@.  The multiplets marked in red do not appear in the OPE, as explained in the main text.  In addition, $j$ must be even for the $(A, 0)$ and $(A, +)$ multiplets and odd for $(A, 2)$.  The $\mathfrak{so}(3, 2) \oplus \mathfrak{so}(8)_R$ quantum numbers are those of the superconformal primary in each multiplet.}
\label{opemult}
\end{table}

Note that a three-point function of two $1/2$-BPS multiplets with a third multiplet of any type is completely determined by the contribution of the superconformal primaries. It then follows that if a superconformal primary has zero OPE coefficient, then so do all its descendants.  Consequently, in Table~\ref{opemult}, the $(B,2)$ multiplets in $[0100]$ and $[0120]$ cannot actually appear in the OPE\@. The reason is that these representations appear in the anti-symmetric product of the OPE, and can therefore contain only odd spin operators, while the superconformal primaries of the above multiplets have even spin. Similarly, $j$ must be even in the $(j+2,j)_{(A,+)}^{[0020]}$ and the (long) $(\Delta,j)_{(A,0)}^{[0000]}$ multiplets, and it must be odd for $(j+2,j)_{(A,2)}^{[0100]}$.

Next, we have to identify the conformal primaries belonging to the superconformal multiplets listed in Table~\ref{opemult}.  For each such superconformal multiplet, we can decompose its corresponding $\mathfrak{osp}(8|4)$ character \cite{Dolan:2008vc} into characters of the maximal bosonic sub-algebra $\mathfrak{so}(3,2) \oplus \mathfrak{so}(8)_R$.  This decomposition is rather tedious, and we describe it in Appendix~\ref{characters}.  Here, let us list the results.  The conformal primaries of the stress-tensor multiplet $(1, 0)_{(B, +)}^{[0020]}$ were already given in Table~\ref{Stress}.  The conformal primaries of all the other multiplets appearing in Table~\ref{opemult} are given in Tables~\ref{Bp}--\ref{long}.   The first column in these tables contains the conformal dimensions and the other columns contain the possible values of the spins in the various R-symmetry channels.  In each table, we only list the operators which could possibly contribute to our OPE, namely only operators with R-symmetry representations in the tensor product \eqref{35times35}, and only even (odd) integer spins for the representations $(a,b)$ with even (odd) $a+b$.

\begin{table}[htpb]
\centering
\begin{tabular}{|c||c|c|c|c|c|c|}
\hline 
$(2,0)^{[0040]}_{(B,+)}$ & \multicolumn{6}{c|}{spins in various $\mathfrak{so}(8)_R$ irreps}  \\
\hline
\multirow{2}{*}{dimension} & ${\bf 1}$ & ${\bf 28}$ & ${\bf 35}_c$ & ${\bf 300}$ & ${\bf 567}_c$ & ${\bf 294}_c$\\  
 & $[0000]$ & $[0100]$ & $[0020]$ & $[0200]$ & $[0120]$ & $[0040]$\\          
\hline
$2$       & --          & --          & --          & --          & --          & $0$    \\
$3$       & --          & --          & --          & --          & $1$         & --     \\
$4$       & --          & --          & $2$         & $0$         & --          & --     \\
$5$       & --          & $1$         & --          & --          & --          & --     \\
$6$       & $0$         & --          & --          & --          & --          & --     \\
\hline
\end{tabular}
\caption{All possible conformal primaries in $\cO_{\mathbf{35}_c}\times \cO_{\mathbf{35}_c}$ corresponding to the $(2,0)^{[0040]}_{(B,+)}$ superconformal multiplet.}\label{Bp}
\end{table}

\begin{table}[htpb]
\centering
\begin{tabular}{|c||c|c|c|c|c|c|}
\hline
$(2,0)^{[0200]}_{(B,2)}$ & \multicolumn{6}{c|}{spins in various $\mathfrak{so}(8)_R$ irreps}  \\
\hline
\multirow{2}{*}{dimension} & ${\bf 1}$ & ${\bf 28}$ & ${\bf 35}_c$ & ${\bf 300}$ & ${\bf 567}_c$ & ${\bf 294}_c$\\  
 & $[0000]$ & $[0100]$ & $[0020]$ & $[0200]$ & $[0120]$ & $[0040]$\\          
\hline
$2$       & --          & --          & --          & $0$         & --          & --    \\
$3$       & --          & $1$         & --          & \red{$0$}   & $1$          & --     \\
$4$       & $0$         & \red{$1$}   & $0,2$       & $0,2$       & \red{$1$}   & $0$     \\
$5$       & \red{$0$}   & $1,3$       & \red{$2$}   & \red{$0$}   & $1$         & --     \\
$6$       & $0,2$       & \red{$1$}   & $2$         & $0$         & --          & --     \\
$7$       & \red{$0$}   & $1$         & --          & --          & --          & --     \\
$8$       & $0$         & --          & --          & --          & --          & --     \\
\hline
\end{tabular}
\caption{All possible conformal primaries in $\cO_{{\bf 35}_c}\times \cO_{{\bf 35}_c}$ corresponding to the $(2,0)^{[0200]}_{(B,2)}$ superconformal multiplet.}\label{B2}
\end{table}

\begin{table}[htpb]
\centering
\begin{tabular}{|c||c|c|c|c|c|c|}
\hline
$(j+2,j)^{[0020]}_{(A,+)}$ & \multicolumn{6}{c|}{spins in various $\mathfrak{so}(8)_R$ irreps}  \\
\hline
\multirow{2}{*}{dimension} & ${\bf 1}$ & ${\bf 28}$ & ${\bf 35}_c$ & ${\bf 300}$ & ${\bf 567}_c$ & ${\bf 294}_c$\\  
 & $[0000]$ & $[0100]$ & $[0020]$ & $[0200]$ & $[0120]$ & $[0040]$\\          
\hline
$j+2$ & --             & --                       & $j$                    & --          & --                  & --    \\
$j+3$ & --             & $j\pm 1$                 & \red{$j$}              & --          & $j+1$               & --    \\
$j+4$ & $j\pm 2$, $j$  & \red{$j\pm 1$}           & $j+2$, $j$             & $j+2$, $j$  & \red{$j+1$}         & $j+2$ \\
$j+5$ & \red{$j+2$}    & $j+3$, $j\pm 1$          & \red{$j+2$}, \red{$j$} & \red{$j+2$} & $j+3$, $j+1$        & --    \\
$j+6$ & $j+2$          & \red{$j+3$}, \red{$j+1$} & $j+4$, $j+2$, $j$             & $j+2$       & --                  & --    \\
$j+7$ & \red{$j+2$}    & $j+3$, $j+1$             & --                     & --          & --                  & --    \\
$j+8$ & $j+2$          & --                       & --                     & --          & --                  & --    \\
\hline
\end{tabular}
\caption{All possible conformal primaries in $\cO_{{\bf 35}_c}\times \cO_{{\bf 35}_c}$ corresponding to the $(j+2,j)^{[0020]}_{(A,+)}$ superconformal multiplet, with $j\ge 2$ even. For $j=0$ one should omit the representations with negative spins as well as $(4,0)^{[0000]}$. }\label{Ap}
\end{table}

\begin{table}[htpb]
\centering
\begin{tabular}{|c||c|c|c|c|c|c|}
\hline
$(j+2,j)^{[0100]}_{(A,2)}$ & \multicolumn{6}{c|}{spins in various $\mathfrak{so}(8)_R$ irreps}  \\
\hline
\multirow{2}{*}{dimension} & ${\bf 1}$ & ${\bf 28}$ & ${\bf 35}_c$ & ${\bf 300}$ & ${\bf 567}_c$ & ${\bf 294}_c$\\  
 & $[0000]$ & $[0100]$ & $[0020]$ & $[0200]$ & $[0120]$ & $[0040]$\\          
\hline
$j+2$ & --               & $j$               & --                  & --           & --              & --    \\
$j+3$ & $j\pm 1$         & \red{$j$}         & $j\pm 1$            & $j+1$        & --              & --    \\
$j+4$ & \red{$j+1$}      & $j\pm 2,j$        & \red{$j\pm 1$}      & \red{$j+1$}  & $j+2,j$         & --    \\
$j+5$ & $j\pm 3, j\pm 1$ & \red{$j\pm 2, j$} & $j+3,j\pm 1$        & $j+3,j\pm 1$ & \red{$j+2,j$}   & $j+1$ \\
$j+6$ & \red{$j+3, j+1$} & $j+4, j\pm 2, j$  & \red{$j+3, j\pm 1$} & \red{$j+1$}  & $j+2,j$         & --    \\
$j+7$ & $j+3,j+1$        & \red{$j+2,j$}     & $j+3, j\pm 1$       & $j+1$        & --              & --    \\
$j+8$ & \red{$j+1$}      & $j+2,j$           & --                  & --           & --              & --    \\
$j+9$ & $j+1$            & --                & --                  & --           & --              & --    \\
\hline
\end{tabular}
\caption{All possible conformal primaries in $\cO_{\mathbf{35}_c}\times \cO_{\mathbf{35}_c}$ corresponding to the $(j+2,j)^{[0100]}_{(A,2)}$ superconformal multiplet, with $j$ odd. For $j=1$ one should omit $(6,0)^{[0000]}$ and representations with negative spin.}\label{A2}
\end{table}

\begin{table}[htpb]
\centering
\begin{tabular}{|c||c|c|c|c|c|c|}
\hline
$(\Delta,j)^{[0000]}_{(A,0)}$ & \multicolumn{6}{c|}{spins in various $\mathfrak{so}(8)_R$ irreps}  \\
\hline
\multirow{2}{*}{dimension} & ${\bf 1}$ & ${\bf 28}$ & ${\bf 35}_c$ & ${\bf 300}$ & ${\bf 567}_c$ & ${\bf 294}_c$\\  
 & $[0000]$ & $[0100]$ & $[0020]$ & $[0200]$ & $[0120]$ & $[0040]$\\          
\hline
$\Delta$   & $j$                & --                     & --               & --         & --             & --    \\
$\Delta+1$ & \red{$j$}          & $j\pm 1$               & --               & --         & --             & --    \\
$\Delta+2$ & $j$                & \red{$j\pm 1$}         & $j\pm 2,j$       & $j$        & --             & --    \\
$\Delta+3$ & \red{$j$}          & $j\pm 3, j\pm 1$       & \red{$j\pm 2,j$} & \red{$j$}  & $j\pm 1$       & --    \\
$\Delta+4$ & $j\pm 4, j\pm 2,j$ & \red{$j\pm 3, j\pm 1$} & $j\pm 2,j$       & $j\pm 2,j$ & \red{$j\pm 1$} & $j$   \\
$\Delta+5$ & \red{$j$}          & $j\pm 3, j\pm 1$       & \red{$j\pm 2,j$} & \red{$j$}  & $j\pm 1$       & --    \\
$\Delta+6$ & $j$                & \red{$j\pm 1$}         & $j\pm 2,j$       & $j$        & --             & --    \\
$\Delta+7$ & \red{$j$}          & $j\pm 1$               & --               & --         & --             & --    \\
$\Delta+8$ & $j$                & --                     & --               & --         & --             & --    \\
\hline
\end{tabular}
\caption{All possible conformal primaries in $\cO_{{\bf 35}_c}\times \cO_{{\bf 35}_c}$ corresponding to the $(\Delta,j)^{[0000]}_{(A,0)}$ (long) superconformal multiplet, with $j$ even, $\Delta\ge j+1$.  The decomposition of this multiplet at unitarity contains a conserved current multiplet, which, in turn, contains higher-spin conserved currents. \label{long}}
\end{table}

Using this information and the Ward identity we can now determine the superconformal blocks. In practice, we expand \eqref{scblocks4pnt} to a high enough order so that we get an overdetermined system of linear equations in the $\lambda_{\cO}^2$.  We can then solve for the OPE coefficients in terms of one overall coefficient.  The fact that we can successfully solve an overdetermined system of equations is a strong consistency check on our computation.  The final expressions are very complicated, and we collect the results in Appendix~\ref{superblocks}. 

As an interesting feature of the superconformal blocks, we find that the OPE coefficients of all the operators which are marked in red in Tables~\ref{Bp}--\ref{long} vanish.  These operators are precisely the super-descendants obtained by acting on the superconformal primary with $\veps^{\alpha\beta}Q_{a\alpha}Q_{b \beta}$ an odd number of times.  This combination of supercharges is odd under parity, while $\cO_{\mathbf{35}_c}$ is even.   There is no a priori reason, however, why an ${\cal N} = 8$ SCFT should be invariant under parity, even though all known examples do have this property.  Our findings show that even if parity is not a symmetry of the full theory, it is a symmetry of the $\cO_{\mathbf{35}_c}\times\cO_{\mathbf{35}_c}$ OPE\@.\footnote{A similar phenomenon occurs in four dimensional $\cN=4$ supersymmetric Yang-Mills theory \cite{Dolan:2001tt}. There, the operators that decouple are the ones which are not invariant under the ``bonus symmetry'' discussed in \cite{Intriligator:1998ig,Intriligator:1999ff}.}

\subsection{Derivation of Superconformal Blocks Using the Results of \cite{Dolan:2004mu}}
\label{DGSblocks}

The superconformal blocks can also be computed using the solution \eqref{wardsol3} of the Ward identity.\footnote{The superconformal blocks of $\cN=2,4$ SCFTs in $d=4$ were derived in this way in \cite{Dolan:2001tt}.}  One first observes that for all multiplets listed in Tables \ref{Bp}--\ref{long}, the $[0040]$ channel receives contributions from a single operator.  The projection of the four-point function onto this channel is then given by a single conformal block. The other channels are related to the $[0040]$ channel by \eqref{wardsol3}, and their conformal block expansion can be determined by using certain recurrence relations obeyed by the conformal blocks.

Let us first write \eqref{wardsol3} in terms of the decomposition into $\mathfrak{so}(8)_R$ representations in \eqref{GExpansion}, 
 \es{A22DGS}{
A_{22} &= \frac{u}{3} \frac{1}{\sqrt{\boldsymbol{\Delta}}} u^2 a \ec\\
A_{21} &=  u \frac{1}{\sqrt{\boldsymbol{\Delta}}} u (v-1) a \ec \\
A_{20} &= \frac{u}{3} \frac{1}{\sqrt{\boldsymbol{\Delta}}} u \left(3(v+1)-u\right) a \ec\\
A_{11} &=  u \frac{1}{\sqrt{\boldsymbol{\Delta}}} \left( (v-1)^2 - \frac{2}{3} u(v+1) + \frac{1}{9}\right) a \ec\\
A_{10} &=  u \frac{1}{\sqrt{\boldsymbol{\Delta}}}  (v-1)\left( (v+1) - \frac{3}{5} u \right) a \ec \\
A_{00} &= \frac{u}{2} \frac{1}{\sqrt{\boldsymbol{\Delta}}}  \left( (v+1)^2 - \frac{1}{2}(v-1)^2 - \frac{3}{7} u(v+1) + \frac{3}{70} u^2 \right) a \ed
 }
For the long multiplet $A_{22}$ is determined (up to an overall coefficient) to be
\begin{align}
A_{22}^{\mathrm{(long)}}(u, v) = \frac{1}{6} g_{\Delta+4,j}(u, v) \,.
\end{align}
Then, for example, the $A_{21}$ channel is given by
\begin{align}
A_{21}^{\mathrm{(long)}}(u, v) = \frac{1}{2} u \frac{1}{\sqrt{\boldsymbol{\Delta}}} \frac{v-1}{u} \sqrt{\boldsymbol{\Delta}}\, \frac{ g_{\Delta+4,j}(u, v)}{u} \ec \label{A21fromA22}
\end{align}
and the other channels are given by similar expressions.   This expression can be expanded in conformal blocks by using recurrence relations derived in \cite{Dolan:2004mu}. We collect these relations\footnote{Appendix \ref{DGSD} also corrects several typos in the equations of \cite{Dolan:2004mu}.} in appendix \ref{DGSD}. The final result matches precisely the long multiplet superconformal block that we found using the method of the previous section. 

It turns out that the superconformal blocks of the short multiplets can be derived by taking limits of the long superconformal block. These limits consist of taking $\Delta$ and $j$ in the long block to certain values below unitarity, i.e.~$\Delta<j+1$. For instance, we can try to obtain the superconformal block of the $(2,0)^{[0040]}_{(B,+)}$ multiplet (see Table~\ref{Bp}) by taking $\Delta\to-2$ and $j\to 0$ in the long superconformal block. In this limit 
 \es{A22Limit}{
   A_{22}^{(\mathrm{long)}} \propto g_{\Delta+4,j} \to g_{2,0} \sim A_{22}^{(B,+)} \,, \qquad
    \text{as $\Delta \to -2$ and $j \to 0$ \,.}
 }

Note that such limits have to be taken with great care for two reasons. The first reason is that some of the conformal blocks $g_{\Delta, j}$ are divergent in this limit, but the coefficients multiplying them vanish, so the limit is finite.  The divergence arises because the conformal blocks $g_{\Delta, j}$, viewed as functions of $\Delta$, have poles below unitarity.  The location and residues of these poles were computed in \cite{Kos:2013tga}.  For example, there is a ``twist-$0$'' pole at $\Delta = j$ given by
\begin{align}
g_{\Delta,j} \sim  -2\,\frac{j(j-1)}{4j^2-1} \frac{g_{j+2,j-2}}{\Delta-j} \,, \qquad \text{as $\Delta\to j$ \,.}
\end{align}
The second reason why the limits have to be taken with care is that the limits $\Delta \to 2$ and $j \to 0$ do not commute, so the result is ambiguous.  We parameterize this ambiguity by taking first $\Delta=-2+c j$ and later sending $j\to 0$.  The constant $c$ is kept arbitrary at this stage.

Taking the above considerations into account, for the  $(2,0)^{[0040]}_{(B,+)}$ multiplet we find\footnote{We use the identity $g_{\Delta,-j-1} = g_{\Delta,j}$, which can be derived from the conformal Casimir equation.}
\begin{align}
-\frac{1}{128}\lim_{j\to 0} \,\lim_{\Delta\to -2 + c j} A_{22}^{(\mathrm{long})} &= \frac{c+1}{c-1} g_{2,0} \ec \\
-\frac{1}{128}\lim_{j\to 0} \, \lim_{\Delta\to -2 + c j} A_{21}^{(\mathrm{long})} &= - \frac{4(c+1)}{3(c-1)} g_{3,1} -  \frac{3}{2(c-1)} g_{1,0} \ec \\
-\frac{1}{128}\lim_{j\to 0} \, \lim_{\Delta\to -2 + c j} A_{20}^{(\mathrm{long})} &= \frac{8(2c-1)(c+1)}{45c(c-1)} g_{4,0} + \frac{3(2c-1)}{4c(c-1)}g_{2,1} \ec \\
-\frac{1}{128}\lim_{j\to 0} \, \lim_{\Delta\to -2 + c j} A_{11}^{(\mathrm{long})} &= \frac{256(c+1)}{675(c-1)} g_{4,2} + \frac{3}{8(c-1)} g_{0,1} \ec \\
-\frac{1}{128}\lim_{j\to 0} \, \lim_{\Delta\to -2 + c j} A_{10}^{(\mathrm{long})} &= - \frac{64(2c-1)(c+1)}{875c(c-1)}g_{5,1} - \frac{2c-1}{4c(c-1)}g_{3,2} - \frac{1}{10(c-1)}g_{1,0} \notag\\
&- \frac{2c-1}{8c(c-1)}g_{1,2} \ec \\
-\frac{1}{128}\lim_{j\to 0} \,  \lim_{\Delta\to -2 + c j} A_{00}^{(\mathrm{long})} &= \frac{128(2c-1)(c+1)}{18375c(c-1)} g_{6,0} + \frac{2c-1}{70c(c-1)}g_{2,1} + \frac{9(2c-1)}{320c(c-1)}g_{2,3} \label{BpExmp}\ed
\end{align}
This result is, in general, inconsistent with unitarity because of the appearance of conformal blocks with negative twists such as $g_{2, 3}$.  These unphysical blocks can be removed in the limit $c \to \infty$.   In this limit, the result matches precisely the $(2,0)^{[0040]}_{(B,+)}$ superconformal block in \eqref{A22Bp}--\eqref{A00Bp}, and we conclude that
\begin{align}
\cG_{2,0}^{(2,2)} = -\frac{1}{128} \lim_{c\to\infty} \lim_{j\to 0} \lim_{\Delta\to -2 + c j} \cG_{\Delta,j}^{(0,0)} \ed
\end{align}

All other short superconformal blocks can be obtained from the long block in a similar fashion.  Hence all the superconformal blocks can be derived from the solution \eqref{wardsol3} of the Ward identity, because we derived the long superconformal block by using this solution and all the short blocks are limits of the long block.   This derivation provides a strong consistency check on the expressions for the superconformal blocks given in Appendix~\ref{superblocks} and on the solution \eqref{wardsol3} of the Ward identity.




\section{Central Charge Computation}\label{centralcharge}

For the numerical bootstrap, we need to specify an input that distinguishes different ${\cal N} =8$ SCFTs.   
As with the 4-d $\cN=4$  case in \cite{Beem:2013qxa},  we use the central charge  $c_T$, defined as the overall coefficient appearing in the two-point function of  the canonically normalized stress tensor\cite{Osborn:1993cr}:
 \es{TmnCorr}{
  \langle T_{\mu\nu}(x) T_{\rho \sigma}(0) \rangle =  \frac{c_T}{64}\left( P_{\mu\rho} P_{\nu\sigma} + P_{\nu \rho} P_{\mu \sigma} - P_{\mu\nu} P_{\rho\sigma} \right) \frac{1}{16 \pi^2 x^2} \,,
 }
where $P_{\mu\nu} \equiv \eta_{\mu\nu} \nabla^2 - \partial_\mu \partial_\nu$.  In \eqref{TmnCorr}, we normalized $c_T$ such that it equals one for a real massless scalar or Majorana fermion. 
For SCFTs preserving $\cN \geq 2$ supersymmetry one can use supersymmetric localization \cite{Kapustin:2009kz, Jafferis:2010un,Hama:2010av} on the three-sphere to compute $c_T$ exactly\cite{Closset:2012vg} .

There are two approaches to using supersymmetric localization to compute $c_T$. 
One way is to compute the squashed sphere partition function $Z_b = e^{-F_b}$\cite{Imamura:2011wg, Hama:2011ea} of the theory with squashing parameter $b$, where $b=1$ corresponds to the round sphere. 
Taking the derivative with respect to the squashing parameter, the central charge can be computed as
$ c_T = \frac{32}{\pi^2} \Re \left. \frac{\partial^2 F_b}{\partial b^2} \right \rvert_{b=1}$\cite{Closset:2012ru}. 
This computation has been carried out in \cite{Closset:2012ru, Nishioka:2013gza, Nishioka:2013haa} in a few simple examples.

Another way of obtaining $c_T$ makes use of having extended supersymmetry.  In our ${\cal N} = 8$ SCFTs, the stress tensor sits in the same $\mathfrak{osp}(8|4)$ multiplet as the $\mathfrak{so}(8)$ R-symmetry current.  But any ${\cal N} = 8$ SCFT can also be thought of as an ${\cal N} = 2$ SCFT by considering an $\mathfrak{osp}(2|4)$ sub-algebra of $\mathfrak{osp}(8|4)$.  From the ${\cal N} = 2$ point of view, the $\mathfrak{so}(8)$ R-symmetry current decomposes into the $\mathfrak{so}(2)$ R-symmetry current as well as several flavor currents.  There are three Abelian flavor currents that commute with one another and with the $\mathfrak{so}(2)$ R-symmetry current.  Together, these four currents generate the Cartan of $\mathfrak{so}(8)$. 

The extended supersymmetry relates $c_T$ to the coefficient appearing in the two-point function of the  Abelian flavor currents.
In general, the flat-space two-point functions of Abelian flavor currents $ j_a^\mu$, with $a$ being a flavor index, takes the form
\es{flavorCurrent}{
\langle j_a^\mu (x)j_b^\nu (0)\rangle =\frac{\tau_{ab}}{16 \pi^2}(\delta^{\mu \nu}\partial^2-\partial^\mu \partial^\nu)\frac{1}{x^2} \,.
}
The normalization in \eqref{flavorCurrent} is such that for a free chiral superfield (where there is only one flavor current $j^\mu$ corresponding to multiplication of the superfield by a phase) we have $\tau = 1$ provided that the chiral superfield carries unit charge under the flavor symmetry.  As explained in  \cite{Closset:2012vg}, the quantity $\tau_{ab}$ can be computed from the $S^3$ partition function corresponding to a supersymmetry-preserving deformation of the ${\cal N} =2$ SCFT\@.   This deformation can be interpreted as a mixing of the R-symmetry with the flavor symmetry, whereby the matter fields are assigned non-canonical R-charges.  The deformed $S^3$ partition function can be computed exactly using the supersymmetric localization results of \cite{Kapustin:2009kz, Jafferis:2010un,Hama:2010av}.

\subsection{Setup of the Computation}
We will follow the second approach for computing $c_T$ exactly in a few ${\cal N} =8$ SCFTs. 
In $\cN = 2$ notation, the matter content of all known $\cN=8$ theories consists of two vector multiplets with gauge group $G_1$ and $G_2$, respectively, and four chiral multiplets that transform in bifundamental representations of $G_1 \times G_2$.  Preserving the marginality of the quartic superpotential, one can consider the most general R-charge assignment parameterized as \cite{Jafferis:2011zi, Freedman:2013oja}
\es{deltas}{
\Delta_{A_1}&=\frac{1}{2}+t_1+t_2+t_3, \qquad \Delta_{A_2}=\frac{1}{2}+t_1-t_2-t_3 \,, \\
\Delta_{B_1}&=\frac{1}{2}-t_1+t_2-t_3,  \qquad \Delta_{B_2}=\frac{1}{2}-t_1-t_2+t_3 \,.
}
This parameterization is chosen such that $\tau_{ab}$ will be diagonal.  

$F$-maximization \cite{Jafferis:2011zi,Klebanov:2011gs,Casini:2012ei} tells us that $\Re F(\Delta) = - \ln \abs{Z(\Delta)}$, where $Z$ is the $S^3$ partition function, is maximized for the superconformal R-charge assignment $\Delta_\alpha=\Delta^* = 1/2$, i.e.~for $t =0$.  Equivalently, $\abs{Z}$ is minimized at $t = 0$, so  
\es{ZMinimization}{
 \frac{\partial \abs{Z}}{ \partial t_a} \bigg |_{t=0} = 0 \,.
 }
As explained in \cite{Closset:2012vg}, the coefficient $\tau_{ab}$ can be computed from the second derivative of $Z$ evaluated at $t=0$:
\es{tauFromZpp}{
\tau_{ab}=   \frac{2}{\pi^2} \Re \frac{1}{Z}\frac{\partial^2 Z }{\partial t_a \partial t_b}  \bigg |_{t=0} \,. 
}
As explained above, $c_T$ should be proportional to $\tau_{ab}$.  The coefficient of proportionality can be fixed through carefully defining representations of the $\mathfrak{osp}(8|4)$ algebra and then decomposing them into their $\mathfrak{osp}(2|4)$ sub-algebra representations.  A quicker way to fix the proportionality factor is from ABJM theory in the large $N$ limit, where $c_T$ is known from supergravity computations \cite{Buchel:2009sk} and $Z$ was computed as a function of $\Delta$ in~\cite{Jafferis:2011zi}.

The three-sphere partition function of R-charge deformed ABJ(M) theories with gauge group $U(M)_{k}\times U(N)_{-k}$ is given by \cite{Jafferis:2010un, Hama:2010av}
 \begin{align}
Z(\Delta)&\propto \int d^M \lambda\,  d^N \mu \,  e^{i \pi k \left[\sum_i\lambda_i^2- \sum_j \mu_j^2 \right] }  \prod_{i \neq j} \sinh [\pi(\lambda_{i}-\lambda_{j})]
 \prod_{i\neq j} \sinh[\pi(\mu_{i}-\mu_{j})] \prod_{\alpha} f_\alpha(\Delta) \,, \notag\\
 f_{\alpha} (\Delta) &\equiv \prod_{i, j} \exp\left[\ell \left(1-\Delta_\alpha + i  \, (-)^\alpha \, \left(\lambda_i-\mu_j\right)\right) \right] \,, \label{ZSphereABJ(M)}
 \end{align}
where $\ell(z)\equiv -z\ln(1-e^{2\pi i z})+\frac{i}{2}\left[ \pi z^2 +\frac{1}{\pi}\text{Li$_2$}(e^{2\pi i z}) \right]-\frac{i\pi}{12}$.  Here, $\alpha$ ranges from $1$ to $4$ and labels the chiral superfields of our theory.

The only $\Delta(t)$ dependence of \eqref{ZSphereABJ(M)} comes from $f_{\alpha} (\Delta)$.  To compute the second derivative required in \eqref{tauFromZpp}, note that 
 \begin{align}
  \frac{1}{\prod_\alpha f_{\alpha}} \frac{\partial^2}{\partial t_1^2} \prod_{\alpha } f_{\alpha}  \Bigg|_{t=0}
     &= 2 \pi^2\sum\limits_{i,j} \textrm{sech}^2 \left[  \pi (\lambda_i - \mu_j) \right] 
       - 4 \pi^2 \left[\sum\limits_{i,j} \tanh \left[  \pi (\lambda_i - \mu_j) \right] \right]^2 \,,\label{fderiv1} \\
   \frac{1}{\prod_\alpha f_{\alpha}} \frac{\partial^2}{\partial t_{2, 3}^2} \prod_{\alpha } f_{\alpha}  \Bigg|_{t=0} &=    
  2 \pi^2 \sum\limits_{i,j} \textrm{sech}^2 \left[  \pi (\lambda_i - \mu_j) \right] \label{fderiv} \,.
 \end{align}
At face value, it looks like \eqref{fderiv1} gives something different from \eqref{fderiv}.  One can check, however, that the extra term present in \eqref{fderiv1} does not contribute to \eqref{tauFromZpp} in the cases we study, as required from ${\cal N} = 8$ supersymmetry. 

Note that sometimes the $\cN=8$ theories that we consider have decoupled sectors.  For instance, the $U(N)_1\times U(N)_{-1}$ ABJM theory has a free $\cN=8$ sector \cite{Bashkirov:2010kz}, which is not visible at the level of the Lagrangian, but must clearly exist if one identifies this theory with the IR limit of $U(N)$ SYM\@.\footnote{On the gravity side this free sector simply corresponds to the center of mass motion of the stack of M2-branes.} In such cases the theory has more than one stress tensor, and our localization computations are only sensitive to the sum of the central charges corresponding to the different decoupled CFTs. 

In particular, we compute the central charges of $U(1)_1\times U(1)_{-1}$, $U(2)_2\times U(1)_{-2}$, $U(2)_1\times U(2)_{-1}$ and $U(2)_2\times U(2)_{-2}$ ABJ(M) theories, which are expected to be equivalent to the IR limit of $\cN=8$ SYM with gauge groups $U(1)$, $SO(3)\simeq SU(2)$, $U(2)\simeq SU(2)\times U(1)$ and $SO(4)\simeq SU(2)\times SU(2)$, respectively. Therefore the $U(2)_1\times U(2)_{-1}$ theory factorizes into a product of the $U(1)_1\times U(1)_{-1}$ and $U(2)_2\times U(1)_{-2}$ theories, while $U(2)_2\times U(2)_{-2}$ ABJM factorizes into two copies of $U(2)_2\times U(1)_{-2}$ ABJ theories. Indeed we find that the central charges computed below for these product CFTs are given by the appropriate sum of central charges corresponding to the irreducible CFTs (see Table \ref{cTValues}).\footnote{We are grateful to O.~Aharony for pointing this out to us.}

\subsection{Large $N$ Limit}\label{section:largeNcT}
First let's consider the theories with supergravity dual descriptions. 
For a theory admitting an  $AdS_{4}$ dual description, the sphere free energy $F$ is proportional to the central charge
\es{cTandF}{
c_T = \frac{64}{\pi^2} F \,.
}
This relation follows from the fact that the central charge in our normalization is $c_T = \frac{32 L^2}{\pi G_4}$ \cite{Buchel:2009sk}, and the $S^3$ free energy is  $F =\frac{\pi L^2}{2G_4}$ \cite{Drukker:2010nc, Jafferis:2011zi}.\footnote{ $L$ is the $AdS_4$ radius, and $G_4$ stands for the 4-dimensional Newton constant.}

Using localization, the large $N$ limit of $F$ is given by the $N^{3/2}$ scaling law\cite{Drukker:2010nc, Jafferis:2011zi}
\es{FlargeN}{
F(\Delta)=\frac{4\pi}{3} \sqrt{2k}N^{3/2} \sqrt{\Delta_{A_1}\Delta_{A_2}\Delta_{B_1}\Delta_{B_2}} \,.
}
Combining this expression with \eqref{cTandF} gives us the central charge of ABJM theories in the large $N$ limit
\es{largeNcT}{
c_T =  \frac{64}{3\pi}\sqrt{2 k} \, N^{3/2} \,.
} 
The  flavor current two-point function coefficient can be computed using \eqref{tauFromZpp}, so
\es{cTvialocalization}{
c_T= 4 \, \tau_{ff} = \frac{8}{\pi^2} \Re\frac{1}{Z}\frac{\partial^2 Z }{\partial \,  t_a^2} \Bigg |_{t=0} \,.
}

$Z(\Delta^*)$ for  ABJ(M) theories have been computed for various values of $N$ and $k$ using Fermi-gas techniques  \cite{Okuyama:2011su, Putrov:2012zi, Hatsuda:2012hm, Awata:2012jb, Honda:2014npa}. 
Thus to compute $c_T$ it suffices to compute $\frac{\partial^2 Z }{\partial \,  t_a^2}  \Big |_{t=0} $, as we do for a few examples in the subsequent sections.
It would be interesting to see if the techniques used for computing $Z(\Delta^*)$  in \cite{Okuyama:2011su, Putrov:2012zi, Hatsuda:2012hm, Awata:2012jb, Honda:2014npa} can be generalized to compute $c_T$ systematically.

\subsection{$U(1) \times U(1)$ ABJM Theory}
In the free $U(1)_k \times U(1)_{-k}$ ABJM theory, the three-sphere partition function \eqref{ZSphereABJ(M)} is
\es{FAbelian}{
Z(\Delta)=\int^{\infty}_{-\infty} \int^{\infty}_{-\infty}d \lambda\,  d\mu \exp \left( i \pi k (\lambda^2-\mu^2)\right)\prod_{\alpha}\exp\left(\ell \left(1-\Delta_\alpha + i  \, (-)^\alpha \, \left(\lambda-\mu\right)\right) \right)\,.
}
The integration can be performed explicitly by changing integration variables to $u_\pm=\lambda \pm \mu$. 
This choice exploits the fact that the product is independent of $u_+$,  giving us a delta-function. 
Performing the integral, we obtain
\es{FAbelian2}{
Z(\Delta)=k^{-1}\exp\left( \sum_\alpha \ell (1-\Delta_\alpha)\right)\,.
}
We see that for $\Delta_\alpha= \Delta^* =1/2$, 
\es{AbelianZderiv}{
Z(\Delta^*)=\frac{1}{4 \, k}, \qquad \left. \frac{\partial^2 Z}{\partial t_a \partial t_b} \right \rvert_{t=0}= \frac{\pi^2}{2 k }\delta_{ab}  \,.
}
Using \eqref{cTvialocalization}, for the abelian theory we get  $c_T =16$.
This result can be directly interpreted from free theory counting. 
There are four chiral multiplets in the theory and for each chiral multiplet there is one complex scalar and one Dirac fermion, each of which contributes two units to the central charge.
In total, we get $16$.

\subsection{$U(2) \times U(2)$ ABJM and $SU(2) \times SU(2)$ BLG Theory}\label{section:N2ABJMBLG}
Now we turn to interacting theories. We consider $U(2)_k \times U(2)_{-k}$ ABJM theory at Chern-Simons level $k=1,2$ and  BLG theories at all $k$. 
This BLG theory can be described as a $SU(2)_k \times SU(2)_{-k}$ CS-matter theory \cite{Bandres:2008vf, VanRaamsdonk:2008ft}. 
The three-sphere partition function is related to that of  $U(2)_k \times U(2)_{-k}$ theory by having the Coulomb branch parameters in the Cartan elements of $U(2)$ sum to zero. 
We see that the central charge result for the two theories are equal.\footnote{See Appendix \ref{N2Detail} for details.} Using \eqref{cTvialocalization} we find the central charge in terms of one integral,
\es{U2final}{
c_T = 32 \left( 2- \frac{I_4}{I_2} \right),\qquad \text{with} \quad I_n \equiv \int_{-\infty}^{\infty} dy  \, y \frac{\tanh^{n}(\pi y)}{ \sinh(\pi k y) } \,.
}

For general $k$, the integral $I_n$ can be evaluated by contour integration as explained in \cite{Okuyama:2011su}.  Depending on whether $k$ is even or odd, one can choose a holomorphic function and a contour that integrates to $I_n$. Summing the residues of the poles gives
\es{I_n Series}{
I_2&= \begin{cases}
\frac{(-1)^{\frac{k-1}{2}}}{\pi } + \sum\limits_{s=1}^{k-1} \frac{(-1)^{s+1}}{2 k^2}(k-2 s) \tan ^2\left(\frac{\pi  s}{k}\right)\,, & \text{if $k$ is odd}\,, \\
-\frac{i^k}{\pi ^2 k} + \sum\limits_{s=1 \atop s\neq k/2}^{k-1}\frac{(-1)^{s+1} }{4 k^3} (k-2 s)^2 \tan ^2\left(\frac{\pi  s}{k}\right)\,, & \text{if $k$ is even} \,, 
\end{cases}\\
I_4&=\begin{cases}
\frac{i^{k+1} \left(3 k^2-8\right)}{6 \pi } + \sum\limits_{s=1}^{k-1} \frac{(-1)^s }{2 k^2}(k-2 s) \tan ^4\left(\frac{\pi  s}{k}\right)\,, & \text{if $k$ is odd} \,, \\
\frac{i^k \left(k^2-8\right)}{6 \pi ^2 k} + \sum\limits_{s=1 \atop s\neq k/2}^{k-1}\frac{(-1)^{s}}{4 k^3}(k-2 s)^2 \tan ^4\left(\frac{\pi  s}{k}\right)\,, & \text{if $k$ is even} \,. 
\end{cases}
}
For ABJM theories with $k= 1, 2$ we get
\es{cTABJMN2}{
c_T^{k=1}=\frac{112}{3} \approx \, 37.333, \qquad \qquad c_T^{k=2}=\frac{128}{3} \approx \, 42.667\,.\\
}

BLG theories have $\cN=8$ superconformal symmetry for any Chern-Simons level $k$.  For $k=3$, the central charge is
 \es{BLG3}{
  c_T = 16\, \frac{31 - 10\,\pi}{3 - \pi} \approx 46.9998 \,.
 }
One can also consider the large $k$ limit, where the theory becomes perturbative.  The central charge in the large $k$ limit is
\es{largekBLG}{
c_T  = 64 \left( 1 - \frac{\pi^2}{k^2}- \frac{13\pi^4}{3k^4}+ \frac{2539 \pi^6}{90 k^6} + O(1/k^8) \right)\,. 
}
The fact that the central charge asymptotes to $64$ can be understood from free theory counting. Four chiral multiplets in a single color factor contribute $16$ to the central charge.  As chiral multiplets are in the bifundamental representation, there are four copies of them, which sums to $64$.

\subsection{$U(2) \times U(1)$ ABJ Theory }

We now compute the central charge explicitly for  $U(2)_k \times U(1)_{-k}$ ABJ theory.\footnote{See section \ref{ABJ21Detail} for details.}  Carrying out the integral for both $Z$ and its second $t$-derivatives, we get
\es{cTABJ21}{
c_T =  32 \, \frac{ \int^{\infty}_{-\infty} dy  \tanh (\pi y)  \csch (k \pi y) \sech^2 ( \pi y) }{\int^{\infty}_{-\infty} dy  \tanh (\pi y) \csch (k \pi y) }\,. 
}
For the $\cN=8$ theories with $k=1, 2$ the central charges are
\es{N8cTABJ21}{
c_T^{k=1} =16\, , \qquad \qquad  c_T^{k=2}=\frac{64}{3} \approx \, 21.333\,.\\
}
The central charge for $k=1$ is consistent with the ABJ duality\cite{Aharony:2008gk, Kapustin:2010mh} as $U(2)_1\times U(1)_{-1}$ ABJ theory is dual to the $U(1)_1\times U(1)_{-1}$  ABJM theory.

\section{Numerics}
\label{numerics}

All ingredients are now in place for our numerical study of the crossing equations~\eqref{13crossing}.  Explicitly, in terms of the functions $A_{ab}(u, v)$ defined in \eqref{GExpansion} and expanded in superconformal blocks in Section~\ref{scblocks} (see also Appendix~\ref{superblocks}), these equations are:
 \es{dbasis}{
      \begin{pmatrix}
      d_1\\d_2\\d_3\\d_4\\d_5\\d_6
    \end{pmatrix} \equiv
    \begin{pmatrix}
      \cF^{+}_{10} + \cF^{+}_{11} + \frac{5}{3}  \cF^{+}_{20} -  \frac{2}{5} \cF^{+}_{21} -  \frac{14}{3} \cF^{+}_{22} \\
      \cF^{+}_{00} - \frac{1}{4} \cF^{+}_{11} -  \frac{20}{21}  \cF^{+}_{20} + \cF^{+}_{21} - \frac{14}{15}  \cF^{+}_{22} \\
      \cF^{-}_{20} + \cF^{-}_{21} + \cF^{-}_{22} \\
      \cF^{-}_{11} + \frac{4}{3} \cF^{-}_{21} + \frac{8}{3} \cF^{-}_{22} \\
      \cF^{-}_{10} + \frac{3}{5}  \cF^{-}_{21} + 3 \cF^{-}_{22}  \\  
      \cF^{-}_{00} - \frac{12}{7} \cF^{-}_{21} + \frac{24}{35}  \cF^{-}_{22}
   \end{pmatrix} = 0\,, 
   }
where we defined 
 \es{curlyF}{
   \cF^{\pm} _{ab} (u,v) \equiv \frac{1}{u} A_{ab} (u,v)  \pm   \frac{1}{v}  A_{ab} (v,u) \,.
 }
Recall that the contribution to $A_{ab}$ coming from each superconformal block takes the form of a linear combination of conformal blocks.  Note that the basis of equations $d_i = 0$ used here is different from the basis $\tilde e_i = 0$ of Section~\ref{RELCROSSING}.  The two bases are related by the linear transformation
\begin{align}
\begin{pmatrix}
d_1\\d_2\\d_3\\d_4\\d_5\\d_6
\end{pmatrix} = \begin{pmatrix}
0 & 1  & 0 & -1 & 0 & 0 \\
0 & -1 & 0 & 0  & 0 & 0 \\
0 &  0 & 0 & 0  & -\frac{1}{4} & \frac{1}{4} \\
\frac{1}{2} & 0 & -1 & 0 & \frac{1}{6} & \frac{1}{3} \\
-\frac{1}{2} & 0 & 0 & 0 & \frac{1}{4} & \frac{1}{4} \\
\frac{1}{8} & 0 & \frac{3}{4} & 0 & \frac{17}{56} & -\frac{5}{28}
\end{pmatrix} \begin{pmatrix}
\tilde{e}_1 \\ \tilde{e}_2 \\ \tilde{e}_3 \\ \tilde{e}_4 \\ \tilde{e}_5 \\ \tilde{e}_6 
\end{pmatrix} \ed \label{eTod}
\end{align}

Crossing equations such as \eqref{dbasis} have been used many times recently to rule out the existence of (S)CFTs whose spectrum of operators satisfies certain additional assumptions.  We will perform several such studies with or without additional assumptions besides locality (i.e.~existence of a stress tensor), unitarity, and invariance under the ${\cal N} = 8$ superconformal algebra $\mathfrak{osp}(8|4)$.  The main observation is that, when expanded in superconformal blocks, the crossing equations \eqref{dbasis} take the form
 \es{dExpanded}{
  d_i = \sum_{{\cal M}\, \in\, \mathfrak{osp}(8|4) \text{ multiplets}} \lambda_{\cal M}^2\,  d_{i, {\cal M}} = 0 \,,
 }
where ${\cal M}$ ranges over all the superconformal multiplets that appear in the OPE of ${\cal O}_{{\bf 35}_c}$ with itself---see Table~\ref{opemult}.  In \eqref{dExpanded}, $d_{i, {\cal M}}$ should be identified with the middle expression in \eqref{dbasis} in which one uses only the contributions to the ${\cal F}_{ab}^\pm$ coming from the superconformal block of the multiplet ${\cal M}$.  

There is in fact a superconformal multiplet that appears in the ${\cal O}_{{\bf 35}_c} \times {\cal O}_{{\bf 35}_c}$ OPE and that was omitted from Table~\ref{opemult}.  It is a rather trivial multiplet that consists solely of the identity operator in the $\mathfrak{so}(8)_R$ singlet channel.  Its superconformal block is given by
 \es{IdBlock}{
    A_{00} = 1 \,, 
 }   
with all other $A_{ab}$ vanishing.  We choose to set the OPE coefficient of this multiplet to $\lambda_\text{Id} = 1$.  This choice is equivalent to fixing the normalization of the operator ${\cal O}_{{\bf 35}_c}$ whose four-point function we want to study.  With the help of the null polarization variables $Y^i$ introduced in Section~\ref{conventions}, we can specify the normalization of ${\cal O}_{{\bf 35}_c}(x) = {\cal O}_{ij}(x) Y^i Y^j$ that corresponds to $\lambda_\text{Id} =1$ by requiring its two-point function to satisfy
 \es{O35Norm}{
  \langle {\cal O}_{{\bf 35}_c}(x_1, Y_1) {\cal O}_{{\bf 35}_c}(x_2, Y_2) \rangle
    = \frac{(Y_1 \cdot Y_2)^2}{\abs{x_1 - x_2}^2} \,.
 }

Importantly, the coefficients $\lambda_{\cal M}^2$ are positive in a unitary SCFT\@.  Their normalization is meaningful only once we specify the normalization of the (super)conformal blocks and that of the operator ${\cal O}_{{\bf 35}_c}$.  In our conventions, if the superconformal primary of ${\cal M}$ has conformal dimension $\Delta$, spin $j$, and transforms as the $(c, d) = [0\, (c-d)\, (2d)\, 0]$ of $\mathfrak{so}(8)_R$, then 
 \es{Normalization}{
   A_{cd}(x, \bar x) \sim \frac{\Gamma(j + 1/2)}{4^{\Delta}  \sqrt{\pi} \, \Gamma(j+1)} x^{\frac{1}{2}\left(\Delta+j\right)}\bar{x}^{\frac{1}{2}\left(\Delta-j\right)} \,, \qquad
    \text{as $x, \bar x \to 0$} \,,
 }
where $\bar x$ is taken to zero first.  (See also Appendix~\ref{confblock}.)

With the normalization described above, we can relate the OPE coefficient of the stress-tensor multiplet $(1, 0)_{(B, +)}^{[0020]}$ (which, for short, will henceforth be referred to as ``stress'') to the central charge $c_T$ discussed in the previous section.  We have\footnote{We stress that $\lambda_\text{stress}$ is not the OPE coefficient of the stress tensor in the ${\cal O}_{{\bf 35}_c} \times {\cal O}_{{\bf 35}_c}$ OPE, but instead the coefficient of the superconformal primary in the stress-tensor multiplet.  The OPE coefficient of the stress tensor is $\lambda_{3, 2} = \lambda_\text{stress}/2$.}
 \es{lambdaStress}{
  \lambda_\text{stress}^2 = \frac{256}{c_T} \,,
 }
where, as in the previous section, we normalized $c_T$ so that $c_T = 1$ for a theory of a free real scalar field or a free Majorana fermion.  In Table~\ref{cTValues} we collect the lowest few values of $c_T$ that we computed in the previous section for known SCFTs with ${\cal N} = 8$ supersymmetry.
 \begin{table}[htdp]
\begin{center}
\begin{tabular}{l|c}
 \multicolumn{1}{c|}{SCFT} & $c_T$ \\
  \hline
  $U(1)_k \times U(1)_{-k}$ ABJM & 16.0000 \\
  $U(2)_2 \times U(1)_{-2}$ ABJ & $21.3333$ \\
  $U(2)_1 \times U(2)_{-1}$ ABJM & $37.3333$ \\
  $U(2)_2 \times U(2)_{-2}$ ABJM & $42.6667$ \\
  $SU(2)_3 \times SU(2)_{-3}$ BLG & $46.9998$ \\
  $SU(2)_4 \times SU(2)_{-4}$ BLG & $50.3575$ \\
  $SU(2)_5 \times SU(2)_{-5}$ BLG & $52.9354$ \\
  \multicolumn{1}{c|}{\vdots} & \vdots
\end{tabular}
\end{center}
\caption{A few values of $c_T$ for known SCFTs.  See Section~\ref{centralcharge} for a derivation as well as analytical formulas for these central charges.}\label{cTValues}
\end{table}%

The approach for excluding (S)CFTs first introduced in \cite{Rattazzi:2008pe} starts with constructing linear functionals of the expressions $d_i$ that are required to vanish by crossing symmetry.  One can construct such linear functionals by considering linear combinations of the $d_i$ and of their derivatives at the crossing-symmetric point $x = \bar x = 1/2$.  Denoting such a functional by $\alpha$, we have
 \es{alphaDef}{
  \alpha(\vec{d}) = \sum_{i} \sum_{m \geq n} \alpha_{i, mn} \left( \partial^m \bar \partial^n d_i \right) \biggr|_{x = \bar x = \frac 12} \,,
 }
where $\alpha_{i, mn}$ are numerical coefficients.  In \eqref{alphaDef}, we restricted the second sum to run only over $m \geq n$ because $\partial^m \bar \partial^n d_i  = \partial^n \bar \partial^m d_i $, as follows from the fact that all conformal blocks are chosen to be invariant under $x \leftrightarrow \bar x$.  Without this restriction, we would be double counting all derivatives with $m \neq n$. 

Note that still not all the terms in the sum \eqref{alphaDef} are linearly independent.  There are two additional sources of linear dependencies between the various terms in \eqref{alphaDef}.  The first such source can be seen from the definitions \eqref{dbasis}--\eqref{curlyF} whereby $d_1$ and $d_2$ are even under $x \to 1-x$ and $\bar x \to 1 - \bar x$, while the other $d_i$ are odd.  Therefore, at the crossing-symmetric point $x = \bar x = 1/2$, we have $ \partial^m \bar \partial^n d_i  = 0 $ for $i = 1, 2$ and $m + n$ odd or $i = 3, 4, 5, 6$ and $m+n$ even.  We should not include these terms that vanish in \eqref{alphaDef}.

The second source of dependencies is more subtle and follows from the discussion in Section~\ref{RELCROSSING}.  Indeed, in Section~\ref{RELCROSSING} we have shown that the derivatives of the $\tilde e_i$ were not all independent.  The linear relation \eqref{eTod} then shows that the derivatives of the $d_i$ are also not all independent.  It is straightforward to check based on the results of Section~\ref{RELCROSSING} that a possibly independent set of derivatives of the $d_i$ consists of the derivatives of $d_2$ as well as the holomorphic derivatives of $d_1$.  There are many other such choices, but we make this one for convenience.

We can now attempt to find linear functionals \eqref{alphaDef} that satisfy certain positivity properties in order to obtain bounds on operator dimensions and OPE coefficients.

\subsection{Obtaining a Lower Bound on $c_T$}
\label{CTBOUND}

In the previous section we have seen that the $U(1) \times U(1)$ ABJM theory at level $k=1, 2$ is free and has $c_T = 16$.  This value can be obtained by adding up the equal unit contributions from the eight real scalars and eight Majorana fermions.  One may then wonder if there exist other ${\cal N} = 8$ SCFTs with $c_T<16$, or, given \eqref{lambdaStress}, with $\lambda_\text{stress}^2 > 16$.  Let us therefore use the bootstrap to find an upper bound on $\lambda_\text{stress}^2$.

The first step is to separate out the contributions from the identity multiplet and from the stress-tensor multiplet in \eqref{alphaDef}.  Since crossing requires $\vec{d} = 0$, we must have
 \es{alphaIdStress}{
  0 = \alpha (\vec{d}) = \alpha(\vec{d}_\text{Id}) + \lambda_\text{stress}^2  \alpha(\vec{d}_\text{stress}) + \sum_{{\cal M} \neq \text{Id}, \text{stress}} \lambda_{\cal M}^2 \alpha(\vec{d}_{\cal M})  \,.
 }
An upper bound on $\lambda_\text{stress}^2$ can be obtained by considering the space of functionals $\alpha$ that satisfy
 \es{CondStress}{
  \alpha(\vec{d}_\text{stress}) = 1 \,, \qquad \text{and} \qquad \alpha(\vec{d}_{\cal M}) \geq 0 \,, \qquad
   \text{for all ${\cal M} \neq \text{Id},\, \text{stress}$} \,.
 }
The conditions \eqref{CondStress} and the equation \eqref{alphaIdStress} imply the bound
 \es{stressBound}{
  \lambda_\text{stress}^2 \leq -\alpha(\vec{d}_\text{Id}) \,.
 }
To obtain the most stringent bound we should minimize $-\alpha(\vec{d}_\text{Id}) $ under the constraints \eqref{CondStress}.

The minimization problem described above needs to be truncated for a numerical implementation.  There are two truncations that should be performed:  one in the number of derivatives used to construct $\alpha$ and one in the range of multiplets ${\cal M}$ that we consider.  Instead of \eqref{alphaDef}, we can consider the truncated version
 \es{alphaTrunc}{
  \alpha_\Lambda (\vec{d}) = \sum_i \sum_{m+n \leq \Lambda} \alpha_{i, mn} \left( \partial^m \bar \partial^n d_i \right) \biggr|_{x = \bar x = \frac 12} \,,
 }
where the sum over $m$ and $n$ should only contain independent terms.  In practice, the cutoff $\Lambda$ that determines the size of our search space will be taken to be $\Lambda = 15$, $17$, or $19$.  We can then minimize $-\alpha_\Lambda(\vec{d}_\text{Id})$ under the constraints
 \es{alphaLambdaConstraints}{
  \alpha_\Lambda(\vec{d}_\text{stress}) &= 1 \,, \\
   \alpha_\Lambda(\vec{d}_{\cal M}) &\geq 0 \,, \qquad \text{for all other }{\cal M} \text{ with $j \leq j_\text{max}$ and $\Delta \geq j + 1$}
 } 
Here, $\Delta$ and $j$ refer to the conformal dimension and spin of the superconformal primary, and $\Delta \geq j+1$ is just the unitarity condition.  The second equation refers to all multiplets ${\cal M}$ other than the identity and the stress-tensor multiplet.  In practice, we found that taking $j_\text{max} = 20$ provides fairly accurate results.

For the long multiplet $(\Delta, j)^{[0000]}_{(A, 0)}$ (henceforth referred to as ``long'') the quantity $ \alpha_\Lambda(\vec{d}_\text{long})$ can further be approximated, for each spin, by a positive function times a polynomial in $\Delta$.  Such expansion is obtained by expanding the conformal blocks that comprise the long superconformal block in a Taylor series around $x = \bar x = 0$ using the recursion formula given in~\cite{Kos:2013tga}, and then approximating some of the poles as a function of $\Delta$ that appear in this expansion in terms of a smaller set of poles, as explained in the Appendix of~\cite{Kos:2013tga}.

The minimization of $-\alpha_\Lambda(\vec{d}_\text{Id})$ under the constraints \eqref{alphaLambdaConstraints} can then be rephrased as a semidefinite programing problem using the method developed in \cite{Poland:2011ey}.  This problem can be solved efficiently by freely available software such as {\tt sdpa\_gmp} \cite{sdpa}.  Implementing it as a dual problem, we obtain $\lambda_\text{stress}^2 \leq 17.02, 16.95, 16.67$, or equivalently, $c_T \geq 15.04, 15.11, 15.35$, for $\Lambda = 15, 17, 19$, respectively.  Clearly, it would be desirable to increase $\Lambda$ further, but we take these numerical results as good evidence that $c_T \geq 16$ in all local unitary SCFTs with ${\cal N} = 8$ supersymmetry.  In the rest of this paper we only study such SCFTs with $c_T \geq 16$.

\subsection{Bounds on Scaling Dimensions of Long Multiplets}
\label{LONGBOUND}

A small variation on the method presented in the Section~\ref{CTBOUND} yields upper bounds on the lowest scaling dimension $\Delta_j^*$ of spin-$j$ superconformal primaries in a long multiplet.  Such superconformal primaries must all be singlets under the $\mathfrak{so}(8)$ R-symmetry---see Table~\ref{opemult}, where the long multiplet is in the last line.  It is worth emphasizing that, as was the case in Section~\ref{CTBOUND}, these bounds do not depend on any assumptions about our ${\cal N} = 8$ SCFTs other than locality and unitarity.

The variation on the method presented in Section~\ref{CTBOUND} is as follows.  Let us fix $c_T$ and look for functionals $\alpha$ satisfying the following conditions:
 \es{CondLong}{
  &\alpha(\vec{d}_\text{Id}) + \frac{256}{c_T}  \alpha(\vec{d}_\text{stress}) = 1 \,, \\
  &\alpha(\vec{d}_{\cal M}) \geq 0 \,, \qquad \text{for all short and semi-short ${\cal M} \notin \{ \text{Id}, \text{stress} \}$} \,, \\
  &\alpha(\vec{d}_{\cal M}) \geq 0 \,, \qquad \text{for all long ${\cal M}$ with $\Delta \geq \Delta_j^*$} \,.
 }
The existence of any such functional $\alpha$ would prove inconsistent all SCFTs with the property that superconformal primaries of spin-$j$ long multiplets all have conformal dimension $\Delta \geq \Delta_j^*$, because if this were the case, then equation \eqref{alphaIdStress} could not possibly hold.  If we cannot find a functional $\alpha$ satisfying \eqref{CondLong}, then we would not be able to conclude anything about the existence of an SCFT for which superconformal primaries of spin-$j$ long multiplets all have conformal dimension $\Delta \geq \Delta_j^*$---such SCFTs may or may not be excluded by other consistency conditions we have not examined.  An instance in which a functional $\alpha$ with the properties \eqref{CondLong} should not exist is if $c_T$ is chosen to be that of an ABJ(M) or a BLG theory and if we only impose restrictions coming from unitarity, namely if we take $\Delta_j^* = j+1$ for all $j$.  Indeed, we should not be able to exclude the ABJ(M) and/or BLG theories, assuming that these theories are consistent as is believed to be the case.

As in the previous section, in order to make the problem \eqref{CondLong} amenable to a numerical study, we should truncate the number of spins used in the second and third lines to $j \leq j_\text{max}$ (where in practice we take $j_\text{max} = 20$) and replace $\alpha$ by $\alpha_\Lambda$ such that our search space becomes finite-dimensional.  We can then use {\tt sdpa\_gmp} to look for functionals $\alpha_\Lambda$ satisfying \eqref{CondLong} for various choices of $\Delta_j^*$.  In practice, we will take $\Lambda = 15$, $17$, and $19$.

We present three numerical studies:
 \begin{enumerate}
   \item We first find an upper bound on the lowest dimension $\Delta_0^*$ of a spin-$0$ long multiplet assuming that all long multiplets with spin $j>0$ are only restricted by the unitarity bound.  In other words, we set $\Delta_j^* = j+1$ for all $j>0$.  This upper bound is plotted as a function of $c_T$ in Figure~\ref{fig:cTDelta0Star} for $\Lambda = 15$ (in light brown), $\Lambda = 17$ (in black), and $\Lambda = 19$ (in orange).  As can be seen from Figure~\ref{fig:cTDelta0Star}, there is very good agreement between the latter two values of $\Lambda$, especially at large $c_T$.
\begin{figure}[t!]
\begin{center}
   \includegraphics[width=0.49\textwidth]{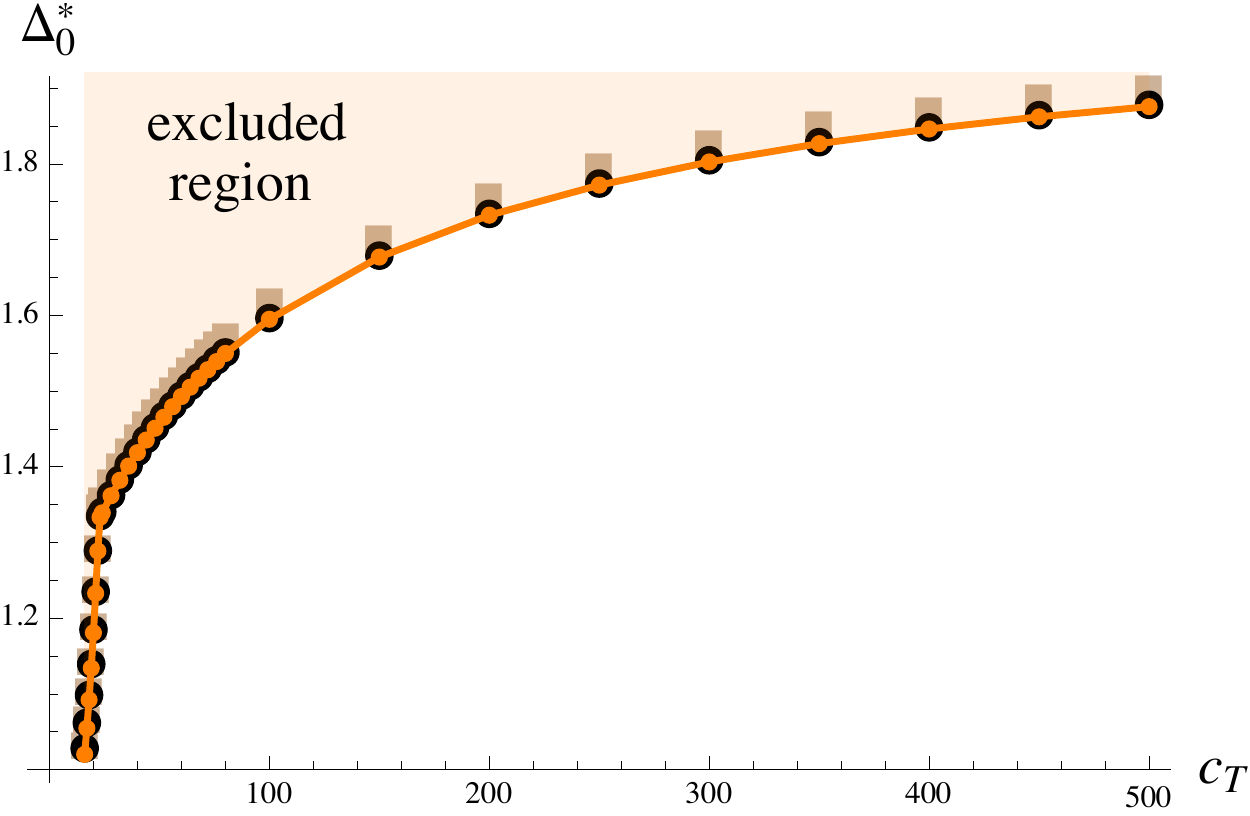}
   \includegraphics[width=0.5\textwidth]{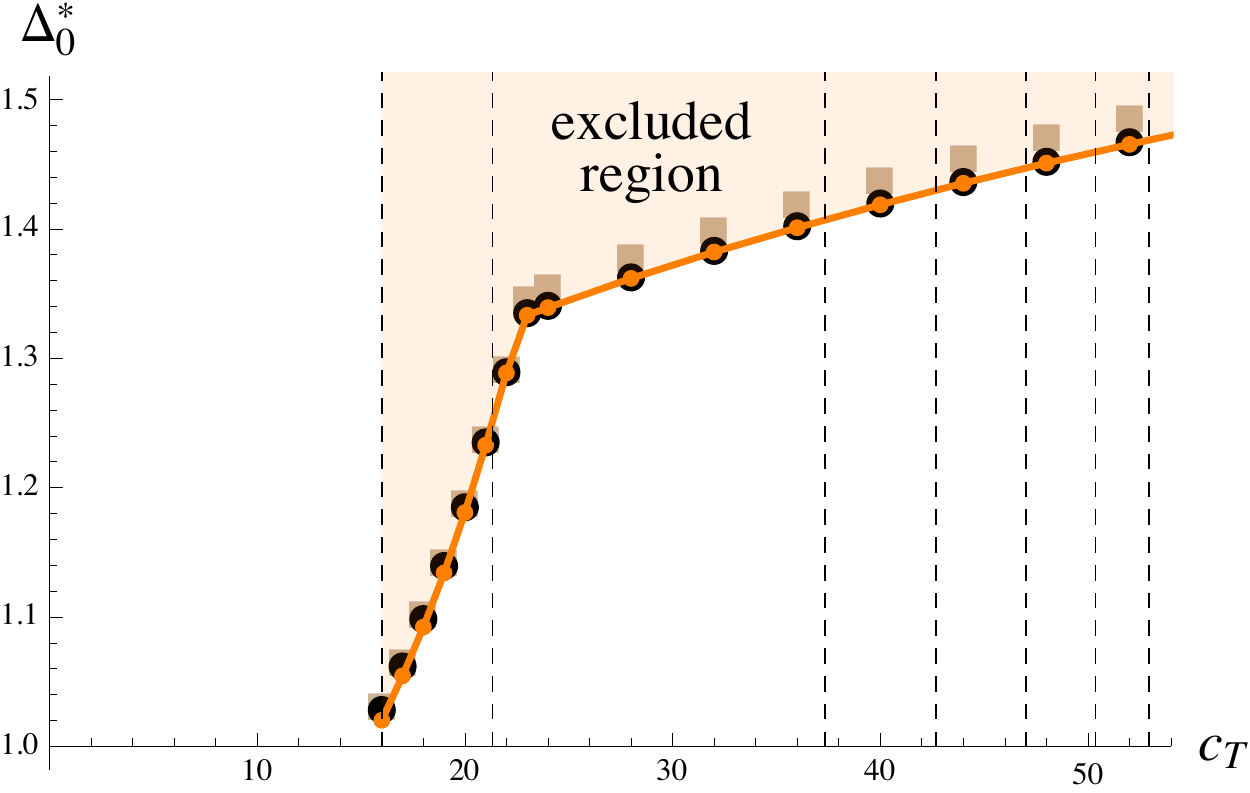}
\caption{Upper bounds on $\Delta_0^*$, which is the smallest conformal dimension of a long multiplet of spin-$0$ appearing in the ${\cal O}_{{\bf 35}_c} \times {\cal O}_{{\bf 35}_c}$ OPE\@.  The long multiplets of spin $j>0$ are only restricted by unitarity.  These bounds are computed with $j_\text{max} = 20$ and $\Lambda = 19$ (orange), $\Lambda = 17$ (black), and $\Lambda = 15$ (light brown).  The plot on the right is a zoomed-in version of the plot on the left.  The dashed vertical lines correspond to the values of $c_T$ in Table~\ref{cTValues}.}
\label{fig:cTDelta0Star}
\end{center}
\end{figure}

The upper bound on $\Delta_0^*$ interpolates monotonically between $\Delta_0^* \lsim 1.02$ at $c_T = 16$ and $\Delta_0^* \lsim 2.03$ as $c_T \to \infty$ when $\Lambda = 19$.   As we will now explain, these bounds are very close to being saturated by the $U(1)_k \times U(1)_{-k}$ ABJM theory at $c_T = 16$ and by the large $N$ $U(N)_k \times U(N)_{-k}$ ABJM theory (or its supergravity dual) at $c_T = \infty$.

Let us denote the real and imaginary parts of the bifundamental scalar matter fields in $U(N) \times U(N)$ ABJM theory with Chern-Simons levels $\pm 1$ or $\pm 2$ by $X_i$, with $i = 1, \ldots, 8$.  In our convention, the $X_i$ transform as the ${\bf 8}_c$ of the emergent $\mathfrak{so}(8)_R$.  The operator ${\cal O}_{ij}$ whose four-point function we have been analyzing transforms in the ${\bf 35}_c$ of $\mathfrak{so}(8)_R$.  It can be written schematically as\footnote{For Chern-Simons levels $k = 1, 2$, the products $X_i X_j$ must be combined with monopole operators into gauge invariant combinations.}
 \es{OijDef}{
  {\cal O}_{ij} = \tr \left[ X_i X_j - \frac{1}{8} \delta_{ij} X_k X^k \right] \,,
 }
up to an overall normalization.  There are two $\mathfrak{so}(8)$ singlets appearing in the ${\cal O}_{ij} \times {\cal O}_{kl}$ OPE as the bottom components of long multiplets  that are worth emphasizing:  the single trace operator ${\cal O}_K = \tr X_k X^k$, which is the analog of the Konishi operator in 4-d ${\cal N} = 4$ SYM, and the double trace operator ${\cal O}_{ij} {\cal O}^{ij}$.  When $N=1$, the theory is free, and ${\cal O}_K$ has scaling dimension $1$, while ${\cal O}_{ij} {\cal O}^{ij}$ has dimension $2$.  In this case $\Delta_0^* = 1$, and therefore this theory almost saturates our numerical bound.  When $N = \infty$, ${\cal O}_K$ is expected to acquire a large anomalous dimension,\footnote{Single trace long multiplets are not part of the supergravity spectrum.  The only single-trace operators that are dual to supergravity fluctuations around $AdS_4 \times S^7$ are part of the half-BPS multiplets $(n/2, 0)_{(B, +)}^{[00n0]}$ with $n \geq 2$ \cite{Biran:1983iy}.} while ${\cal O}_{ij} {\cal O}^{ij}$ still has dimension $2$ by large $N$ factorization.  Therefore, in this case $\Delta_0^* = 2$, and so the large $N$ ABJM theory also almost saturates our numerical bound.  It would be interesting to know whether for intermediate values $16 < c_T < \infty$ ABJM theory is close to saturating the bounds on $\Delta_0^*$ as well.

There is another feature of the bounds in Figure~\ref{fig:cTDelta0Star} that is worth noting:  as a function of $c_T$, the bound on $\Delta_0^*$  has a kink.  The location of the kink is approximately at $c_T \approx 22.8$ and $\Delta_0^* \approx 1.33$.  We do not know of any SCFT with $\mathfrak{osp}(8|4)$ symmetry at this particular value of $c_T$.  The known such SCFTs in this region are marked with dashed lines in Figure~\ref{fig:cTDelta0Star}.  At this point it is hard to know if the kink in Figure~\ref{fig:cTDelta0Star} has any physical meaning.  

From a fit at large values of $c_T$ we obtain $\Delta_0^* \gsim 2.03 - 94.6 / c_T + \ldots$.  See Figure~\ref{fig:logloglargecT}.  
\begin{figure}[t!]
\begin{center}
   \includegraphics[width=0.7\textwidth]{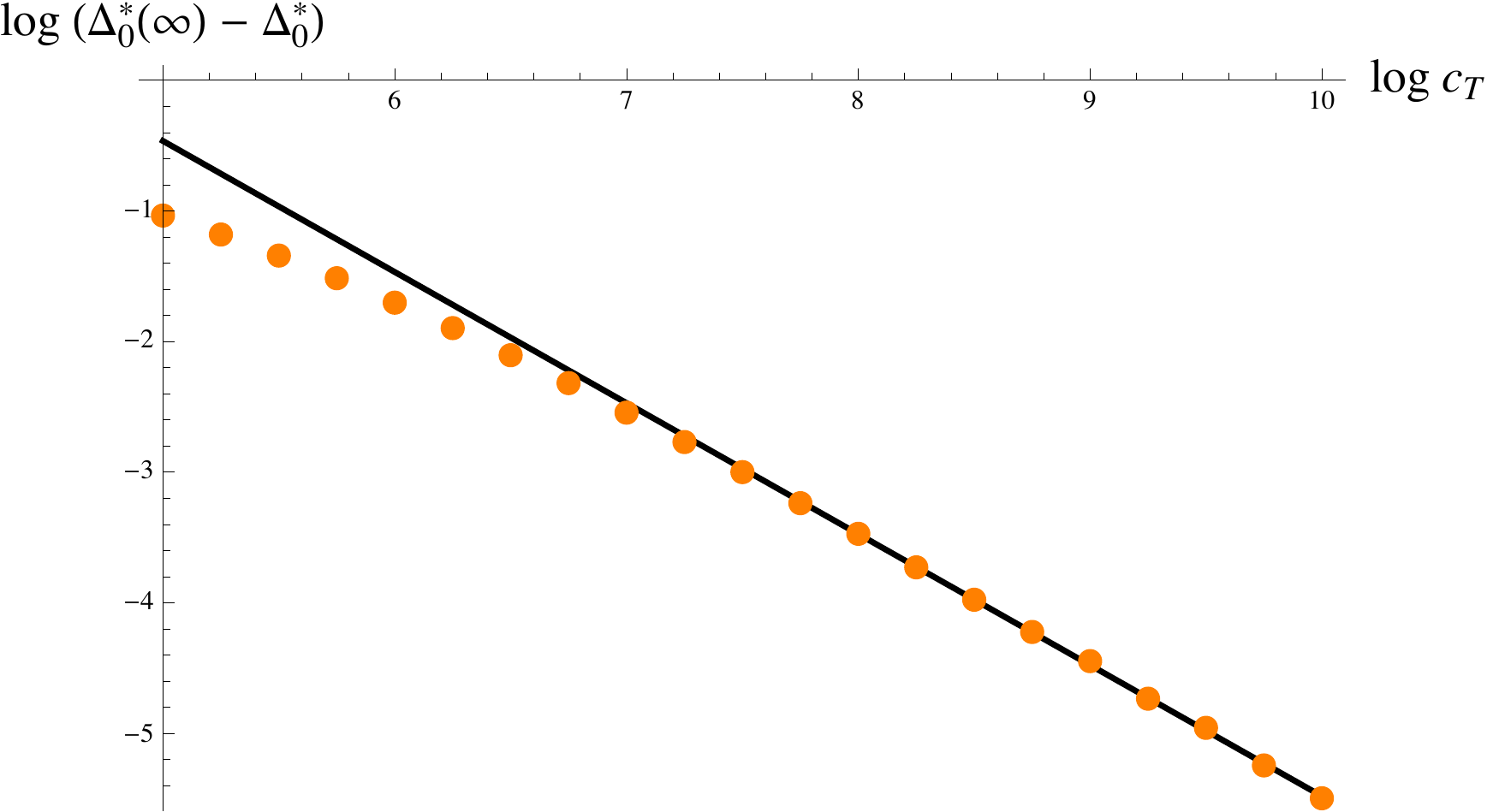}
\caption{Upper bounds on $\Delta_0^*$ (the smallest conformal dimension of a spin-$0$ long multiplet appearing in the ${\cal O}_{{\bf 35}_c} \times {\cal O}_{{\bf 35}_c}$ OPE) for large values of $c_T$.  The bounds are computed with $j_\text{max} = 20$ and $\Lambda = 19$.  The long multiplets of spin $j>0$ are only restricted by unitarity.  The best fit for the last ten points (shown in black) is $\log (\Delta_0^*(\infty) - \Delta_0^* ) = 4.55 - 1.00 \log c_T$.}
\label{fig:logloglargecT}
\end{center}
\end{figure}
In particular, the first subleading term at large $c_T$ scales as $1/c_T$.  Such a behavior is also what would be expected from supergravity.  Indeed, in radial quantization, the anomalous dimension of the double trace operator ${\cal O}_{ij} {\cal O}^{ij}$ takes the form of a binding energy, and, within supergravity, one expects such binding energies to be of the order of the effective $4$-d Newton constant $G_4 \propto 1/c_T$ (see Section~\ref{section:largeNcT}).\footnote{We thank I.~Klebanov for a discussion on this issue.}

 \item Our second numerical study is similar to the first.  Instead of obtaining an upper bound on $\Delta_0^*$, we now obtain an upper bound on $\Delta_2^*$, which is the lowest scaling dimension of a spin-$2$ long multiplet.  We obtain the bound on $\Delta_2^*$ under the assumption that long multiplets of spin $j \neq 2$ are only restricted by the unitarity condition.  In other words, we set $\Delta_j^* = j+ 1$ for all $j \neq 2$.  In Figure~\ref{fig:cTDelta2Star}, we plot the upper bound on $\Delta_2^*$ as a function of $c_T$ for $\Lambda = 15$ (in light brown), $\Lambda = 17$ (in black), and $\Lambda = 19$ (in orange).  The convergence as a function of $\Lambda$ is poorer than in the $\Delta_0^*$ case, but it is still reasonably good throughout, especially at large $c_T$.
\begin{figure}[t!]
\begin{center}
   \includegraphics[width=0.49\textwidth]{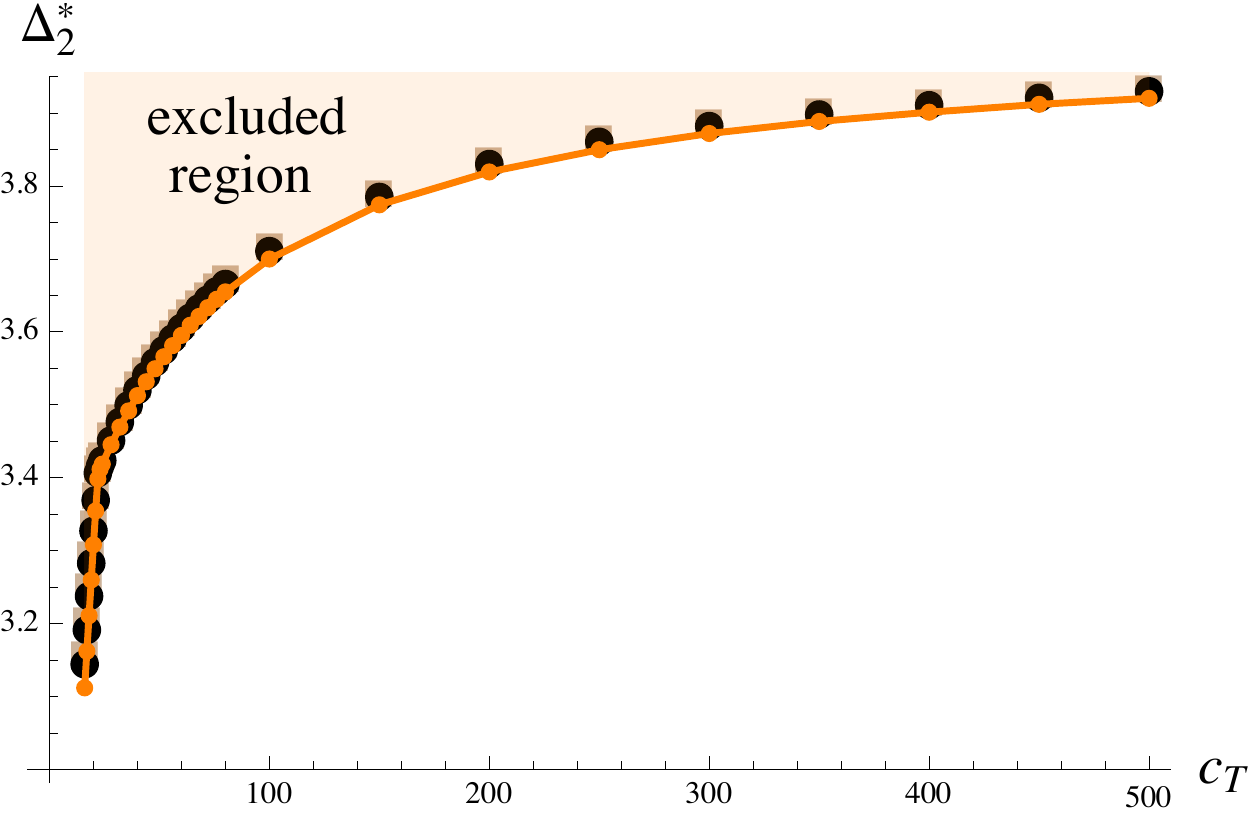}
   \includegraphics[width=0.5\textwidth]{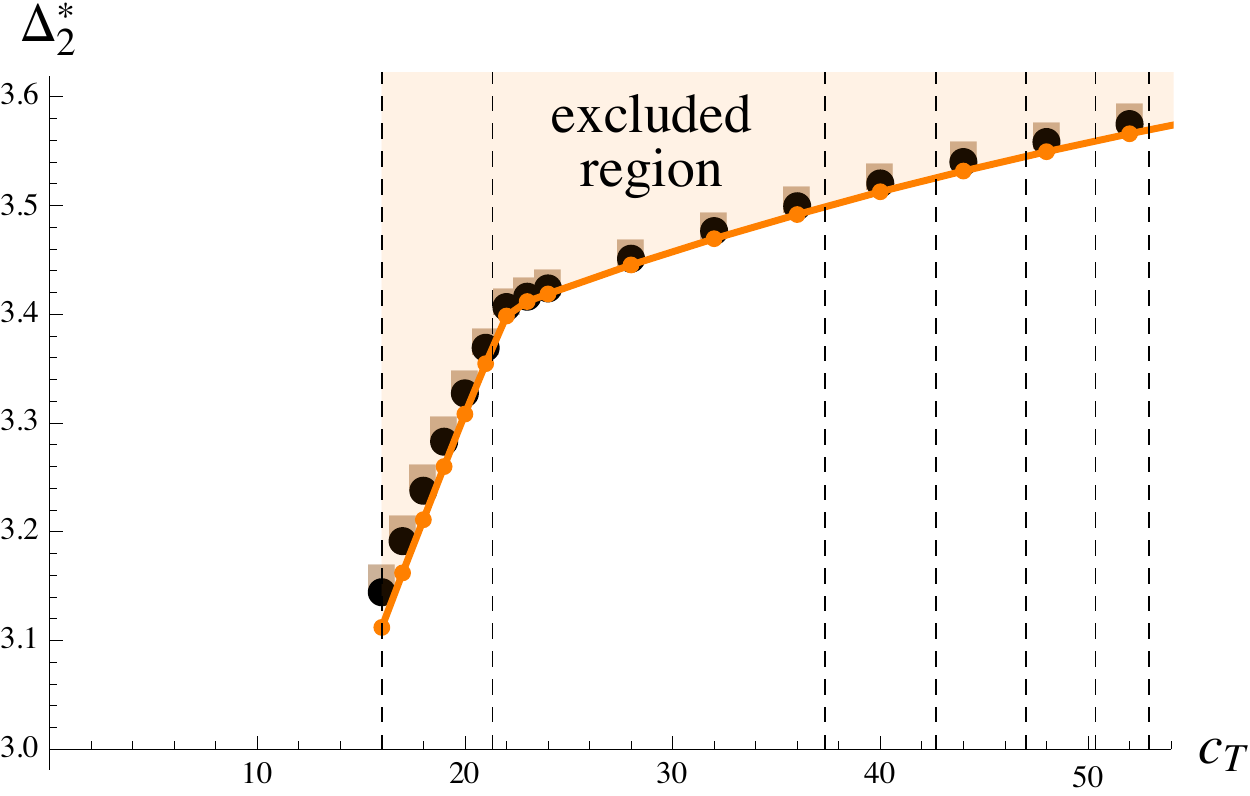}
\caption{Upper bounds on $\Delta_2^*$, which is the smallest conformal dimension of a long multiplet of spin-$2$ appearing in the ${\cal O}_{{\bf 35}_c} \times {\cal O}_{{\bf 35}_c}$ OPE\@.  The long multiplets of spin $j\neq 2$ are only restricted by unitarity.  These bounds are computed with $j_\text{max} = 20$ and $\Lambda = 19$ (orange), $\Lambda = 17$ (black), and $\Lambda = 15$ (light brown).  The plot on the right is a zoomed-in version of the plot on the left.  The dashed vertical lines correspond to the values of $c_T$ in Table~\ref{cTValues}.}
\label{fig:cTDelta2Star}
\end{center}
\end{figure}

A main feature of the plot in Figure~\ref{fig:cTDelta2Star} is that it interpolates monotonically between $\Delta_2^* \lsim 3.11$ at $c_T = 16$ and $\Delta_2^* \lsim 4.006$ at $c_T = \infty$.  It is likely that as one increases $\Lambda$, the bound at $c_T = 16$ will become stronger still, since at this value of $c_T$ the bound obtained when $\Lambda = 19$ is still noticeably different from that obtained when $\Lambda = 17$ and convergence has not yet been achieved.

As was the case for the bounds on $\Delta_0^*$, the bounds on $\Delta_2^*$ are also almost saturated by ABJM theory at $c_T = 16$ and $c_T = \infty$.  Indeed, two of the spin-$2$ $\mathfrak{so}(8)$ singlets that appear in the ${\cal O}_{ij} \times {\cal O}_{kl}$ OPE as bottom components of long multiplets are the single trace operator $\tr X_k \partial_\mu \partial_\nu X^k$ and the double trace operator ${\cal O}_{ij} \partial_\mu \partial_\nu {\cal O}^{ij}$.  For $U(1) \times U(1)$ ABJM theory, they have scaling dimensions $3$ and $4$, respectively;  in ABJM theory at infinite $N$, the first has a large anomalous dimension, while the second has scaling dimension $4$ because of large $N$ factorization.  Therefore, the $N=1$ theory has $\Delta_2^* = 3$, while the large $N$ theory has $\Delta_2^* = 4$, in agreement with our numerical bounds.

Note that just as in the $\Delta_0^*$ case, our upper bound on $\Delta_2^*$ in Figure~\ref{fig:cTDelta2Star} also exhibits a kink for $c_T \approx 22.8$.  Within our numerical precision, this kink is in the same location as that in Figure~\ref{fig:cTDelta0Star}.

 \item Our last numerical study yields combined upper bounds on $\Delta_0^*$ and $\Delta_2^*$ under the assumption that all long multiplets with spin $j>2$ are restricted only by the unitarity bound, i.e.~$\Delta_j^* = j+1$ for all $j>2$.  In Figure~\ref{fig:box02Plot} we provide such combined upper bounds only for a few values of $c_T$ corresponding to the ABJ(M) / BLG theories for which we computed $c_T$ in Section~\ref{centralcharge}.  
 
As can be seen from Figure~\ref{fig:box02Plot}, the combined bounds take the form of a rectangle in the $\Delta_0^*$-$\Delta_2^*$ plane, suggesting that these bounds are set by a single ${\cal N} = 8$ SCFT, if such an SCFT exists.  A similar feature is present for the ${\cal N} =4$ superconformal bootstrap in 4-d \cite{Beem:2013qxa}.

Note that for $c_T = \infty$, the combined $\Delta_0^*$-$\Delta_2^*$ bound comes very close to the values $(\Delta_0^*, \Delta_2^*) = (2, 4)$ of the large $N$ ABJM theory.

\begin{figure}[t!]
\begin{center}
   \includegraphics[width=\textwidth]{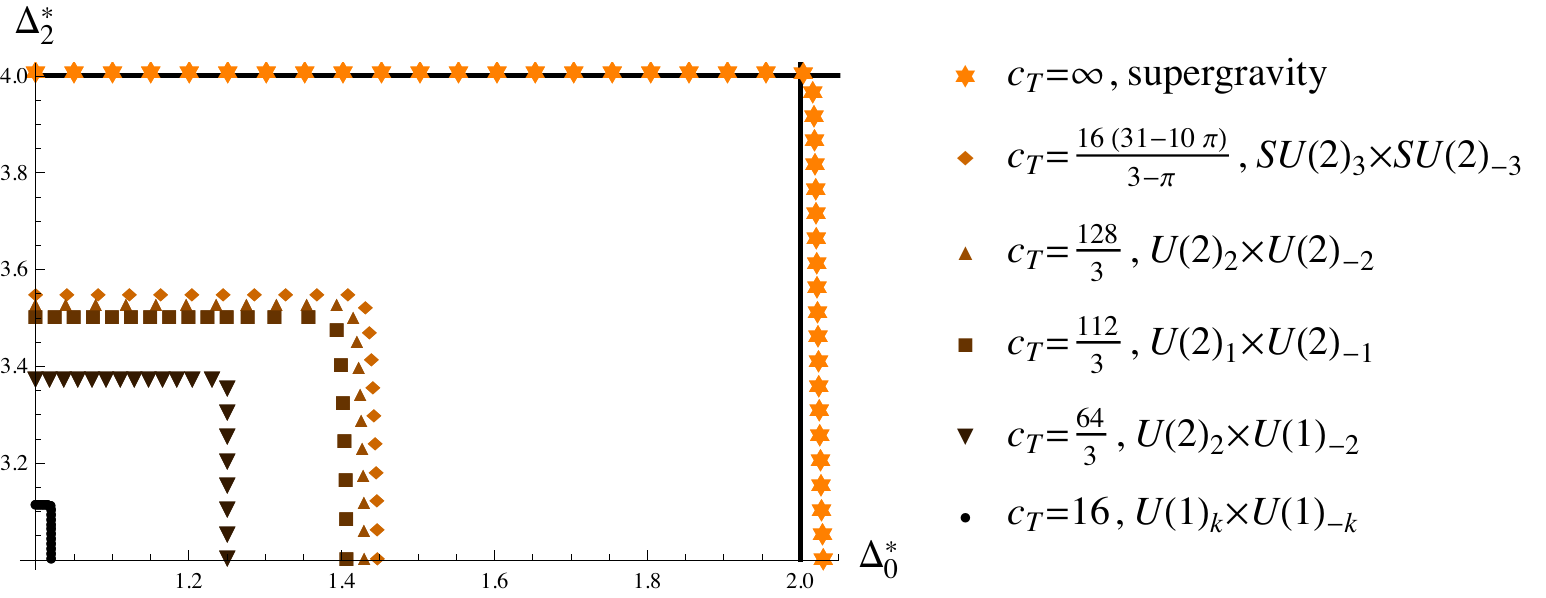}
\caption{Combined upper bounds on $\Delta_0^*$ and $\Delta_2^*$, which are the smallest scaling dimensions of spin-$0$ and spin-$2$ long multiplets appearing in the ${\cal O}_{{\bf 35}_c} \times {\cal O}_{{\bf 35}_c}$ OPE\@.  The long multiplets of spin $j>2$ are only restricted by unitarity.  The bounds are computed with $j_\text{max} = 20$ and $\Lambda = 19$.  The solid lines correspond to the expected scaling dimensions in ABJM theory at large $N$.}
\label{fig:box02Plot}
\end{center}
\end{figure}

\end{enumerate}

\subsection{Bounds on OPE Coefficients}
\label{OPEBOUND}

We can also obtain upper bounds on various OPE coefficients, just as we did in Section~\ref{CTBOUND} for $\lambda_\text{stress}^2$.  In this section we will only do so for the protected multiplets $(2, 0)_{(B, +)}^{[0040]}$ and $(2, 0)_{(B, 2)}^{[0200]}$, which for brevity will henceforth be denoted by $(B, +)$ and $(B, 2)$, respectively.  

An upper bound on $\lambda_{(B, +)}^2$, for instance, can be found by considering functionals $\alpha$ satisfying 
 \es{CondOPE}{
  &\alpha(\vec{d}_{(B, +)}) = 1 \,, \\
  &\alpha(\vec{d}_{\cal M}) \geq 0 \,, \qquad \text{for all short and semi-short ${\cal M} \notin \{ \text{Id}, \text{stress}, (B, +)  \}$} \,, \\
  &\alpha(\vec{d}_{\cal M}) \geq 0 \,, \qquad \text{for all long ${\cal M}$ with $\Delta \geq \Delta_j^*$} \,.
 }
If such a functional $\alpha$ exists, then \eqref{alphaIdStress} implies that 
 \es{UpperOPE}{
  \lambda_{(B, +)}^2 \leq - \alpha (\vec{d}_\text{Id})  - \frac{256}{c_T} \alpha( \vec{d}_\text{stress} ) \,,
 }
provided that all long multiplets $(\Delta, j)$ satisfy $\Delta \geq \Delta_j^*$.  (Choosing the unitarity values $\Delta_j^* = j+1$ provides no restrictions on the set of ${\cal N} = 8$ SCFTs for which the inequality \eqref{UpperOPE} holds.)  To obtain the most stringent upper bound on $\lambda_{(B, +)}^2$, one should then minimize the RHS of \eqref{UpperOPE} under the constraints \eqref{CondOPE}.  A similar prescription obtained by replacing $(B, +)$ by $(B, 2)$ in \eqref{CondOPE}--\eqref{UpperOPE} yields an upper bound on $\lambda_{(B, 2)}^2$.  As in the previous sections, one should consider a truncated version $\alpha_\Lambda$ of $\alpha$ and restrict the set of spins of the superconformal multiplets to a finite number such as $j \leq j_\text{max} = 20$.

As a warm-up, let us start with the $c_T = \infty$ limit and see how sensitive the bounds on $\lambda_{(B, +)}^2$ and $\lambda_{(B, 2)}^2$ are on the values of $\Delta_j^*$ that we choose.  Requiring only unitarity means setting $\Delta_j^* = j+1$ for all $j$.  When $c_T = \infty$ we know, however, that there exists an ${\cal N} = 8$ SCFT (namely ABJM theory with $k=1, 2$ and $N = \infty$) for which $\Delta_j^* = j+2$.  In Table~\ref{ChangeBound} we show the upper bounds on $\lambda_{(B, +)}^2$ and $\lambda_{(B, 2)}^2$ that we obtain under the assumption that $\Delta_j^* = j+1$ for $j<J$ and $\Delta_j^* = j+2$ for $j\geq J$ as we vary $J$.  
\begin{table}[htdp]
\begin{center}
\begin{tabular}{r|r|r}
 $J$ & $\lambda_{(B, +)}^2$ bound & $\lambda_{(B, 2)}^2$ bound \\
 \hline
 $0$ & $5.42443$ & $11.1221$ \\
 $2$ & $5.33344$ & $10.6672$ \\
 $4$ & $5.33344$ & $10.6672$ \\
 $6$ & $5.33338$ & $10.6669$ \\
 $8$ & $5.33338$ & $10.6669$ \\
 $10$ & $5.33337$ & $10.6668$ \\
 $12$ & $5.33337$ & $10.6668$ \\
 $14$ & $5.33336$ & $10.6668$

\end{tabular}
\caption{Upper bounds on OPE coefficients for the $(2, 0)_{(B, +)}^{[0040]}$ and $(2, 0)_{(B, 2)}^{[0200]}$ mutliplets.  These bounds are computed for $c_T = \infty$ and under the assumption that $\Delta_j^* = j+1$ for $j\geq J$ and $\Delta_j^* = j+2$ for $j<J$.}\label{ChangeBound}
\end{center}

\end{table}%
The bounds for $J = 0$ are the least restrictive and they hold in any SCFT with ${\cal N} = 8$ supersymmetry.  As we increase $J$, the bounds converge to 
 \es{LimitingValues}{
   \lambda_{(B, +)}^2 \to \frac{16}{3} \,, \qquad \lambda_{(B, 2)}^2 \to \frac{32}{3} \,, \qquad \text{(as $J \to \infty$ for $c_T = \infty$)}
 }
extremely quickly.  The limiting values in \eqref{LimitingValues} can be derived analytically using large $N$ factorization.  In the large $N$ limit, they correspond to the double-trace operators ${\cal O}_{ij} {\cal O}_{kl}$ projected onto the $[0040]$ (symmetric traceless) and $[0200]$ irreps of $\mathfrak{so}(8)_R$.

\begin{figure}[t!]
\begin{center}
   \includegraphics[width=0.49\textwidth]{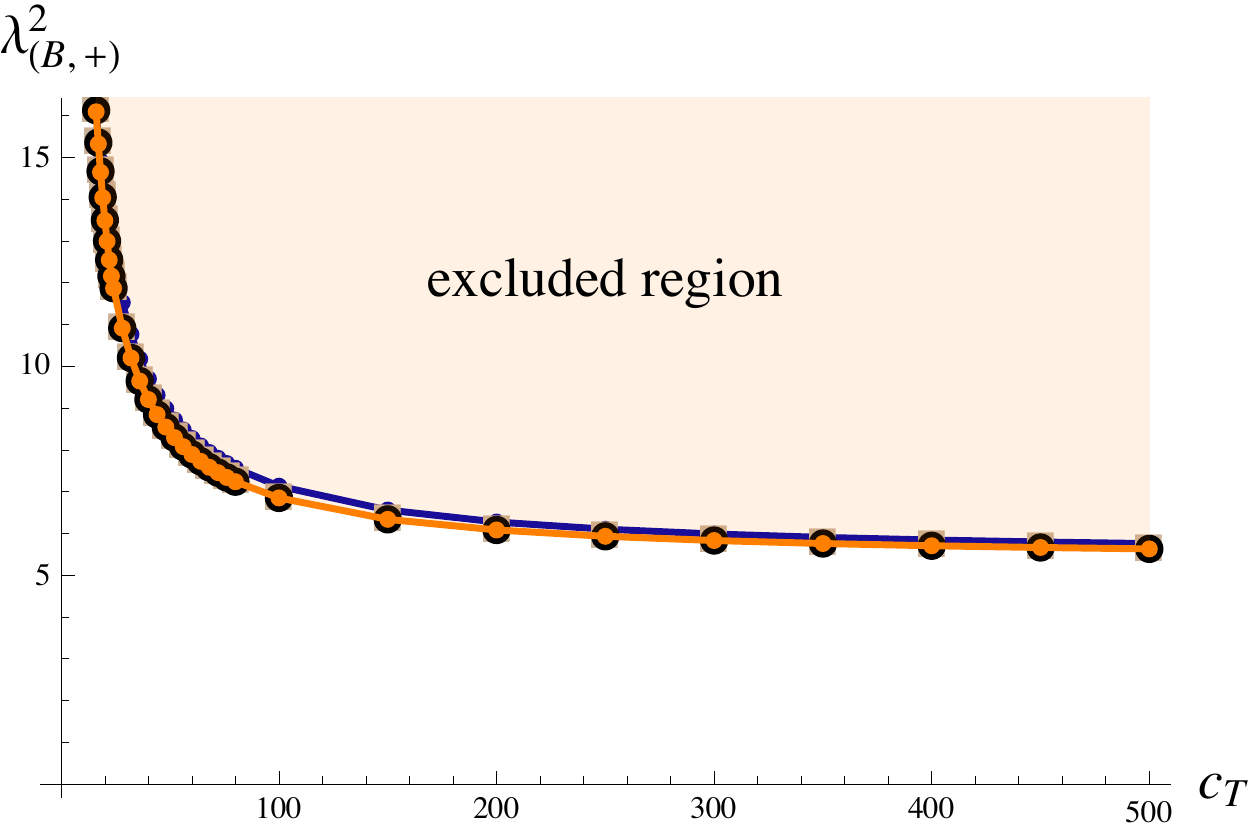}
   \includegraphics[width=0.5\textwidth]{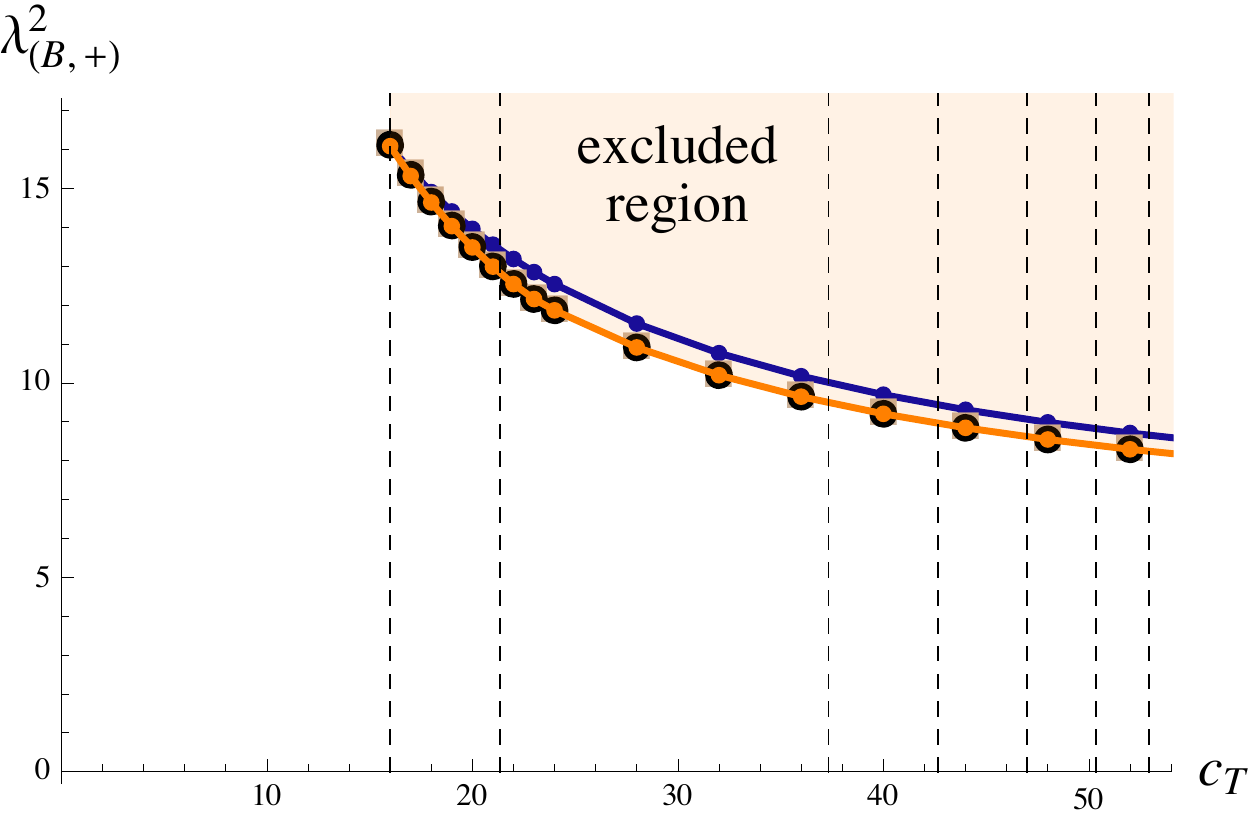}
\caption{Upper bounds on $\lambda_{(B, +)}^2$ using only the unitarity assumption (in blue) or a more restrictive assumption on scaling dimensions of long multiplets of spin-$0$ in orange.  (See main text.)  These bounds are computed with $j_\text{max} = 20$ and $\Lambda = 19$.  For the more restrictive bounds, we also show the corresponding values computed with $\Lambda= 17$ (in black) and $\Lambda = 15$ (in light brown).  The plot on the right is a zoomed-in version of the plot on the left.  The dashed vertical lines correspond to the values of $c_T$ in Table~\ref{cTValues}.}
\label{fig:BpCombined}
\end{center}
\end{figure}

\begin{figure}[t!]
\begin{center}
   \includegraphics[width=0.49\textwidth]{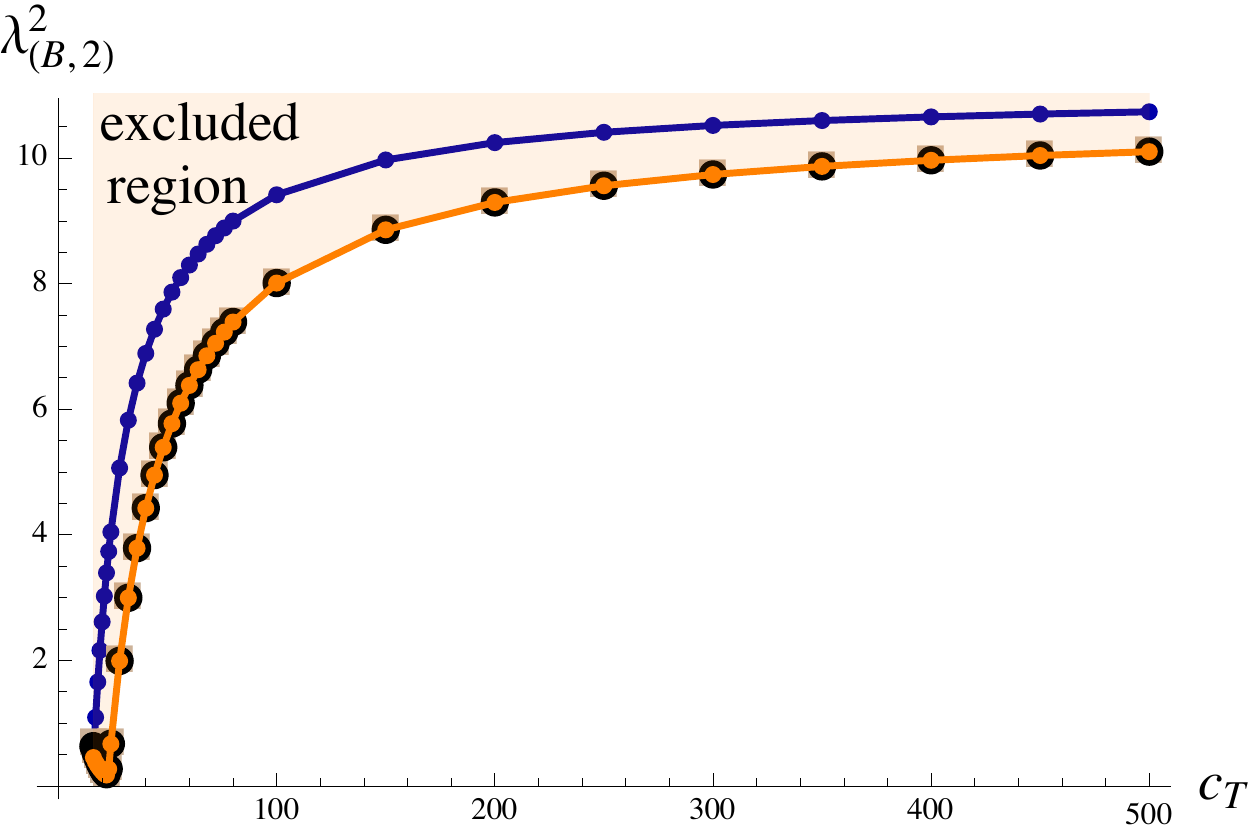}
   \includegraphics[width=0.5\textwidth]{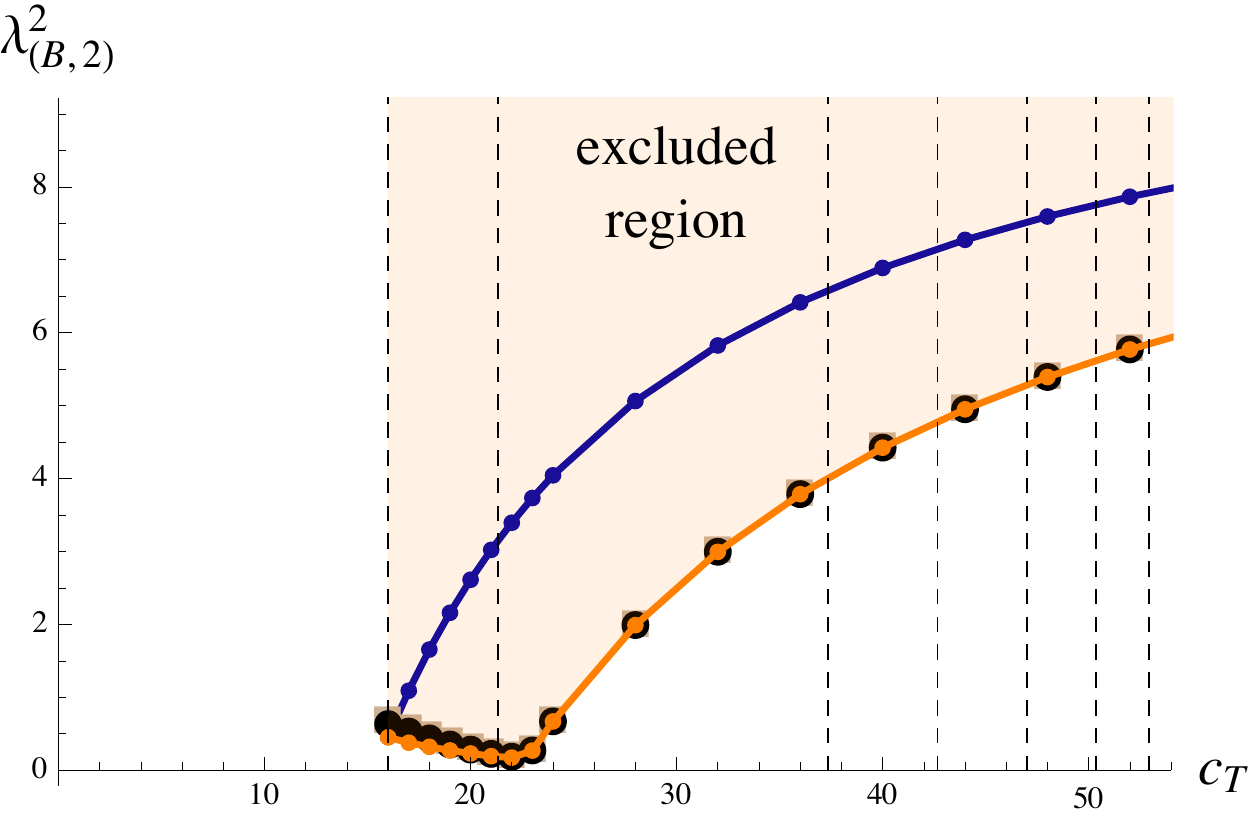}
\caption{Upper bounds on $\lambda_{(B, 2)}^2$ using only the unitarity assumption (in blue) or a more restrictive assumption on scaling dimensions of long multiplets of spin-$0$ in orange.  (See main text.)  These bounds are computed with $j_\text{max} = 20$ and $\Lambda = 19$.  For the more restrictive bounds, we also show the corresponding values computed with $\Lambda= 17$ (in black) and $\Lambda = 15$ (in light brown).  The plot on the right is a zoomed-in version of the plot on the left.  The dashed vertical lines correspond to the values of $c_T$ in Table~\ref{cTValues}.}
\label{fig:B2Combined}
\end{center}
\end{figure}

In Figures~\ref{fig:BpCombined} and~\ref{fig:B2Combined}, we show upper bounds on $\lambda_{(B, +)}^2$ and $\lambda_{(B, 2)}^2$ for a wide range of $c_T$.  The bounds plotted in blue correspond to $\Delta_j^* = j+1$ for all $j$ and hold for any ${\cal N}=8$ SCFTs.  The bounds plotted in orange are more restrictive.  They are obtained with $\Delta_j^* = j+1$ for all $j>0$ and $\Delta_0^*$ chosen approximately by the bounds given in Figure~\ref{fig:cTDelta0Star}.  At large $c_T$, these latter bounds approach approximately the limits in \eqref{LimitingValues}.  At $c_T = 16$, the upper bound for $\lambda_{(B, +)}^2$ is approximately $16$, while that for $\lambda_{(B, 2)}^2$ is very small.  In the $U(1) \times U(1)$ ABJM theory at CS level $k = 1, 2$ one can show analytically that $\lambda_{(B, +)}^2 = 16$ while $\lambda_{(B, 2)}^2 = 0$.  The latter value follows from the fact that there  are simply no $(B, 2)$ multiplets, because the projection of $X_i X_j X_k X_l$ onto the $[0200]$ irrep involves anti-symmetrizations of the $X_i$, which in this case commute.

\begin{figure}[t!]
\begin{center}
   \includegraphics[width=0.49\textwidth]{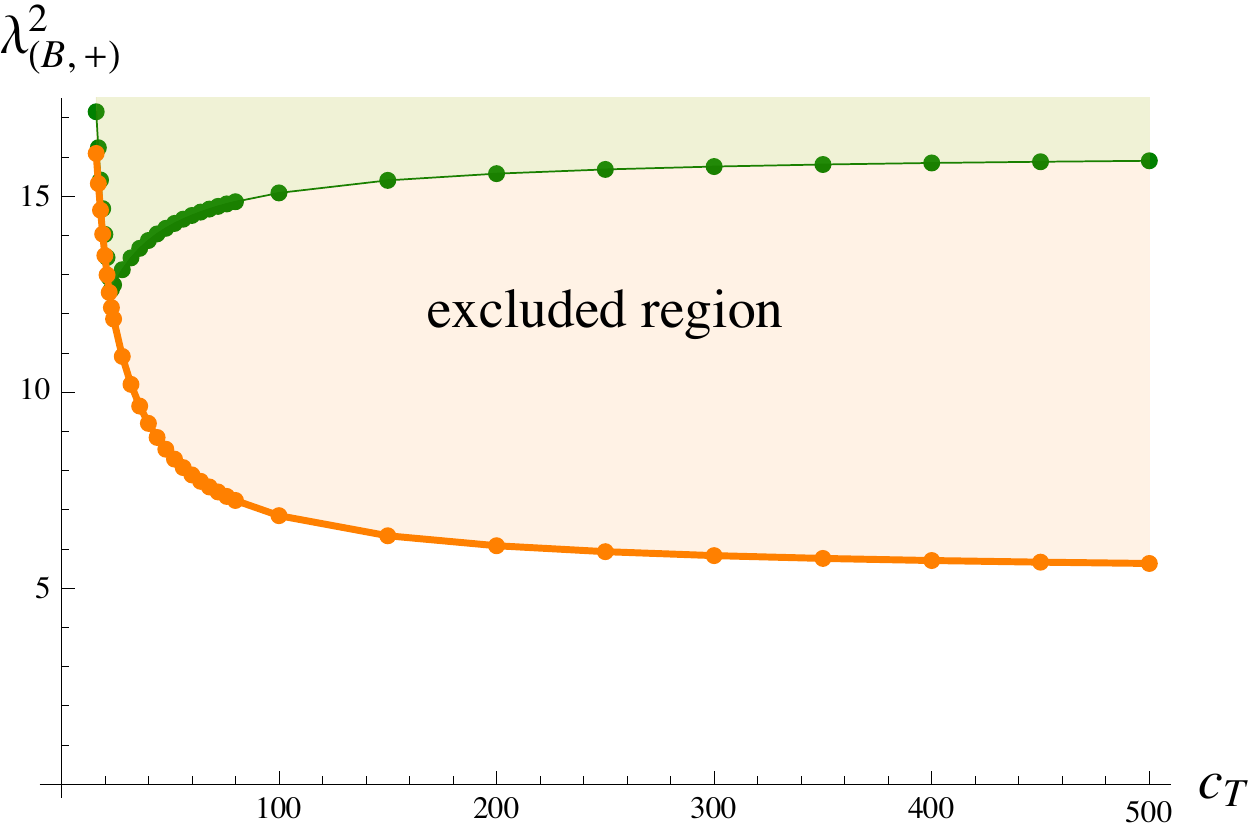}
   \includegraphics[width=0.49\textwidth]{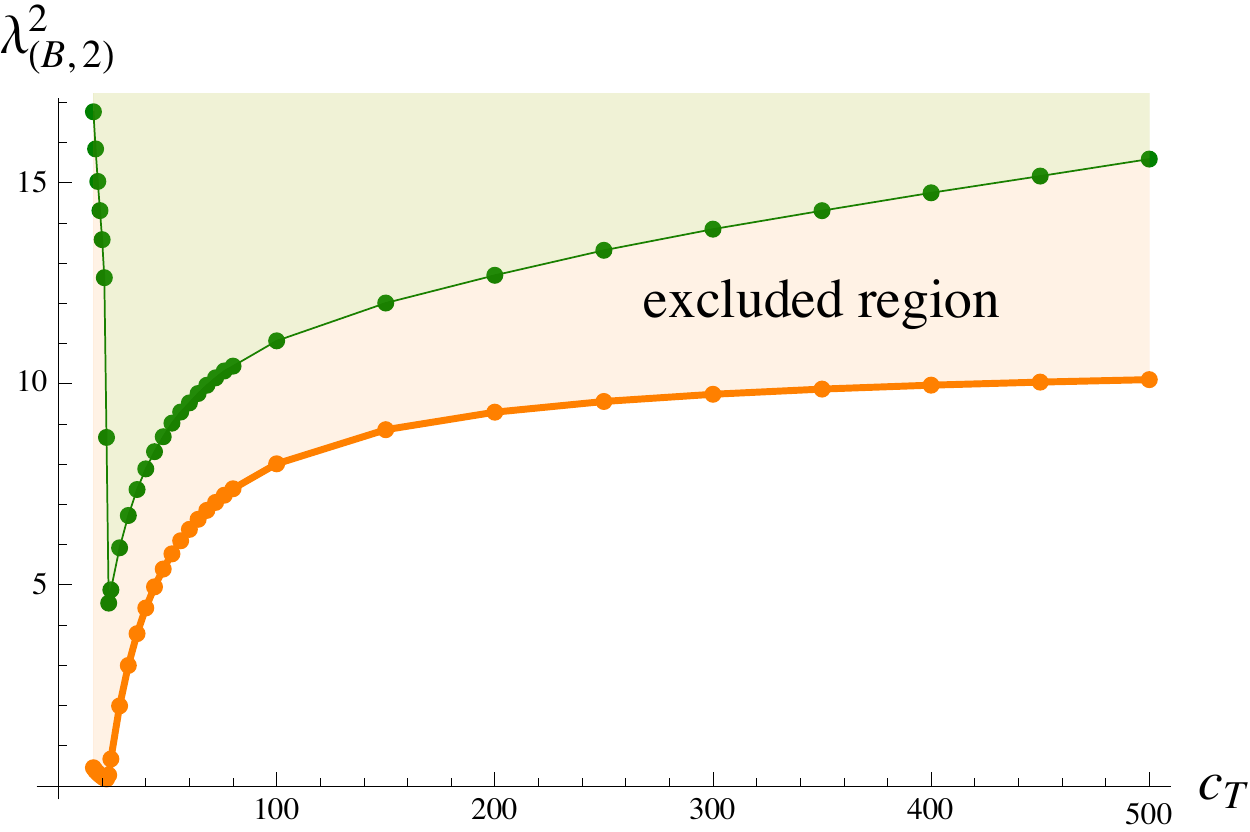}
\caption{Upper bounds on $\lambda_{(B, +)}^2$ and $\lambda_{(B, 2)}^2$ computed either using derivatives w.r.t.~$x$ and $\bar x$ of $d_2$ (see \eqref{dbasis}) and holomorphic derivatives of $d_1$ (in orange), or using only derivatives of $d_2$ (in green).  These bounds are obtained with $j_\text{max} = 20$, $\Lambda = 19$, and a more restrictive set of assumptions on $\Delta_0^*$ only.}
\label{fig:B2BpEqsPlot}
\end{center}
\end{figure}

Lastly, in Figure~\ref{fig:B2BpEqsPlot} we show a comparison plot between upper bounds on $\lambda_{(B, +)}^2$ and $\lambda_{(B, 2)}^2$ that differ in how the functionals $\alpha$ are constructed.  The bounds in orange correspond to constructing $\alpha$ from derivatives w.r.t.~$x$ and $\bar x$ of the quantity $d_2$ defined in \eqref{dbasis} as well as holomorphic derivatives of $d_1$.  The bounds in green are obtained only using derivatives w.r.t.~$x$ and $\bar x$ of $d_2$.  As can be seen from Figure~\ref{fig:B2BpEqsPlot}, the holomorphic derivatives of $d_1$ do carry additional information not contained in $d_2$.

\section{Discussion}
\label{CONCLUSIONS}

Our conformal bootstrap analysis provides us with true non-perturbative information about $\cN=8$ SCFTs. Generically these theories are strongly coupled, and the conformal bootstrap is possibly the only available method to study them. Indeed, except for the $U(1)\times U(1)$ ABJM theory (which is trivial) and BLG theory at large $k$ (which has no known gravity description), all known $\cN=8$ SCFTs are strongly interacting. In addition, while the large $N$ limit of the ABJM theory can be studied through its weakly coupled supergravity dual, it is hard to obtain detailed information directly from the field theory side. 

The operator spectrum and OPE coefficients of theories that saturate the bounds provided by the numerical bootstrap can be determined numerically \cite{ElShowk:2012hu}. It is therefore interesting to contemplate whether the known $\cN=8$ theories saturate (or come close to saturating) our numerical bounds. Note that since $\cN=8$ theories are not expected to have continuous parameters, it seems plausible that there are unique $\cN=8$ SCFTs corresponding to particular given values of $c_T$. In such cases, if the numerical bounds were optimal, they would be saturated by those unique theories. In contrast, 4-d $\cN=4$ SCFTs have a continuous coupling, and therefore there is a continuous family of theories for any given value of the central charge. 

Our results show that the bounds are indeed very close to being saturated for large $c_T$ by the large $N$ ABJM theory, and for $c_T=16$ by the (free) $U(1)\times U(1)$ ABJM theory. However, for values of $c_T$ corresponding to large $k$ BLG theories the bounds seem far from being saturated. Moreover, it is hard to determine whether the bounds we obtained for intermediate values of $c_T$ are saturated by ABJ(M) theories without any additional independent information on those theories. 

An additional feature of our numerical studies is the appearance of a kink in the bounds on operator dimensions as a function of $c_T$, for $c_T\approx 22.8$. We have seen that these bounds are approximately saturated by certain single trace operators in the free ($c_T=16$) theory, while they are saturated by different, double trace operators for $c_T\to\infty$. It is possible that the region near the kink corresponds to a situation in which these two operators become nearly degenerate.\footnote{Indeed, in other numerical bootstrap studies, kinks were shown to correspond to abrupt changes in the operator spectrum \cite{El-Showk:2014dwa}.} However, note that we do not know of any $\cN=8$ SCFTs with this particular value of $c_T$.

Our results can be generalized in various ways.  For example, the Ward identity for $1/2$-BPS multiplets other than the stress-tensor multiplet\footnote{The solution to the Ward identity is slightly different in those cases (see \cite{Dolan:2004mu}).}  is also given by \eqref{GenWard}. It should be straightforward to use our methods to determine the superconformal blocks and the relations between the various crossing equations for four-point functions of those multiplets. This information could then be used to study other correlation functions using the numerical bootstrap method.\footnote{In the 4-dimensional $\cN=4$ theory such a study was performed in \cite{Alday:2014qfa}.}

In addition, our results for the superconformal blocks show that the $\cO_{\mathbf{35}_c}\times\cO_{\mathbf{35}_c}$ OPE has parity symmetry. While all the known $\cN=8$ SCFTs have parity symmetry, we do not know of a proof that this must always be the case. Since the Ward identities for the four-point functions of other $1/2$-BPS operators are identical to the one we studied, it should be possible to generalize this result for the OPE of any $1/2$-BPS operators in $\cN=8$ theories.

In 4-dimensional $\cN=4$ SCFTs the contributions coming from short multiplets to four-point functions of $1/2$-BPS operators can be determined analytically. One way to fix these contributions is by proving a non-renormalization theorem \cite{Eden:2000bk}, showing that they are the same as in the free theory. In interacting 3-dimensional $\cN=8$ theories there are no continuous couplings that one can tune to obtain a free theory. One could then argue that the absence of continuous couplings implies that the short multiplet contributions cannot be determined in the same way as in four dimensions.

However, in 4-d $\cN=4$ SCFTs one can also fix the short multiplet contributions by using only superconformal invariance and crossing symmetry, without ever referring to a free theory or Lagrangian description.\footnote{These contributions are parameterized by the central charge.} It is possible that such contributions to the four-point functions of $1/2$-BPS operators in 3-d $\cN=8$ theories could also be fixed in this fashion. In this work we have solved the differential relations between the crossing equations, which were implied by superconformal invariance, in a series expansion around the crossing symmetric point. While this solution was sufficient for the purpose of implementing the numerical bootstrap, it is possible that with a more thorough analysis of those equations, one would be able to determine the contributions from short operators in three dimensions as well. We hope to return to this interesting question in the future.

\subsection*{Acknowledgments}
\label{s:acks}

We thank Ofer Aharony, Igor Klebanov, Mark Mezei, and David Simmons-Duffin for useful discussions, as well as Vinod Gupta and Sumit Saluja for their help with using the Princeton University departmental computing resources.  
The work of SMC, SSP, and RY was supported in part by the US NSF under Grant No.~PHY-1418069.  The work of JL~was supported in part by a Samsung Scholarship.

\appendix

\section{$\mathfrak{so}(d, 2)$ Conformal Blocks}
\label{confblock}

It was shown in \cite{Dolan:2003hv} that conformal blocks in any dimension can be written as series expansions in two variable Jack polynomials. Jack polynomials\footnote{Strictly speaking these are polynomials only if both $\lambda_1$ and $\lambda_2$ in \eqref{jack} are integers. In the expansions of conformal blocks we use \eqref{jack} with $\lambda_1-\lambda_2=0,1,2,\ldots$, but $\lambda_1$ can take non-integer values.} can be defined using Gegenbauer polynomials as
\begin{align}
P_{\lambda_1 \lambda_2}^{(\veps)}(x,\bar{x}) &= \frac{(\lambda_1-\lambda_2)!}{(2\veps)_{\lambda_1-\lambda_2}} (x\bar{x})^{\frac{1}{2}\left(\lambda_1+\lambda_2\right)} C_{\lambda_1-\lambda_2}^{(\veps)}\left(\frac{x+\bar{x}}{2(x\bar{x})^{1/2}}\right) \ecq \veps=\frac{d-2}{2} \ed \label{jack}
\end{align}

The conformal blocks can then be written as
\begin{align}
g_{\Delta,j}(x,\bar{x}) = \sum_{m,n\ge 0} r_{m,n}(\Delta,j) P^{(\veps)}_{\frac{1}{2}\left(\Delta+j\right)+m \ec \frac{1}{2}\left(\Delta-j\right)+n}(x,\bar{x}) \ed
\end{align}

Define 
\begin{align}
r_{mn} &= \Big(\frac{1}{2}\left(\Delta+j\right)\Big)_m^2 \,\, \Big(\frac{1}{2}\left(\Delta-j\right) -\veps\Big)_n^2 \,\,\hat{r}_{mn} \ed
\end{align}
The coefficients $\hat{r}_{mn}$ can be computed using the recursion relation
\begin{align}
&\left( m(m+\Delta+j-1) + n(n+\Delta-j-2\veps-1) \right) \hat{r}_{mn} \notag\\
&= \frac{j+m-n-1+2\veps}{j+m-n-1+\veps}\hat{r}_{m-1\,n} 
+ \frac{j+m-n+1}{j+m-n+1+\veps}\hat{r}_{m\,n-1} \ed
\end{align}

Our normalization convention is fixed by taking $r_{00} = 1/4^{\Delta}$. With this convention we have
\begin{align}
g_{\Delta,j}(x,\bar{x}) \sim \frac{(\veps)_j}{4^{\Delta}(2\veps)_j} x^{\frac{1}{2}\left(\Delta+j\right)}\bar{x}^{\frac{1}{2}\left(\Delta-j\right)}\,, \qquad \text{as $x, \bar{x}\to 0$} \ec \label{normconv}
\end{align}
where $\bar{x}$ is taken to zero first. This normalization is adapted to the $r,\eta$ coordinates\footnote{This $r$ coordinate should not be confused with that introduced in Section~\ref{conventions}.} of \cite{Hogervorst:2013sma} which are related to $x$ and $\bar{x}$ by
\begin{align}
r^2 = \frac{x\bar{x}}{\left(1+\sqrt{1-x}\right)^2 \left(1+\sqrt{1-\bar{x}}\right)^2 } \,, \qquad \eta = \frac{1 - \sqrt{(x-1)(\bar{x}-1)}}{\sqrt{x\bar{x}}} \ed
\end{align}
The normalization \eqref{normconv} is equivalent to
\begin{align}
g_{\Delta,j}(x,\bar{x}) \sim r^{\Delta}\,,  \qquad \text{as $r \to 0$} \,.
\end{align}
In practice, to approximate conformal blocks in our numerics we use the recursion relations of \cite{Hogervorst:2013sma, Kos:2013tga}.

\section{Characters of $\mathfrak{osp}(8|4)$}
\label{characters}

In this section we will review the character formulas of $\mathfrak{osp}(8|4)$, which were computed in \cite{Dolan:2008vc}, as well as their decomposition under  $\mathfrak{osp}(8|4)\rightarrow \mathfrak{so}(3,2)\oplus\mathfrak{so}(8)$. This decomposition was used in Section~\ref{scblocks} to determine which conformal primaries reside in each supermultiplet appearing in the $\cO_{{\bf 35}_c}\times \cO_{{\bf 35}_c}$ OPE, and, in particular, to derive Table~\ref{Stress} and Tables~\ref{Bp}--\ref{long}. 

The $\mathfrak{osp}(8|4)$ characters are defined by
\begin{align}
\chi_{(\Delta;j;\mathrm{r})}(s,x,y) \equiv \mathrm{Tr}_{\cR_{(\Delta;j;\mathrm{r})}} \left(s^{2D} x^{2J_3} y_1^{H_1}\cdots y_4^{H_4}\right)\ec
\end{align}
where $\Delta$, $j$, and $\mathrm{r}=(r_1\ec\ldots\ec r_4)\in\frac{1}{2}\mathbb{Z}^4$ are, respectively, the conformal dimension, spin and $\mathfrak{so}(8)_R$ highest weights defining the $\mathfrak{osp}(8|4)$ representation. Moreover, $H_i$ and $J_3$ are the Cartan generators of $\mathfrak{so}(8)_R$ and the $\mathfrak{su}(2)$ Lorentz algebra, respectively, and $D$ is the dilatation operator.  The Dynkin labels are related to $(r_1,\ldots,r_4)$ by 
\begin{align}
[a_1 \, a_2 \, a_3 \, a_4] = [r_1-r_2 \ec r_2-r_3 \ec r_3+r_4 \ec r_3-r_4] \ed
\end{align}

The characters are most easily computed by first computing the Verma module characters. Verma modules are infinite (reducible) representations obtained from highest weights by acting unrestrictedly with lowering ladder operators. For instance, the $\mathfrak{su}(2)$ and $\mathfrak{so}(8)$ Verma module characters are given by
\begin{align}
C_j(x) &= \frac{x^{j+1}}{x-x^{-1}} \ec  \label{su2V}\\
C_{\mathrm{r}}(y) &=\frac{\prod_{j=1}^{4}y_j^{r_j+4-j}}{\Delta(y+y^{-1})} \ec  \label{so8V}\\
\Delta(y) &\equiv \prod_{1 \leq i<j \leq 4} (y_i-y_j)\ed
\end{align}

The characters of irreducible representations are obtained from the Verma module characters by Weyl symmetrization, which projects out all the null states in the Verma module. For $\mathfrak{su}(2)$ and $\mathfrak{so}(8)$, these symmetrizations are given, respectively, by
\begin{align}
\mathfrak{W}^{\cS_2} f(x) &= f(x) + f(x^{-1}) \ec \\
\mathfrak{W}^{\cS_4\ltimes (\cS_2)^3} f(y) &= \sum_{\substack{\epsilon_1\ec\ldots\ec\epsilon_3=\pm 1\\ \prod\epsilon_i=1}} \, \sum_{\sigma\in\cS_4} f(y_{\sigma(1)}^{\epsilon_1}\ec\ldots\ec y_{\sigma(4)}^{\epsilon_4}) \ed
\end{align}

Indeed, acting with $\mathfrak{W}^{\cS_2}$ and $\mathfrak{W}^{\cS_4\ltimes (\cS_2)^3}$ on \eqref{su2V} and \eqref{so8V}, one obtains the standard expressions for the $\mathfrak{su}(2)$ and $\mathfrak{so}(8)$ characters, 
\begin{align}
\chi_j(x) &= \mathfrak{W}^{\cS_2} C_j(x) = \frac{x^{j+1}-x^{-j-1}}{x-x^{-1}}\ec\\
\chi_{\mathrm{r}}(y) &= \mathfrak{W}^{\cS_4 \ltimes (\cS_2)^3} C_r(y) \notag\\
&= \left(\det\left[y_i^{r_j+4-j}+y_i^{-r_j-4+j}\right] + \det\left[y_i^{r_j+4-j}-y_i^{-r_j-4+j}\right]\right)/2\Delta(y+y^{-1}) \ed\label{Rchar}
\end{align}

Defining $\mathfrak{W}=\mathfrak{W}^{\cS_2}\mathfrak{W}^{\cS_4\ltimes(\cS_2)^3}$, the $\mathfrak{osp}(8|4)$ characters are given by
\begin{align}
\chi^{(i,n)}_{(\Delta;j;r_1,\ldots,r_1,r_{n+1},\ldots,r_4)}(s,x,y) &= s^{2\Delta} P(s,x) \mathfrak{W}\left( C_{2j}(x) C_{\mathrm{r}}(y) \cR^{(i,n)}(s,x,y)\prod_{\epsilon=\pm 1} \bar{\cQ}_4(s^{-1}y,x^{\epsilon})\right) \ec \label{char2}\\
\chi^{(i,+)}_{(\Delta;j;r,r,r,r)}(s,x,y) &= s^{2\Delta}P(s,x) \mathfrak{W} \left( C_{2j}(x) C_{\mathrm{r}}(y) \cR^{(i,+)}(s,x,y) \prod_{\epsilon=\pm 1} \bar{\cQ}_3(s^{-1}y,x^{\epsilon})\right) \label{charP}\ec
\end{align}
where
\begin{align}
\cR^{(i,n)} &= \begin{cases}
\cQ_0(sy,x)\cQ_n(sy,x^{-1}) & i=A \ec \\
\cQ_n(sy,x)\cQ_n(sy,x^{-1}) & i=B \ec
\end{cases}\\
\cR^{(i,+)} &= \begin{cases}
\cQ_0(sy,x)(1+sy_4^{-1}x)(1+sy_4^{-1}x^{-1}) & i=A \ec\\
(1+sy_4^{-1}x)(1+sy_4^{-1}x^{-1}) & i=B\ec
\end{cases}\\
\cQ_n(y,x) &= \prod_{j=n+1}^{4}(1+y_jx) \,, \qquad  \bar{\cQ}_n(y,x) = \prod_{j=1}^{n}(1+y_j^{-1}x) \ec \\
P(s,x) &= \frac{1}{1-s^4}\sum_{n=0}^{\infty} s^{2n}\chi_{2n}(x) \ed
\end{align}

The function $P(s,x)$ in \eqref{char2} and \eqref{charP} is related to the $\mathfrak{so}(3,2)$ characters $\cA_{\Delta,j}$, computed in \cite{Flato:1978qz,Dolan:2005wy}:
\begin{align}
\cA_{\Delta,j} &= \mathrm{Tr}_{(\Delta,j)}\left(s^{2D}x^{2J_3}\right) = s^{2\Delta}\chi_{2j}(x) P(s,x) \ed \label{confchar}
\end{align}
Note that since conformal representations decompose at unitarity as
\begin{align}
(\Delta,j)\xrightarrow{\Delta\rightarrow j+1} (j+1,j)_{\mathrm{short}} + (j+2,j-1)\ec
\end{align}
the $\mathfrak{so}(3,2)$ character of a spin-$j$ conserved current is actually $\cA_{j+1,j}-\cA_{j+2,j-1}$.

In order to expand the $\mathfrak{osp}(8|4)$ characters as a sum of products of conformal characters \eqref{confchar} times R-symmetry characters \eqref{Rchar}, we need to disentangle the $s$, $x$ and $y$ dependence in \eqref{char2}, and \eqref{charP}. Explicitly, it is straightforward to show that\footnote{Note that for the $B$ series $\Delta=r_1$, while for the $A$ series $\Delta=r_1+j+1$ except for the long multiplet $(A,0)$ for which $\Delta\geq r_1+j+1$.}
\begin{align}
\chi^{(A,+)}_{(\Delta;j;r,r,r,r)}(s,x,y) &= s^{2\Delta} P(s,x) \sum_{a_1\ec\ldots\ec a_4=0}^2 \,  \sum_{\bar{a}_1\ec\ldots\ec\bar{a}_4=0}^{1}  s^{a_1+\cdots+a_4+\bar{a}_1+\cdots+\bar{a}_4} \chi_{2j+\bar{a}_1+\cdots+\bar{a}_4}(x) \notag\\
&\times\left(\prod_{i=1}^4\chi_{j_{a_i}}(x)\right) \, \chi_{(r+\bar{a}_1-a_1\ec\ldots\ec r+\bar{a}_4-a_4)}(y) \ec \label{Apchar}\\
\chi^{(B,+)}_{(\Delta;0;r,r,r,r)}(s,x,y) &= s^{2\Delta} P(s,x) \sum_{a_1\ec\ldots\ec a_4=0}^2 s^{a_1+\cdots+a_4} \left(\prod_{i=1}^4\chi_{j_{a_i}}(x)\right) \, \chi_{(r-a_1\ec\ldots\ec r-a_4)}(y) \ec
\end{align}
\begin{align}
\chi^{(A,n)}_{(\Delta;j;r_1,\ldots,r_1,r_{n+1},\ldots,r_4)}(s,x,y) &= s^{2\Delta} P(s,x) \sum_{a_1\ec\ldots\ec a_4=0}^2 \sum_{\bar{a}_{n+1},\ldots,\bar{a}_4=0}^{2} \sum_{\bar{a}_{1},\ldots,\bar{a}_n=0}^{1} s^{a_1+\cdots+a_4+\bar{a}_1+\cdots+\bar{a}_4} \chi_{2j+\bar{a}_1+\cdots+\bar{a}_n}(x) \notag\\
&\times \left(\prod_{i=n+1}^4\chi_{j_{\bar{a}_i}}(x) \right) \left(\prod_{i=1}^4 \chi_{j_{a_i}}(x)\right)  \, \chi_{(r_1+\bar{a}_1-a_1\ec\ldots\ec  r_4+\bar{a}_4-a_4)}(y)\ec \\
\chi^{(B,n)}_{(\Delta;0;r_1,\ldots,r_1,r_{n+1},\ldots,r_4)}(s,x,y) &= s^{2\Delta}P(s,x) \sum_{a_1\ec\ldots\ec a_4\ec\bar{a}_{n+1},\ldots,\bar{a}_4=0}^{2} s^{a_1 + \cdots + a_4 + \bar{a}_{n+1} + \cdots + \bar{a}_4} \left(\prod_{i=n+1}^4\chi_{j_{\bar{a}_i}}(x) \right) \notag\\
&\times \left(\prod_{i=1}^4 \chi_{j_{a_i}}(x)\right)  \, \chi_{(r_1-a_1,\ldots,r_1-a_n,r_{n+1}+\bar{a}_{n+1}-a_{n+1},\ldots,r_4+\bar{a}_4-a_4)}(y) \ec\label{Bnchar}
\end{align}
where $j_a\equiv a\,\,(\!\!\!\!\mod{2})$.

The products of the $\mathfrak{su}(2)$ characters in \eqref{Apchar}--\eqref{Bnchar} are easily transformed into sums of such characters by decomposing $\mathfrak{su}(2)$ tensor products. After doing so, we see that \eqref{Apchar}--\eqref{Bnchar} become sums over $\mathfrak{so}(3,2)\oplus\mathfrak{so}(8)$ characters, as desired.\footnote{Sometimes the $\mathfrak{so}(8)$ characters in \eqref{Apchar}--\eqref{Bnchar} appear with negative Dynkin labels. One can then try to use the identity
\begin{align}
\chi_{\mathrm{r}^{\omega}}(y) = (-)^{\ell(\omega)} \chi_{\mathrm{r}}(y) \ec\notag
\end{align} 
to obtain a character with non-negative Dynkin labels. In this identity $\omega\in\cS_4\ltimes (\cS_2)^3$ is a Weyl transformation, $\mathrm{r}^{\omega} = \omega(\mathrm{r}+\rho)-\rho$ is a Weyl reflection, $\rho=(3,2,1,0)$ is the Weyl vector, and $(-)^{\ell(\omega)}$ is the signature of the Weyl transformation. If there is no Weyl transformation such that $\mathrm{r}^{\omega}$ correspond to non-negative integer Dynkin labels, then $\chi_{\mathrm{r}}=0$.}

\section{Superconformal Blocks}
\label{superblocks}
Let us write our results for the superconformal blocks in order of increasing complexity.  In all the supermultiplets, we normalize the coefficient of the superconformal primary to one. The results are presented in terms of the R-symmetry channels $A_{ab}(u,v)$, which were defined in \eqref{GExpansion}.

For $(1,0)_{(B,+)}^{[0020]}$, corresponding to the stress-tensor multiplet, we have
\begin{align}
A_{11}(u,v) &= g_{1,0}(u,v) \ec \label{A11stress}\\
A_{10}(u,v) &= -g_{2,1}(u,v) \label{A10stress}\ec\\
A_{00}(u,v) &= \frac{1}{4} g_{3,2}(u,v) \label{A00stress}\ed
\end{align}

The superconformal blocks corresponding to $(2,0)_{(B,+)}^{[0040]}$ are
\begin{alignat}{3}
A_{22}(u,v) &= g_{2,0}(u,v) \label{A22Bp}\ec\\
A_{21}(u,v) &= -\frac{4}{3} g_{3,1}(u,v) \label{A21Bp}\ec\\
A_{20}(u,v) &= \frac{16}{45} g_{4,0}(u,v) \label{A20Bp}\ec\\
A_{11}(u,v) &= \frac{256}{675} g_{4,2}(u,v) \label{A11Bp}\ec\\
A_{10}(u,v) &= -\frac{128}{875}g_{5,1}(u,v) \label{A10Bp}\ec\\
A_{00}(u,v) &= \frac{256}{18375} g_{6,0}(u,v) \label{A00Bp}\ed
\end{alignat}

For $(2,0)_{(B,2)}^{[0200]}$, the superconformal blocks are 
\begin{align}
A_{22}(u,v) &= \frac{8}{9}g_{4,0}(u,v) \ec\\
A_{21}(u,v) &= -\frac{8}{3} g_{3,1}(u,v) - \frac{192}{175} g_{5,1}(u,v) \ec\\
A_{20}(u,v) &= g_{2,0}(u,v)+\frac{16}{63} g_{4,0}(u,v)+ \frac{64}{45} g_{4,2}(u,v)+\frac{256}{1225} g_{6,0}(u,v) \ec\\
A_{11}(u,v) &= \frac{32}{135} g_{4,0}(u,v)+\frac{512}{945} g_{4,2}(u,v)+\frac{8192}{25725}g_{6,2}(u,v) \ec\\
A_{10}(u,v) &= -\frac{12}{35} g_{3,1}(u,v)-\frac{128}{525} g_{5,1}(u,v)-\frac{2304}{6125} g_{5,3}(u,v) - \frac{1024}{11319} g_{7,1}(u,v) \ec
\end{align}
\begin{align}
A_{00}(u,v) &= \frac{16}{735} g_{4,0}(u,v) + \frac{512}{56595}g_{6,0}(u,v)+\frac{1024}{25725}g_{6,2}(u,v) + \frac{5120}{539539}g_{8,0}(u,v) \ed
\end{align}

For $(j+2,j)_{(A,+)}^{[0200]}$, we find
\begin{align}
A_{22}(u,v) &= \frac{16}{3} g_{j+4,j+2}(u,v) \ec\\
A_{21}(u,v) &= - 4 g_{j+3,j+1}(u,v) - \frac{32 (j+2) (j+3)^2}{(2 j+5)^2 (2 j+7)} g_{j+5,j+1}(u,v) \notag\\ 
             & - \frac{64 (j+3)^4}{\left(4 j^2+24 j+35\right)^2}g_{j+5,j+3}(u,v) \ec\\
A_{20}(u,v) &= \frac{4 (j+1)}{2 j+3} g_{j+4,j}(u,v) + \frac{32 (j+2) (j+3)}{3 (2 j+3) (2 j+7)} g_{j+4,j+2}(u,v) \notag\\ 
             &+ \frac{64 (j+3)^3 (j+4)^2}{(2 j+5) (2 j+7)^3 (2 j+9)} g_{j+6,j+2}(u,v) \ec\\
A_{11}(u,v) &= g_{j+2,j}(u,v) + \frac{16 (j+1) (j+2) (j+3)}{3 (2 j+3)^2 (2 j+7)}g_{j+4,j}(u,v) \notag \\
             &+ \frac{64 (j+2)^2 (j+3)^2 }{9 (2 j+3)^2 (2 j+7)^2} g_{j+4,j+2}(u,v) + \frac{48 (j+1) (j+2)(j+3)^2 (j+4)^2 }{(2 j+3) (2 j+5)^2 (2 j+7)^2 (2 j+9)} g_{j+6,j}(u,v) \notag\\
             &+ \frac{256 (j+2)(j+3)^4 (j+4)^2 }{3 (2 j+3) (2 j+5) (2 j+7)^4 (2 j+9)} g_{j+6,j+2}(u,v) + \frac{256 (j+3)^4 (j+4)^4 }{(2 j+5)^2 (2 j+7)^4 (2 j+9)^2} g_{j+6,j+4}(u,v) \ec\\
A_{10}(u,v) &= - \frac{j}{2 j+1} g_{j+3,j-1}(u,v) - \frac{12 (j+1) (j+3)}{5 (2 j+1) (2 j+7)}    g_{j+3,j+1}(u,v) \notag\\ 
             &- \frac{6 j (j+1) (j+3)^2}{(2 j+1) (2 j+3) (2 j+5) (2 j+7)} g_{j+5,j-1}(u,v) \notag\\
             &- \frac{48 (j+2) (2 j (j+5) (4 j (j+5)+35)+137) (j+3)^2}{5 (2 j+1) (2 j+3) (2 j+5)^2 (2 j+7)^2 (2 j+9)} g_{j+5,j+1}(u,v) \notag\\
             &- \frac{192 (j+2)(j+3)^4 (j+4) }{5 (2 j+3) (2 j+5)^2 (2 j+7)^2 (2 j+9)} g_{j+5,j+3}(u,v) \notag\\
             &- \frac{96 (j+2) (j+3)^3 (j+4)^2 (j+5)^2}{(2 j+5)^2 (2 j+7)^3 (2 j+9)^2 (2 j+11)} g_{j+7,j+1}(u,v) \notag\\
             &- \frac{256 (j+3)^4 (j+4)^3 (j+5)^2 }{(2 j+5)^2 (2 j+7)^3 (2 j+9)^3 (2 j+11)} g_{j+7,j+3}(u,v) \ec
\end{align}
\begin{align}
A_{00}(u,v) &= \frac{3 (j-1) j}{32 j^2-8} g_{j+4,j-2}(u,v) + \frac{4 j (j+1) (j+3)}{7 (2 j-1) (2 j+3) (2 j+7)} g_{j+4,j}(u,v) \notag\\
             &+ \frac{72 (j+1) (j+2) (j+3) (j+4) }{35 (2 j+1) (2 j+3) (2 j+7) (2 j+9)} g_{j+4,j+2}(u,v) \notag\\
             &+ \frac{64 (j+2) (j+3)^3 (j+4)^2 (j+5)}{7 (2 j+3) (2 j+5) (2 j+7)^3 (2 j+9) (2 j+11)} g_{j+6,j+2}(u,v) \notag\\
             &+ \frac{96 (j+3)^3 (j+4)^3 (j+5)^2 (j+6)^2 }{(2 j+5) (2 j+7)^3 (2 j+9)^3 (2 j+11)^2 (2 j+13)} g_{j+8,j+2}(u,v) \ed
\end{align}

The blocks for $(j+2,j)_{(A,2)}^{[0100]}$ are given by
\begin{align}
A_{22}(u,v) &= \frac{32 (j+2)}{6 j+15} g_{j+5,j+1}(u,v) \ec\\
A_{21}(u,v) &= -\frac{8 (j+1) }{2 j+3} g_{j+4,j}(u,v) - \frac{32 (j+2)^2}{(2 j+3) (2 j+5)} g_{j+4,j+2}(u,v) \notag\\
&- \frac{48 (j+1) (j+2) (j+4)^2 }{(2 j+3) (2 j+5) (2 j+7) (2 j+9)} g_{j+6,j}(u,v) - \frac{128 (j+2)^2 (j+3) (j+4)^2 }{(2 j+3) (2 j+5) (2 j+7)^2 (2 j+9)} g_{j+6,j+2}(u,v)  \ec\\
A_{20}(u,v) &=  4 g_{j+3,j+1}(u,v) + \frac{6 j (j+1) }{4 j (j+2)+3} g_{j+5,j-1}(u,v) + \frac{64 (j+2) \left(j^2+5 j+3\right) }{3 (2 j+5) \left(4 j^2+20 j+9\right)} g_{j+5,j+1}(u,v) \notag\\
&+ \frac{64 (j+2)^2 (j+3)^2 }{(2 j+3) (2 j+5)^2 (2 j+7)} g_{j+5,j+3}(u,v) + \frac{96 (j+2)(j+3) (j+4)^2 (j+5)^2 }{(2 j+5) (2 j+7)^2 (2 j+9)^2 (2 j+11)} g_{j+7,j+1}(u,v) \ec\\
A_{11}(u,v) &= \frac{2 j}{2 j+1} g_{j+3,j-1}(u,v) + \frac{16 (j+1) (j+2)}{3 (2 j+1) (2 j+5)} g_{j+3,j+1}(u,v) \notag\\
&+ \frac{8 j (j+1)(j+3) (j+4) }{(2 j+1) (2 j+3) (2 j+5) (2 j+9)} g_{j+5,j-1}(u,v) \notag\\
&+ \frac{32 (j+2)^2 (j+3) (j (j+5) (52 j (j+5)+445)+822) }{9 (2 j+1) (2 j+3) (2 j+5)^3 (2 j+7) (2 j+9)} g_{j+5,j+1}(u,v) \notag\\
&+ \frac{256 (j+2)^2 (j+3)^3 (j+4)}{3 (2 j+3) (2 j+5)^3 (2 j+7) (2 j+9)} g_{j+5,j+3}(u,v) \notag\\
&+ \frac{80 j (j+1) (j+2) (j+4)^2 (j+5)^2}{(2 j+1) (2 j+3) (2 j+5) (2 j+7) (2 j+9)^2 (2 j+11)} g_{j+7,j-1}(u,v) \notag\\
&+ \frac{128 (j+1) (j+2)^2 (j+3) (j+4)^2 (j+5)^2 }{(2 j+1) (2 j+5)^2 (2 j+7)^2 (2 j+9)^2 (2 j+11)} g_{j+7,j+1}(u,v) \notag\\
&+ \frac{512 (j+2)^2 (j+3)^2 (j+4)^3 (j+5)^2 }{(2 j+3) (2 j+5)^2 (2 j+7)^2 (2 j+9)^3 (2 j+11)} g_{j+7,j+3}(u,v) \ec
\end{align}
\begin{align}
A_{10}(u,v) &= -g_{j+2,j}(u,v) - \frac{3 (j-1) j}{8 j^2-2} g_{j+4,j-2}(u,v) - \frac{4 (j+1) (j+2)^2 (44 j (j+4)-75) }{5 (2 j-1) (2 j+3)^2 (2 j+5) (2 j+9)} g_{j+4,j}(u,v) \notag\\
&- \frac{48 (j+2)^2 (2 j (j+5) (4 j (j+5)+35)+137)}{5 (2 j+1) (2 j+3)^2 (2 j+5) (2 j+7) (2 j+9)} g_{j+4,j+2}(u,v) \notag\\
&- \frac{10 j \left(j^2-1\right) (j+4)^2}{(2 j-1) (2 j+1) (2 j+3) (2 j+7) (2 j+9)} g_{j+6,j-2}(u,v) \notag\\
&- \frac{72 (j+1) (j+2) (2 j (j+5)-3) (j+4)^2}{5 (2 j-1) (2 j+3) (2 j+5) (2 j+7) (2 j+9) (2 j+11)} g_{j+6,j}(u,v) \notag\\
&- \frac{64 (j+2)^2 (j+3)^3 (44 j (j+6)+145) (j+4)^2}{5 (2 j+1) (2 j+3) (2 j+5)^2 (2 j+7)^3 (2 j+9) (2 j+11)} g_{j+6,j+2}(u,v) \notag\\
&- \frac{256 (j+2)^2 (j+3)^2 (j+4)^4}{(2 j+3) (2 j+5)^2 (2 j+7)^3 (2 j+9)^2} g_{j+6,j+4}(u,v) \notag\\
\phantom{A_{1,0}(u,v)}&- \frac{160 (j+1) (j+2) (j+3) (j+5)^2 (j+6)^2 (j+4)^2}{(2 j+3) (2 j+5) (2 j+7)^2 (2 j+9)^2 (2 j+11)^2 (2 j+13)} g_{j+8,j}(u,v) \notag\\
&- \frac{384 (j+2)^2 (j+3)(j+4)^3 (j+5)^2 (j+6)^2 }{(2 j+3) (2 j+5) (2 j+7)^2 (2 j+9)^3 (2 j+11)^2 (2 j+13)} g_{j+8,j+2}(u,v) \ec
\end{align}
\begin{align}
A_{00}(u,v) &=  \frac{j}{8 j+4} g_{j+3,j-1}(u,v) + \frac{4 (j+1) (j+4) }{7 (2 j+1) (2 j+9)} g_{j+3,j+1}(u,v) \notag\\
&+ \frac{5 (j-2) (j-1) j }{8 (2 j-3) (2 j-1) (2 j+1)} g_{j+5,j-3}(u,v) + \frac{6 j (j+2) \left(j^2-1\right) }{7 (2 j+5) \left(8 j^3+4 j^2-18 j-9\right)} g_{j+5,j-1}(u,v) \notag\\
&+ \frac{144 (j+2)^2 (j+3) (j (j+5) (4 j (j+5)+5)-14)}{35 (2 j-1) (2 j+1) (2 j+3) (2 j+5) (2 j+7) (2 j+9) (2 j+11)} g_{j+5,j+1}(u,v) \notag\\
&+ \frac{64 (j+1) (j+2)^2 (j+3)^2 (j+4) }{7 (2 j+1) (2 j+3) (2 j+5)^2 (2 j+7) (2 j+9)} g_{j+5,j+3}(u,v) \notag\\
&+ \frac{96 (j+2) (j+3)^2 (j+5)^2 (j+6) (j+4)^2 }{7 (2 j+5)^2 (2 j+7)^2 (2 j+9)^2 (2 j+11) (2 j+13)} g_{j+7,j+1}(u,v) \notag\\
&+ \frac{64 (j+2)^2 (j+3)^2 (j+4)^3 (j+5)^2}{(2j+3)(2 j+5)^2 (2 j+7)^2 (2 j+9)^3 (2j+11)} g_{j+7,j+3}(u,v) \notag\\
&+ \frac{160 (j+2) (j+3) (j+5)^2 (j+6)^2 (j+7)^2 (j+4)^3 }{(2 j+5) (2 j+7)^2 (2 j+9)^3 (2 j+11)^2 (2 j+13)^2 (2 j+15)} g_{j+9,j+1}(u,v)\ed
\end{align}

Finally, for the long multiplet $(\Delta,j)_{(A,0)}^{[0000]}$ we find
\begin{align}
A_{22}(u,v) &= \frac{128 (\Delta -j+1) (\Delta -j-1) (\Delta +j) (\Delta +j+2) }{3 (\Delta -j+2) (\Delta -j) (\Delta +j+1) (\Delta +j+3)}g_{\Delta +4,j}(u,v) \ec \label{A22long}
\end{align}
\begin{align}
A_{21}(u,v) &= - \frac{64 j (\Delta -j+1) (\Delta -j-1) (\Delta +j) }{(2 j+1) (\Delta -j+2) (\Delta-j ) (\Delta +j+1)} g_{\Delta +3,j-1}(u,v) \notag\\
&- \frac{64 (j+1) (\Delta -j-1) (\Delta +j) (\Delta +j+2) }{(2 j+1) (\Delta-j ) (\Delta +j+1) (\Delta +j+3)} g_{\Delta +3,j+1}(u,v) \notag\\
&- \frac{256 (\Delta +3)^2 j}{(2 \Delta +5) (2 \Delta +7) (2 j+1)}  \notag\\
&\times \frac{ (\Delta -j+3) (\Delta -j+1) (\Delta -j-1) (\Delta +j) (\Delta +j+2) }{ (\Delta -j+4) (\Delta -j+2) (\Delta-j ) (\Delta +j+1) (\Delta +j+3)} g_{\Delta +5,j-1}(u,v) \notag\\
&- \frac{256 (\Delta +3)^2 (j+1) }{(2 \Delta +5) (2 \Delta +7) (2 j+1)}  \notag\\
&\times \frac{(\Delta -j+1) (\Delta -j-1) (\Delta +j) (\Delta +j+2) (\Delta +j+4) }{(\Delta -j+2) (\Delta-j ) (\Delta +j+1) (\Delta +j+3) (\Delta +j+5)} g_{\Delta +5,j+1}(u,v)\ec 
\label{A21long}\\
A_{20}(u,v) &= \frac{16 (\Delta -j-1) (\Delta +j)}{(\Delta -j) (\Delta +j+1)} g_{\Delta +2,j}(u,v) \notag\\
&+ \frac{64 (j-1) j (\Delta -j+3) (\Delta -j+1) (\Delta -j-1) (\Delta +j) }{\left(4 j^2-1\right) (\Delta -j+4) (\Delta -j+2) (\Delta-j ) (\Delta +j+1)} g_{\Delta +4,j-2}(u,v) \notag\\
&+ \frac{8 (\Delta -j+1) (\Delta -j-1) (\Delta +j) (\Delta +j+2) \left(\frac{3}{2 \Delta +3}-\frac{3}{2 \Delta +7}+\frac{4 (8 j (j+1)-3)}{4 j (j+1)-3}\right) }{3 (\Delta -j+2) (\Delta -j) (\Delta +j+1) (\Delta +j+3)} g_{\Delta +4,j}(u,v) \notag\\
&+ \frac{64 (j+1) (j+2) (\Delta -j-1) (\Delta +j) (\Delta +j+2) (\Delta +j+4) }{(2 j+1) (2 j+3) (\Delta-j ) (\Delta +j+1) (\Delta +j+3) (\Delta +j+5)} g_{\Delta +4,j+2}(u,v) \notag\\
&+ \frac{256 (\Delta +3)^2 (\Delta +4)^2 }{(2 \Delta +5) (2 \Delta +7)^2 (2 \Delta +9) }  \notag\\
&\times \frac{(\Delta -j+3) (\Delta -j+1) (\Delta -j-1) (\Delta +j) (\Delta +j+2) (\Delta +j+4) }{(\Delta -j+4) (\Delta -j+2) (\Delta -j) (\Delta +j+1) (\Delta +j+3) (\Delta +j+5)}g_{\Delta +6,j}(u,v)\ec\label{A20long}\\
A_{11}(u,v) &= \frac{32 (j-1) j (\Delta -j+1) (\Delta -j-1)}{(2 j-1) (2 j+1) (\Delta -j+2) (\Delta-j )} g_{\Delta +2,j-2}(u,v) \notag\\
&+ \frac{64 j (j+1) (\Delta -j-1) (\Delta +j) }{3 (2 j-1) (2 j+3) (\Delta-j ) (\Delta +j+1)} g_{\Delta +2,j}(u,v) \notag\\
&+ \frac{32 (j+1) (j+2) (\Delta +j) (\Delta +j+2) }{\left(4 j^2+8 j+3\right) (\Delta +j+1) (\Delta +j+3)} g_{\Delta +2,j+2}(u,v) \notag\\
&+ \frac{512 (\Delta +2) (\Delta +3) j (j+1) (\Delta -j+1) (\Delta -j-1) (\Delta +j) (\Delta +j+2)}{9 (2 \Delta +3) (2 \Delta +7) (2 j-1) (2 j+3) (\Delta -j+2) (\Delta-j ) (\Delta +j+1) (\Delta +j+3)} g_{\Delta +4,j}(u,v) \notag\\
&+ \frac{256 (\Delta +2) (\Delta +3) (j+1) (j+2) }{3 (2 \Delta +3) (2 \Delta +7) (2j+1)(2j+3) } \notag\\
&\times \frac{(\Delta -j-1) (\Delta +j) (\Delta +j+2) (\Delta +j+4) }{(\Delta-j ) (\Delta +j+1) (\Delta +j+3) (\Delta +j+5)} g_{\Delta +4,j+2}(u,v) \notag\\
&+ \frac{256 (\Delta +2) (\Delta +3) (j-1) j (\Delta -j+3) (\Delta -j+1) (\Delta -j-1) (\Delta +j) }{3 (2 \Delta +3) (2 \Delta +7) \left(4 j^2-1\right) (\Delta -j+4) (\Delta -j+2) (\Delta-j ) (\Delta +j+1)} g_{\Delta +4,j-2}(u,v)\notag
\end{align}
\begin{align}
\phantom{A_{11}(u,v)} &+ \frac{512 (\Delta +3)^2 (\Delta +4)^2 (j-1) j }{(2 \Delta +5) (2 \Delta +7)^2 (2 \Delta +9) (2 j-1) (2 j+1) } \notag\\
&\times \frac{(\Delta -j+5) (\Delta -j+3) (\Delta -j+1) (\Delta -j-1) (\Delta +j) (\Delta +j+2)}{(\Delta -j+6) (\Delta -j+4) (\Delta -j+2) (\Delta-j ) (\Delta +j+1) (\Delta +j+3)}g_{\Delta +6,j-2}(u,v)+\notag\\
&+\frac{1024 (\Delta +3)^2 (\Delta +4)^2 j (j+1)  }{3 (2 \Delta +5) (2 \Delta +7)^2 (2 \Delta +9) (2 j-1) (2 j+3) } \notag\\
&\times \frac{(\Delta -j+3) (\Delta -j+1) (\Delta -j-1) (\Delta +j) (\Delta +j+2) (\Delta +j+4)}{(\Delta -j+4) (\Delta -j+2) (\Delta-j ) (\Delta +j+1) (\Delta +j+3) (\Delta +j+5)} g_{\Delta +6,j}(u,v) \notag\\
&+ \frac{512 (\Delta +3)^2 (\Delta +4)^2 (j+1) (j+2) }{(2 \Delta +5) (2 \Delta +7)^2 (2 \Delta +9) (2 j+1) (2 j+3) } \notag\\
&\times \frac{(\Delta -j+1) (\Delta -j-1) (\Delta +j) (\Delta +j+2) (\Delta +j+4) (\Delta +j+6)}{(\Delta -j+2) (\Delta-j ) (\Delta +j+1) (\Delta +j+3) (\Delta +j+5) (\Delta +j+7)} g_{\Delta +6,j+2}(u,v)\ec \label{A11long}\\
A_{10}(u,v) &= - \frac{8 j (\Delta -j-1) }{(2 j+1) (\Delta-j )} g_{\Delta +1,j-1}(u,v) - \frac{8 (j+1) (\Delta +j)}{(2 j+1) (\Delta +j+1)} g_{\Delta +1,j+1}(u,v) \notag\\
&- \frac{32 (j-2) (j-1) j (\Delta -j+3) (\Delta -j+1) (\Delta -j-1)}{(2 j-3) (2 j-1) (2 j+1) (\Delta -j+4) (\Delta -j+2) (\Delta-j )} g_{\Delta +3,j-3}(u,v) - \notag\\
&-\frac{96 j}{5 (2 j-3) (2 j+1) (2 j+3)(2 \Delta +1) (2 \Delta +7) } \notag\\
&\times \frac{ (\Delta -j+1) (\Delta-j-1) (\Delta+j ) \left((8 \Delta  (\Delta +4)+19) j^2-13 \Delta  (\Delta +4)-34\right) g_{\Delta +3,j-1}(u,v)}{(\Delta -j+2) (\Delta-j ) (\Delta +j+1) } \notag\\
&- \frac{96 (j+1) \left((8 \Delta  (\Delta +4)+19) j^2+2 (8 \Delta  (\Delta +4)+19) j-5 (\Delta +1) (\Delta +3)\right) }{5 (2 j-1) (2 j+1) (2 j+5) (2 \Delta +1) (2 \Delta +7)} \notag\\
&\times \frac{(\Delta -j-1) (\Delta+j ) (\Delta+j +2)}{(\Delta-j ) (\Delta +j+1) (\Delta +j+3)} g_{\Delta +3,j+1}(u,v) \notag\\
&- \frac{32 (j+1) (j+2) (j+3) (\Delta+j ) (\Delta+j +2) (\Delta+j +4)}{(2 j+1) (2 j+3) (2 j+5) (\Delta+j +1) (\Delta +j+3) (\Delta +j+5)} g_{\Delta +3,j+3}(u,v) \notag\\
&- \frac{128 (j-2) (j-1) j (\Delta +3)^2}{(2 j-3) (2 j-1) (2 j+1)  (2 \Delta +5) (2 \Delta +7)}  \notag\\
&\times \frac{(\Delta-j+5) (\Delta -j+3) (\Delta-j+1) (\Delta-j-1) (\Delta +j)}{(\Delta-j+6) (\Delta -j+4) (\Delta -j+2) (\Delta-j ) (\Delta+j +1)} g_{\Delta +5,j-3}(u,v) \notag\\
&- \frac{384 j \left((8 \Delta  (\Delta +6)+59) j^2-13 \Delta  (\Delta +6)-99\right) (\Delta +3)^2}{5 (2 j-3) (2 j+1) (2 j+3) (2 \Delta +3) (2 \Delta +5) (2 \Delta +7) (2 \Delta +9)}  \notag\\
&\times \frac{(\Delta -j+3) (\Delta -j+1) (\Delta -j-1) (\Delta+j ) (\Delta+j +2)}{(\Delta -j+4) (\Delta -j+2) (\Delta-j ) (\Delta +j+1) (\Delta +j+3)} g_{\Delta +5,j-1}(u,v) \notag\\
&- \frac{384 (j+1) \left((8 \Delta  (\Delta +6)+59) j^2+2 (8 \Delta  (\Delta +6)+59) j-5 (\Delta +2) (\Delta +4)\right)}{5 (2 j-1) (2 j+1) (2 j+5) (2 \Delta +3) (2 \Delta +5) (2 \Delta +7) (2 \Delta +9)}  \notag\\
&\times \frac{(\Delta -j+1) (\Delta -j-1) (\Delta +3)^2 (\Delta +j) (\Delta +j+2) (\Delta+j +4)}{(\Delta -j+2) (\Delta-j ) (\Delta +j+1) (\Delta +j+3) (\Delta +j+5)} g_{\Delta +5,j+1}(u,v) \notag
\end{align}
\begin{align}
\phantom{A_{10}(u,v)} &- \frac{128 (j+1) (j+2) (j+3) (\Delta +3)^2}{(2 j+1) (2 j+3) (2 j+5) (2 \Delta +5) (2 \Delta +7)} \notag\\
&\times \frac{(\Delta-j-1) (\Delta+j ) (\Delta+j +2) (\Delta +j+4) (\Delta+j +6)}{(\Delta-j ) (\Delta +j+1) (\Delta +j+3) (\Delta +j+5) (\Delta +j+7)} g_{\Delta +5,j+3}(u,v)\notag\\
&- \frac{512 j (\Delta +4)^2 (\Delta +5)^2 (\Delta+j ) (\Delta +j+2) (\Delta +j+4)(\Delta +3)^2}{(2 j+1) (2 \Delta +5) (2 \Delta +7)^2 (2 \Delta +9)^2 (2 \Delta +11)} \notag\\
&\times \frac{(\Delta-j+5) (\Delta -j+3) (\Delta -j+1) (\Delta-j-1)}{(\Delta -j+6) (\Delta -j+4) (\Delta -j+2) (\Delta-j ) (\Delta +j+1) (\Delta +j+3) (\Delta +j+5)} g_{\Delta +7,j-1}(u,v) \notag\\
&-\frac{(\Delta -j+3) (\Delta -j+1) (\Delta-j-1)(\Delta+j ) (\Delta +j+2) (\Delta +j+4) (\Delta +j+6)}{(\Delta -j+4) (\Delta -j+2) (\Delta-j ) (\Delta+j +1) (\Delta +j+3) (\Delta +j+5) (\Delta +j+7)}\notag\\
&\times \frac{512 (j+1) (\Delta +4)^2 (\Delta +5)^2 (\Delta +3)^2}{(2 j+1) (2 \Delta +5) (2 \Delta +7)^2 (2 \Delta +9)^2 (2 \Delta +11)}  g_{\Delta +7,j+1}(u,v)  \ec \label{A10long}\\
A_{00}(u,v) &= g_{\Delta ,j}(u,v)+\frac{16 (\Delta-j-1) \Delta  (\Delta +3) (\Delta+j ) g_{\Delta +2,j}(u,v)}{7 (\Delta-j) (\Delta+j +1) (2 \Delta -1) (2 \Delta +7)} \notag\\
&+ \frac{16 (j-3) (j-2) (j-1) j (\Delta-j-5) (\Delta-j+3) (\Delta-j+1) (\Delta-j-1) g_{\Delta +4,j-4}(u,v)}{(2 j-5) (2 j-3) (2 j-1) (2 j+1) (\Delta-j-6) (\Delta-j+4) (\Delta-j+2) (\Delta-j)} \notag\\
&+ \frac{64 (j-2) (j-1) j (j+1) (\Delta-j+3) (\Delta-j+1) (\Delta-j-1) (\Delta+j ) g_{\Delta +4,j-2}(u,v)}{7 (2 j-5) (2 j-1) (2 j+1) (2 j+3) (\Delta-j+4) (\Delta-j+2) (\Delta-j) (\Delta+j +1)} \notag\\
&+ \frac{288}{35 (2 \Delta +1) (2 \Delta +3) (2 \Delta +7) (2 \Delta +9) (2 j-3) (2 j-1) (2 j+3) (2 j+5)} \Big[\notag\\
& 8 \Delta ^2 (\Delta +5)^2 j (j+1) (4 j (j+1)-13)+40 \Delta  (\Delta +5) j (j+1) (7 j (j+1)-24)\notag\\
&+ 3 (15 (\Delta +1) (\Delta +2) (\Delta +3) (\Delta +4)+j (j+1) (191 j (j+1)-702))\Big] \notag\\
&\times \frac{(\Delta-j+1) (\Delta-j-1) (\Delta +j) (\Delta +j+2)}{(\Delta-j+2) (\Delta-j ) (\Delta +j+1) (\Delta +j+3)} g_{\Delta +4,j}(u,v) \notag \\
&+ \frac{64 j (j+1) (j+2) (j+3) (\Delta-j-1) (\Delta+j ) (\Delta+j +2) (\Delta+j +4)}{7 (2 j-1) (2 j+1) (2 j+3) (2 j+7) (\Delta-j) (\Delta+j +1) (\Delta+j +3) (\Delta+j +5)}g_{\Delta +4,j+2}(u,v) \notag\\
&+ \frac{16 (j+1) (j+2) (j+3) (j+4) }{(2 j+1) (2 j+3) (2 j+5) (2 j+7) } \notag\\
&\times \frac{(\Delta+j ) (\Delta+j +2) (\Delta+j +4) (\Delta+j +6) }{ (\Delta+j +1) (\Delta+j +3) (\Delta+j +5) (\Delta+j +7)} g_{\Delta +4,j+4}(u,v) \notag\\
&+ \frac{256 (\Delta +2) (\Delta +3)^2 (\Delta +5)   (\Delta +4)^2}{7 (2 \Delta +3) (2 \Delta +5) (2 \Delta +7)^2 (2 \Delta +9) (2 \Delta +11)} \notag\\
&\times \frac{(\Delta-j+3) (\Delta-j+1) (\Delta-j-1) (\Delta+j ) (\Delta+j +2) (\Delta+j +4)}{(\Delta-j+4) (\Delta-j+2) (\Delta-j) (\Delta+j +1) (\Delta+j +3) (\Delta+j +5)} g_{\Delta +6,j}(u,v) \notag\\
&+ \frac{(\Delta-j+1) (\Delta-j-1)(\Delta-j+3) (\Delta-j +5) (\Delta+j ) (\Delta+j +2) (\Delta+j +4) (\Delta+j +6) }{(\Delta-j-6) (\Delta-j+4) (\Delta-j+2) (\Delta-j) (\Delta+j +1) (\Delta+j +3) (\Delta+j +5) (\Delta+j +7))} \notag\\
&\times \frac{256 (\Delta +3)^2 (\Delta +5)^2 (\Delta +6)^2  (\Delta +4)^2}{ (2 \Delta +5) (2 \Delta +7)^2 (2 \Delta +9)^2 (2 \Delta +11)^2 (2 \Delta +13)} g_{\Delta +8,j}(u,v) \ed \label{A00long}
\end{align}

\section{Recurrence Relations}
\label{DGSD}
In this section we collect various recurrence relations that were derived in \cite{Dolan:2004mu} and used in section \ref{DGSblocks} to derive the superconformal blocks.  We also correct various mistakes in Appendix~D of \cite{Dolan:2004mu}, some of which lead to inconsistencies with known results in four dimensions.

Define
\begin{align}
F_{r,s}(x,\bar{x}) &\equiv D(\Delta,j,r,s) g_{\Delta+r+s,j+r-s}(x,\bar{x}) \ec\\
D(\Delta,j,r,s)  &\equiv (-4)^{r+s} \frac{(j+2\veps)_{r-s}}{(j+\veps)_{r-s}} B^{\sgn r}_{\frac{1}{2}(\Delta+j),r} B^{\sgn s}_{\frac{1}{2}(\Delta-j)-\veps,s} \notag\\
&\times A_{j+1, -\frac{1}{2}\left(\sgn(r-s)-1\right)(r-s)} A_{2-\Delta, -\frac{1}{2}\left(\sgn(r+s)+1\right)(r+s)} \ec
\end{align}
where
\begin{align}
A_{\lambda,t} \equiv \frac{(\lambda+\veps)_t(\lambda+\veps-1)_t}{(\lambda)_t(\lambda+2\veps-1)_t} \ecq\! B^+_{\lambda,t} \equiv 16^{-t} \frac{(\lambda)_t(\lambda+\veps-1)_t}{(\lambda-\frac{1}{2})_t(\lambda+\frac{1}{2})_t} \ecq\! B^-_{\lambda,t} \equiv \frac{(\lambda)_t}{(\lambda+1-\veps)_t} \ed
\end{align}

The following recurrence relations hold
\begin{align}
u^{2\veps}\Delta_{\veps} \frac{v-1}{u} \Delta_{\veps}^{-1} g_{\Delta,j} &= F_{-1,0} + F_{0,-1} + F_{0,1} + F_{1,0} \ec\\
\frac{1}{2}u^{2\veps}\Delta_{\veps} \frac{v+1}{u} \Delta_{\veps}^{-1} g_{\Delta,j} &= F_{-1,-1} + F_{-1,1} + F_{1,-1} + F_{1,1} \notag\\
&+ \frac{1}{4}\left(1-\frac{1}{2}\veps(\veps-1)\left(\cA_{j+1} + \cA_{2-\Delta} - (2\veps-1)(2\veps-3)\cA_{j+1}\cA_{2-\Delta}\right)\right)F_{0,0} \ec\\
u^{2\veps}\Delta_{\veps} \frac{(v-1)^2}{u^2} \Delta_{\veps}^{-1} g_{\Delta,j} &= F_{-2,0} + F_{0,-2} + F_{0,2} + F_{2,0} \notag\\
&+ 2\left(1-\veps(\veps-1)\cA_{j+1}\right)\left(F_{-1,-1,}+F_{1,1}\right) + 2\left(1-\veps(\veps-1)\cA_{2-\Delta}\right)\left(F_{-1,1}+F_{1,-1}\right) \notag\\
&+ C_{\Delta,j} F_{0,0} \ec\\
\frac{1}{2}u^{2\veps}\Delta_{\veps} \frac{v^2-1}{u^2} \Delta_{\veps}^{-1} g_{\Delta,j} &= F_{-2,-1} + F_{-1,-2} + F_{-2,1} + F_{1,-2} + F_{-1,2} + F_{2,-1} + F_{1,2} + F_{2,1} \notag\\
&+ a_{\Delta,j}F_{0,-1} + a_{\Delta,-j-2\veps}F_{-1,0} + a_{2\veps+2-\Delta,j} F_{1,0} + a_{2\veps+2-\Delta,-j-2\veps} F_{0,1} \ec
\end{align}
\begin{align}
\frac{1}{2}u^{2\veps}\Delta_{\veps} \frac{(v+1)^2}{u^2} \Delta_{\veps}^{-1} g_{\Delta,j} &= F_{-2,-2} + F_{-2,2} + F_{2,-2} + F_{2,2} \notag\\
&+ \frac{1}{8}\left(1-(2\veps-1)(2\veps-3)\cB_{l+\veps+1-\Delta}\right)\left(F_{-2,0}+F_{2,0}\right) \notag\\
&+ \frac{1}{8}\left(1-(2\veps-1)(2\veps-3)\cB_{\Delta+j-\veps-1}\right)\left(F_{0,-2}+F_{0,2}\right) \notag\\
&+ e_{\Delta,j}F_{-1,-1} + e_{-j+3,1-\Delta}F_{-1,1} + e_{-j+1,1-\Delta} F_{1,-1} + e_{2+\Delta,j}F_{1,2} \notag\\
&+ D_{\Delta,j}^{(\veps)}F_{0,0} \ed
\end{align}
Here, we used the definitions
\begin{align}
\cA_{\lambda} = \frac{1}{(\lambda+\veps)(\lambda+\veps-2)} \ecq \cB_{\lambda} = \frac{1}{(\lambda+\veps+2)(\lambda+\veps-2)} \ecq \cC_{\lambda} = \frac{1}{(\lambda+\veps+1)(\lambda+\veps-2)} \ec
\end{align}
and
\begin{align}
a_{\Delta,j} &= \frac{1}{8}\left(3-\frac{3}{2}\veps(\veps-1)\left(\cC_{j+1}+\cC_{2-\Delta}\right)-(2\veps-1)(2\veps-3)\cB_{\Delta+j-\veps-1} \right.\notag\\
&\left.+ \veps(\veps-1)(2\veps-1)(2\veps-3)\left( \cC_{j+1}\cC_{2-\Delta} + \frac{1}{2}\cB_{\Delta+j-\veps-1}\left(\cC_{j+1}+\cC_{2-\Delta}-10\cC_{j+1}\cC_{2-\Delta}\right)\right)\right) \ec\\
e_{\Delta,j} &= \frac{1}{2}\left( 1 - \frac{1}{2}\veps(\veps-1)\left(\cA_{j+1}+\cB_{2-\Delta} - (2\veps-1)(2\veps-3)\cA_{j+1}\cB_{2-\Delta}\right)\right) \ec \\
C_{\Delta,j} &= \frac{1}{4}\left( 1 - \frac{1}{2}(2\veps-1)(2\veps-3)\left(\cB_{j+\veps+1-\Delta} + \cB_{\Delta+j-\veps-1}\right)\right. \notag\\
&+ 2\veps(\veps-1)\left(\veps(\veps-1)\cA_{j+1}\cA_{2-\Delta} - \cA_{j+1} - \cA_{2-\Delta}\right) \notag\\
&\times\left.\left( \frac{1}{2} - (2\veps-1)(2\veps-3)\left( \frac{1}{4}\cB_{j+\veps+1-\Delta} + \frac{1}{4}\cB_{\Delta+j-\veps-1} - 3\cB_{j+\veps+1-\Delta}\cB_{\Delta+j-\veps-1}\right)\right)\right)
\end{align}

For $D_{\Delta,j}^{(\veps)}$ we can write the results for specific dimensions:
\begin{align}
D_{\Delta,j}^{(1/2)} &= \frac{9}{512 (2 \Delta -7) (2 \Delta +1)}+\frac{124 j(j+1)-77}{512 (2 \Delta -5) (2 \Delta -1) (2 j-1) (2 j+3)} \notag\\
&+\frac{40 (4 j (j+1)-17) (j+1) j+327}{128 (2 j-3) (2 j-1) (2 j+3) (2 j+5)} \ec
\end{align}
\begin{align}
D_{\Delta,j}^{(1)} &= \frac{1}{64} \left(\frac{4 j^2-3}{(\Delta -j-1) (\Delta -j-5) (\Delta +j-3) (\Delta +j+1)} + \frac{2}{(\Delta -j-5) (\Delta +j+1)}+5\right) \ec\\
D_{\Delta,j}^{(3/2)} &= \frac{1}{512} \left( \frac{20 (8 j (j+3) (4 j (j+3)-5)-37)}{(2 j-1) (2 j+1) (2 j+5) (2 j+7)}  - \frac{3 (140 (j+3) j+127)}{(2 \Delta -7) (2 \Delta -3) (2 j+1) (2 j+5)} \right.\notag\\
&\left. - \frac{15}{4 (\Delta -5) \Delta +9}\right) \ec\\
D_{\Delta,j}^{(2)} &= \frac{3 (j-1) (j+3) (4 j (j+2)-3)}{512 (j+1)^2 (j+2) (\Delta -j-3) (\Delta +j-3)} - \frac{3 (j+1) (j+5) (4 (j+6) j+29)}{512 (j+2) (j+3)^2 (\Delta -j-7) (\Delta +j+1)}  \notag\\
&- \frac{j (j+4) (5 (j+4) j+8)}{32 (\Delta -4) (\Delta -2) (j+1)^2 (j+3)^2}+\frac{5 ((j+4) j+1)}{64 (j+1) (j+3)} \ed
\end{align}

\section{Details of Central Charge Computation}
\label{cTdetail}
Here we lay out details of computations in Section~\ref{centralcharge}. 
The theories we consider have a natural parity transformation that flips the sign of $k$, so we choose $k>0$ without loss of generality.

 Recall that the three-sphere partition function of ABJ(M) theories \eqref{ZSphereABJ(M)} is given by \cite{Kapustin:2009kz, Marino:2011nm}
 \begin{align}
Z(\Delta)&= \cN_{M,N} \int d^M \lambda\,  d^N \mu \,  e^{i \pi k \left[\sum_i\lambda_i^2- \sum_j \mu_j^2 \right] }  \prod_{i < j} \left[2\sinh [ \pi(\lambda_{i}-\lambda_{j})]\right]^2
 \prod_{i< j}\left[2 \sinh[2 \pi(\mu_{i}-\mu_{j})] \right]^2\notag \\& \quad \times \prod_{\alpha} f_\alpha(\Delta) \, , \label{AppZSphereABJ(M)}
 \end{align}
where 
\es{Zdefs}{
\cN_{M,N} &\equiv \frac{i^{-(M^2-N^2) \sgn(k)/2}}{M! \, N!}, \\
 f_{\alpha} (\Delta) &\equiv  \prod_{i, j} \exp\left(\ell \left(1-\Delta_\alpha + i \, (-)^\alpha \, \left(\lambda_i-\mu_j\right)\right) \right)\,, \\
 \ell(z) &\equiv -z\ln(1-e^{2\pi i z})+\frac{i}{2}\left( \pi z^2 +\frac{1}{\pi}\text{Li$_2$}(e^{2\pi i z}) \right)-\frac{i\pi}{12}\,. 
}
The function $\ell(z)$ satisfies $\ell'(z) = - \pi z \cot (\pi z)$.  Using the property of $\ell(z)$ at $\Delta=\Delta^* =1/2$, one can show that
\es{ellId}
{
 \ell(z) + \ell(z^*) = - \ln \left(2 \cosh (\pi \theta)  \right) \qquad \text{for} \quad z= \frac{1}{2} + i \theta\,.
}
When $\Delta = \Delta^*$, the product over $f_{\alpha} (\Delta) $ in \eqref{AppZSphereABJ(M)} can be written simply as
\es{productfalpha}{
  \prod_\alpha f_{\alpha} (\Delta^*) = \prod_{i,j} \frac{1}{\left[ 2\cosh \pi ( \lambda_i - \mu_j ) \right]^2} \, .
}
For ABJM theories with $M=N$, we use the Cauchy identity
\es{CauchyID}{
\frac{\prod_{i<j} \sinh\left[ \pi(\lambda_i-\lambda_j)\right]  \sinh\left[\pi(\mu_i-\mu_j)\right] }{\prod_{i,j}  \cosh \left[ \pi(\lambda_i-\mu_j)\right] } = \sum_{ \rho \in S_N} (-)^\rho  \prod_i^N \frac{1}{ \cosh  \left[ \pi(\lambda_i-\mu_{\rho(i)})\right]}\,,
}
to rewrite the three-sphere partition function of undeformed ABJM theory as
\es{ZSphereABJM}{
Z(\Delta^*)
&=\frac{1}{2^{2N} \,  N! } \int d^N \lambda \, d^N \mu\sum_{ \rho \in S_N} (-)^\rho \prod_i^N \frac{ \exp(i \pi k(\lambda_i^2- \mu_i^2)) }{ \cosh  \left[\pi(\lambda_i-\mu_i)\right] \cosh  \left[ \pi(\lambda_i-\mu_{\rho(i)})\right]}\,.}

\subsection{$U(2) \times U(2)$ ABJM Theory }\label{N2Detail}
We now provide more details on the computation given in Section~\ref{section:N2ABJMBLG}.  In the end, we want to compute  $\partial^2 F/\partial t^2_a \Big|_{t=0}=-Z^{-1}\partial^2 Z/\partial t^2_a \Big|_{t=0}$ so the overall normalization of $Z$ is irrelevant, and we will not keep track of it. From \eqref{AppZSphereABJ(M)}, we have
\es{N2Part}{
Z(\Delta)&=\int d\lambda_1 d\lambda_2d\mu_1d\mu_2\exp\left[ i \pi k \left( \lambda_1^2+\lambda_2^2-\mu_1^2-\mu_2^2\right)\right] \\
&\quad\qquad\times\sinh^2 \left[ \pi \left(\lambda_1-\lambda_2\right) \right] \sinh^2 \left[\pi\left( \mu_1-\mu_2 \right)\right]\prod_\alpha f_\alpha(\Delta)\,.
}
Schematically, we get
\es{scheme}{
\frac{\partial^2 Z}{\partial t^2_a }\Bigg|_{t=0}=\int d\lambda_1 d\lambda_2d\mu_1d\mu_2\, Z_{\textrm{int}}(\Delta_*)\cD_{a} \,, \qquad
  {\cal D}_a \equiv   \frac{1}{\prod_\alpha f_{\alpha}} \frac{\partial^2}{\partial t_a^2} \prod_{\alpha } f_{\alpha} \Bigg|_{t=0} \,,
}
where ${\cal D}_a$ can be read off from \eqref{fderiv1}--\eqref{fderiv}, and $Z_{\textrm{int}}(\Delta_*) $ is the integrand of \eqref{ZSphereABJM} . We will first ignore the second term in \eqref{fderiv1}, which is shown later to cancel anyway. Defining the following variables 
\es{subsA}{
x=\lambda_1-\mu_2, \quad y=\lambda_1-\mu_1, \quad z=\lambda_2-\mu_2, \quad w=\lambda_1+\mu_1 \,,
}
$w$ only appears in the exponential giving us a $\delta$-function that sets $y+z=0$. Combining both terms for $a \neq 1$,
\es{U2ZderivA}{
 \frac{\partial^2 Z }{\partial t_a^2}  \Bigg |_{t=0} &=\frac{ \pi ^2}{8 k }\int dx dy  \, e^{ 2 i \pi k x y } \, 
 \sech^2(\pi y) \left[ \sech^4(\pi y)-\sech^4(\pi  x )\right]\, \\
&=\frac{\pi^2}{8 k^2}\left( 1- 2 k^2 \int^{\infty}_{-\infty} dy \, y \sech^4(\pi y) \csch(\pi ky)\right)\,\\
&=\frac{\pi^2}{4}\ \left(2\int^{\infty}_{-\infty} dy \, y\tanh^2(\pi y) \csch(\pi ky)-\int^{\infty}_{-\infty} dy \, y\tanh^4(\pi y) \csch(\pi ky) \right)\,.
}
Similarly for  $Z(\Delta^*)$:
\es{U2ZA}{
Z(\Delta^*)&=\frac{ 1}{32 k }\int dx dy  \, e^{ 2 i \pi k x y } \, 
 \sech^2(\pi y) \left[ \sech^2(\pi y)-\sech^2(\pi  x )\right]\, \\
&=\frac{1}{32 k^2}\left( 1- 2 k^2 \int^{\infty}_{-\infty} dy \, y \sech^2(\pi y)  \csch(\pi ky)\right)\,\\
&=\frac{1}{16}\ \int ^{\infty}_{-\infty}dy \, y\tanh^2(\pi y) \csch(\pi ky) \,.
}
Thus we get the central charge give in \eqref{U2final}
\es{cTU2}{
c_T=  \frac{8}{\pi^2} \Re \frac{1}{Z(\Delta^*)}\frac{\partial^2 Z }{\partial t_{a} \partial t_{a}}  \Big |_{t=0} &  = 32 \left( 2- \frac{I_4}{I_2} \right),}
with $ I_n \equiv \int_{-\infty}^{\infty} dy  \, y \tanh^{n}(\pi y) \csch(\pi k y)$ and $a=2,3$.

Lastly, we show that the second term in \eqref{fderiv1} does not contribute to $\tau_{11}$.  Up to normalization, in terms of the variables defined in \eqref{subsA}, its contribution would be
\es{asymmN2}{ \left[ \sum^2\limits_{i,j} \tanh \left[ \pi (\lambda_i - \mu_j) \right] \right]^2 = -2 \left[\tanh (\pi x)+\tanh (\pi y)+\tanh (\pi  z)+\tanh (\pi  (-x+y+z))\right]^2\,.
}
Due to $w$ independence we again obtain a delta function integration setting $y+z=0$. Since $\tanh(-x)= - \tanh x$, one sees that \eqref{asymmN2} vanishes.

\subsection{Relating three-sphere partition function of ABJM to BLG}\label{ABJMtoBLG}
First, start with the $S^3$ partition function \eqref{N2Part} of $U(2)_k \times U(2)_{-k}$ ABJM theory\footnote{The computations in this appendix are similar to the ones performed in \cite{Honda:2012ik}.}
\es{ZDeltaU2}{
Z(\Delta^*)_{U(2)}=\frac{1}{32}\int &d\lambda_1 d\lambda_2d\mu_1d\mu_2\exp\left[i\pi k\left( \lambda_1^2+\lambda_2^2-\mu_1^2-\mu_2^2\right)\right] \\
&\quad  \times\left[ \sech^2\left[\pi(\lambda_1-\mu_1)\right]\sech^2\left[\pi(\lambda_2-\mu_2)\right]-\prod^2_{ij}\sech\left[\pi(\lambda_i-\mu_j)\right]\right]. \\
}
To get the $S^3$ partition function of the $SU(2)_k \times SU(2)_{-k}$ BLG theory one needs to factor out the diagonal $U(1)\times U(1)$ contribution. Introducing the following variables
\es{UtoSUcov}
{
x = \lambda_1 + \lambda_2 + \mu_1 + \mu_2, \quad y =  \lambda_1 + \lambda_2 - \mu_1 - \mu_2, \quad \lambda_- = \lambda_1 - \lambda_2, \quad \mu_- = \mu_1 - \mu_2\,,
} 
the three-sphere partition function becomes
\es{ZDeltaU2Again}{
Z(\Delta^*)_{U(2)}&= \frac{1}{256}\int dx \,dy \,d\lambda_- \,d\mu_- \exp\left[i \pi k \left(x \, y + \lambda_-^2 -\mu_-^2\right)/ 2 \right] \cF(y, \lambda_-, \mu_-) \,,}
where $\cF$ is an $x$ independent function given by
\es{calFDef}{
\cF&(y, \lambda_-, \mu_-)\equiv \sech\left[\pi  (y+ \eta_+)/2\right] \sech\left[ \pi  (y-\eta_-)/2\right]\\
&\times  \left[\sech\left(\pi  (y+ \eta_+)/2\right) \sech\left( \pi  (y-\eta_-)/2\right)-\sech\left( \pi (y + \eta_- )/2\right) \sech\left(\pi (y-\eta_-)/2\right)\right]\,,\\
}
with $ \eta_\pm \equiv  \lambda_- \pm \mu_-$. Since the only $x$  dependence is in the exponential factor,  integrating $x$ gives a $\delta$-function which sets $y=0$. 
Then we get
\es{ZDeltaStarU2}{
Z(\Delta^*)_{U(2)}
&= \frac{1}{16 k}\int d\lambda_- d\mu_- \exp\left[2 i \pi k \left( \lambda_-^2 -\mu_-^2\right) \right]\frac{\sinh^2 ( 2 \pi \lambda_- ) \sinh^2 (2 \pi \mu_-)} {\cosh^4\left[\pi  (\lambda_- -\mu_-)\right] \cosh^2\left[ \pi  (\lambda_-+\mu_-)\right]}\,.
}
Recall that the $SU(2)$ theory three-sphere partition function is obtained by restricting the Cartan elements of the $U(2)$ three-sphere partition function
\es{ZDeltaStarSU2}{
&Z(\Delta^*)_{SU(2)}\\
&\qquad =\frac{1}{8}\int  d\lambda_1 \, d\lambda_2\, d\mu_1\, d\mu_2 \,  d\alpha \, d\beta  \, e^{2 \pi i\left[ \alpha \,(\lambda_1+\lambda_2) + \beta \, (\mu_1 +\mu_2)\right]}\,  e^{i\pi k\left( \lambda_1^2+\lambda_2^2-\mu_1^2-\mu_2^2\right)}\\
&\qquad \qquad\times   \left[ \sech^2\left[\pi(\lambda_1-\mu_1)\right]\sech^2\left[\pi(\lambda_2-\mu_2)\right]-\prod^2_{ij}\sech\left[\pi(\lambda_i-\mu_j)\right]\right]  \\
&\qquad=\frac{1}{8}\int d\lambda_1 \, d\mu_1 \exp\left[ 2 i \pi k \left( \lambda_1^2 -\mu_1^2\right) \right] \frac{\sinh^2 ( 2 \pi \lambda_1) \sinh^2 (2 \pi \mu_1)}  {\cosh^4\left[\pi  (\lambda_1 -\mu_1)\right] \cosh^2\left[\pi  (\lambda_1+\mu_1)\right]}\,.
}
Thus we find\footnote{This procedure generalizes to $Z(\Delta^*)_{SU(N)} = N k \, Z(\Delta^*)_{U(N)}$ \cite{MPUnpublished}.}
\es{ZDeltaStarSU2RelU2}{
Z(\Delta^*)_{SU(2)} = 2 k \, Z(\Delta^*)_{U(2)}\,.
}
Similarly, one can look at how the second derivative of $Z$ with respect to $t$ is related between the $U(2)$ and $SU(2)$ theories.  These derivatives are obtained by multiplying the previous integrand by $2 \pi^2 \sum_{i,j} \textrm{sech}^2 \left[  \pi (\lambda_i - \mu_j) \right]$.  Using a similar procedure of reducing to the form of $SU(2)_k \times SU(2)_{-k}$ theory one can show that 
\es{DerivativeZSU2}{
\frac{\partial^2  }{\partial t_a^2} Z(\Delta^*)_{SU(2)} = 2 k \, \frac{\partial^2  }{\partial t_a^2} Z(\Delta^*)_{U(2)} \,.
}
From \eqref{cTvialocalization} we conclude that the central charges of $U(2)_{k} \times U(2)_{-k}$ theories and $SU(2)_{k} \times SU(2)_{-k}$ theories are equal.

\subsection{$U(2) \times U(1)$  ABJ Theory}\label{ABJ21Detail}
Recall the three-sphere partition function \eqref{AppZSphereABJ(M)} for ABJ(M) theories with gauge group $U(M)_{k}\times U(N)_{-k}$:
\es{ZSphereABJ}{
Z(\Delta^*)&={\mathcal N}_{M,N}\int d^M \lambda\,  d^N \mu \,  e^{i \pi k \left[\sum_i\lambda_i^2- \sum_j \mu_j^2 \right] }  \frac{ \prod_{i < j} \left[2\sinh [ \pi(\lambda_{i}-\lambda_{j})]\right]^2
 \prod_{i< j}\left[2 \sinh[2 \pi(\mu_{i}-\mu_{j})] \right]^2}{\prod_i^M\prod_j^N  \left[ 2 \cosh ( \pi(\lambda_i-\mu_j)) \right]^2}\,. }
 For the $U(2)_k \times U(1)_{-k}$ theory, we have
\es{ZDeltaStarU2U1}{
Z(\Delta^*)&=\frac{{\mathcal N}_{2,1}}{4}\int d^2 \lambda d \mu_1 \,  \exp\left[ i \pi k (\lambda_1^2+\lambda_2^2- \mu_1^2)\right] \left[\frac{ \sinh(\pi(\lambda_1-\lambda_2))}{\cosh ( \pi(\lambda_1-\mu_1))\cosh ( \pi(\lambda_2-\mu_1))} \right]^2\,.
 }
Using the following variables
\es{subsABJ21}{
x=\lambda_1-\mu_1\,, \qquad y=\lambda_2- \mu_1 \,,\qquad \text{and} \qquad z=\lambda_2+\mu_1\,,
}
the three-sphere partition function becomes
\es{ZDeltaStarU2U1Again}{Z(\Delta^*)&=\frac{{\mathcal N}_{2,1}}{8}\int dx \, dy\,  dz \,  \exp\left[ i \pi k \left(z^2 /4 +\left(x+ y/2 \right)z+ x^2 -xy +y^2/4\right) \right]  \cH_1(x,y)\,,
 }
 where $\cH_1(x,y) \equiv  \left[ \tanh(\pi x) - \tanh(\pi y) \right]^2$. The integral in $z$ is Gaussian and gives 
 \es{ZDeltaStarH1}{Z(\Delta^*)&= \frac{{\mathcal N}_{2,1}}{4} \sqrt{\frac{i}{k}} \int dx \, dy \,   e^{- 2\pi i k \, x \, y}    \cH_1(x,y)\,.
 }
Only the cross-term in $\cH_1(x,y)$, namely $-2\tanh(\pi x)\tanh(\pi y)$, contributes to this integral:
 \es{ZDeltaStarH1FT}{
 Z(\Delta^*) &= \frac{{\mathcal N}_{2,1}}{2} \frac{\,  i^{3/2}}{k^{1/2}} \int^{\infty}_{-\infty}  dy \,  \frac{\tanh (\pi y)}{ \sinh (\pi k  y)}
 =\frac{1}{4 \sqrt{k}} \int^{\infty}_{-\infty} dy \,  \tanh (\pi y) \csch( \pi k y) \,,
}
This expression reproduces the results in \cite{Awata:2012jb, Honda:2014npa}.

For $\partial^2 Z/\partial t^2_a \Big|_{t=0}$, the only change is that we get $\cH_2(x,y)$ instead of $\cH_1(x,y)$
  \es{ZDeltaStarH2}{
   \left. \frac{\partial^2 Z}{\partial t_a^2} \right \rvert_{t=0}&= \frac{{\mathcal N}_{2,1}}{4} \sqrt{\frac{i}{k}}\ \int dx \, dy \, e^{- 2\pi i k \, x \, y}  \cH_2(x,y)\, .
 }
where $\cH_2(x,y)  \equiv  2\pi^2 \, \left[ \tanh(\pi x) - \tanh(\pi y) \right]^2 \left[\sech^2(\pi x)+\sech^2(\pi y)\right]$. Expanding $\cH_2$ we see that there are three categories of terms that are doubly degenerate due to the symmetry that interchanges $x$ and $y$. The first category is of the form $ h_2^{(1)}\equiv\tanh ^2(\pi  y)\,  \sech^2(\pi  y)$, which does not contribute to the integral. The second category is of the form $ h_2^{(2)}=\ \tanh ^2(\pi  x) \, \sech^2(\pi  y)$.  The $x$ integral is just a Fourier transform and in this case it gives
\es{ZDeltaStarH2FT}{
 \int dx \, dy \,  e^{- 2\pi i k \, x \, y} h_2^{(2)} (x,y)  = 
2 k \int^{\infty}_{-\infty} dy   \, y \tanh ^2(\pi  y) \, \csch (\pi  k y)\,.
}
The last category consists of $h_2^{(3)} \equiv -2 \tanh (\pi  x)\,  \tanh (\pi  y)\,  \sech^2(\pi  y)$.  Performing the $x$ integral yields
\es{ZDeltaStarH3}{
 \int dx \, dy \,  e^{- 2\pi i k \, x \, y} h_2^{(3)} (x,y)
= 2 i \int^{\infty}_{-\infty} dy \,   \tanh (\pi  y) \, \sech^2(\pi  y) \, \csch(\pi  k y)\,.
}
Combining all these terms we get (for $a=2,3$)
\es{DerZDeltaStarU2U1}{
 \left. \frac{\partial^2 Z}{\partial t_{a}^2} \right \rvert_{t=0}&=  \frac{{\mathcal N}_{2,1}}{4} \sqrt{\frac{i}{k}}\ \int dx \, dy \,  e^{- 2\pi i k \, x \, y} \left(4\pi^2\right (h_2^{(1)} (x,y)+ h_2^{(2)} (x,y)+ h_2^{(3)} (x,y))) \,\\ 
 &=  \frac{\pi^2} { k^{1/2} }  \int^{\infty}_{-\infty} dy   \, \left[ \frac{\tanh (\pi y)}{\cosh^2 (\pi y)} - i k y \tanh^2 (\pi y) \right]  \csch (\pi  k y)  \,.  
}
Thus the central charge of $U(2)_k \times U(1)_{-k}$ ABJ theory \eqref{cTABJ21} is 
\es{cTU2U1}{
c_T =  \frac{8}{\pi^2} \Re \frac{1}{Z(\Delta^*)}\frac{\partial^2 Z }{\partial t_{a} \partial t_{a}}  \Big |_{t=0}  = 32 \, \frac{ \int^{\infty}_{-\infty} dy  \tanh (\pi y)  \csch (k \pi y) \sech^2 ( \pi y)  }{\int^{\infty}_{-\infty} dy  \tanh (\pi y) \csch (k \pi y) }\,,
}
again for $a=2,3$.

Finally, for $a=1$, we should check the extra term from  $  \frac{\partial^2}{\partial t_1^2 } \prod\limits_{\alpha} f_{\alpha} |_{t=0}$ does not contribute to the central charge computation. Following the same procedure as above, we get
  \es{D1MinusD2}{
\cD_{1}- \cD_{2}&= \frac{{\mathcal N}_{2,1}}{4} \sqrt{\frac{i}{k}} \int dx \, dy \,   e^{- 2\pi i k \, x \, y} \, \cH_3(x,y)\,,
  }
  where $\cH_3(x,y) = 8\pi^2  \left[ \tanh^2(\pi x) - \tanh^2(\pi y) \right]^2$. As with the $\cH_1$ example, we see that only the cross term contributes giving us
    \es{D1MinusD2Again}{\cD_{1}- \cD_{2}&= 8\pi^2 k  \, {\mathcal N}_{2,1}  \sqrt{\frac{i}{k}} \int^{\infty}_{-\infty}  dy \,  y \, \frac{ \tanh ^2(\pi  y)}{ \sinh (\pi  k y)} = - 4\pi^2 i  \sqrt{k} \,  I_2 \,.
  }
where $I_2$ is defined in \eqref{U2final} and the result is given by \eqref{I_n Series}. Notice that this extra term is non zero, nevertheless it is purely imaginary.  It has no effect on the central charge as \eqref{cTvialocalization} depends only on the real part.

\bibliographystyle{ssg}
\bibliography{N8D3Bootstrap}

\end{document}